%% file: 00_main.tex
\documentclass[acmtog]{acmart}
\acmSubmissionID{296}

\usepackage{float}
\usepackage{stfloats}
\usepackage{animate}
\usepackage{dtof}
\usepackage{siunitx}
\DeclareSIUnit{\nothing}{\relax}
\usepackage{amsmath}

\newcommand\blfootnote[1]{%
  \begingroup
  \renewcommand\thefootnote{}\footnote{#1}%
  \addtocounter{footnote}{-1}%
  \endgroup
}

\usepackage{url}                             % for nicer URLs in bibliography
\usepackage[nameinlink,capitalise]{cleveref} % easy cross-referencing (option noabbrev)
\crefname{section}{Sec.}{Secs.}
\crefname{table}{Tab.}{Tabs.}
\usepackage{microtype}                       % nicer typesetting, usually saves space
\usepackage{wrapfig}                         % for wrapping text around figures
\usepackage{booktabs}                        % better tables
\usepackage{tabularx}
\usepackage{cancel}                          % canceling terms
\usepackage{mathtools}                       % for \underbracket
\usepackage{xfrac}                           % nicer fractions
\usepackage{nicefrac}                        % another set of nicer fractions
\usepackage[shortlabels,inline]{enumitem}    % inline enums
\setlist[enumerate]{label=\arabic*.}
\usepackage{ellipsis}                        % fix ellipsis spacing
\usepackage{xspace}                          % fix macro spacing
\usepackage{bm}                              % provide bold symbols

\setcopyright{acmlicensed}
\acmJournal{TOG}
\acmYear{2023} \acmVolume{42} \acmNumber{6} \acmArticle{271} \acmMonth{12} \acmPrice{15.00}\acmDOI{10.1145/3618335}

\title{Doppler Time-of-Flight Rendering}

\author{Juhyeon Kim}
\affiliation{%
  \institution{Dartmouth College}
  \country{USA}
}
\email{juhyeon.kim.gr@dartmouth.edu}
\orcid{0000-0002-6218-3426}

\author{Wojciech Jarosz}
\affiliation{%
 \institution{Dartmouth College}
  \country{USA}
}
\email{wojciech.k.jarosz@dartmouth.edu}
\orcid{0000-0002-1652-0954}

\author{Ioannis Gkioulekas}
\affiliation{%
 \institution{Carnegie Mellon University}
  \country{USA}
}
\email{igkioule@cs.cmu.edu}
\orcid{0000-0001-6932-4642}

\author{Adithya Pediredla}
\affiliation{%
  \institution{Dartmouth College}
  \country{USA}
}
\email{adithya.k.pediredla@dartmouth.edu}
\orcid{0000-0002-6623-020X}

\begin{document}

\include{sections/teaser}
    \include{sections/abstract}




    \maketitle
    \input{sections/01_introduction_v1.3.tex}
    \input{sections/02_related_works_v1.1.tex}

    \input{sections/03_doppler_rendering_v3.5.tex}

    \input{sections/04_time_domain_antithetic_sampling_v3.3.tex}
    \input{sections/05_spatial_correlation_v3.6.tex}
    \input{sections/06_results_v3.5.tex}
    \input{sections/07_applications_v3.2.tex}
    \input{sections/conclusions.tex}

    \bibliographystyle{ACM-Reference-Format}
    \bibliography{strings-full,rendering-bibtex,additional}

\end{document}

%% file: sections/teaser.tex
\begin{teaserfigure}
    \centering
    \hspace*{3ex}\includegraphics[width=\columnwidth]{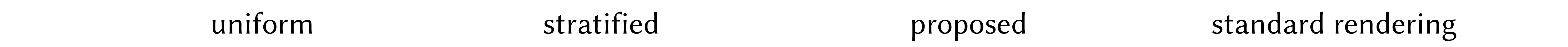}
    \includegraphics[width=0.06\columnwidth]{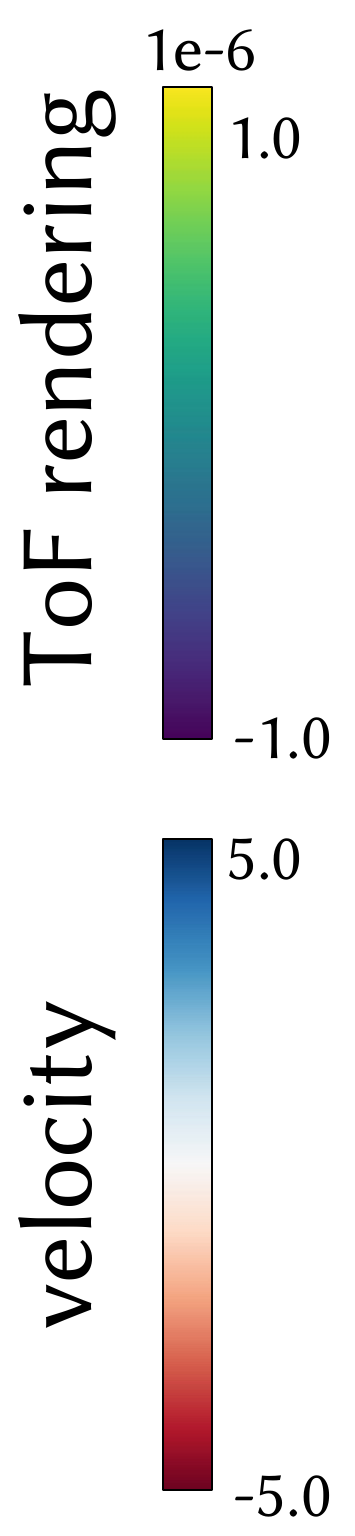}
    \animategraphics[width=0.90\columnwidth,autoplay,loop,poster=25]{10}{figures/video/larger_interval3/frame_}{0}{49}
    \hspace*{3ex}
    \includegraphics[width=\columnwidth]{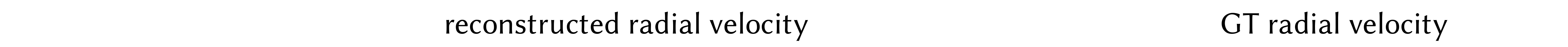}
   % \animategraphics[width=0.9\columnwidth,autoplay,loop,poster=35]{10}{figures/video/linear_interpolate_shorter_interval2/domino/merged/frame_}{0}{74}
    % \animategraphics[width=\columnwidth,autoplay,loop]{10}{figures/videos/bowling/merged/frame_}{24}{72}
    % \animategraphics[width=\columnwidth,autoplay,loop]{10}{figures/videos/domino/merged/frame_}{0}{148}
  % \includegraphics[width=\textwidth]{figures/0_teaser/teaser_v1.8.pdf}
  \caption{We propose an unbiased and efficient rendering algorithm for Doppler time-of-flight cameras.
  %
  % \protect\footnotemark\
  %
  Compared to naive sampling algorithms (left two columns), ours uses antithetic sampling with path correlation, and results in rendered images with orders of magnitude lower variance (top row). This makes it feasible to use rendering to evaluate radial velocity reconstruction algorithms (bottom row). (A teaser animation can be viewed in Adobe Acrobat Reader.)}
  \label{fig:teaser}
\end{teaserfigure}

%% file: sections/abstract.tex
\begin{abstract}
We introduce Doppler time-of-flight (D-ToF) rendering, an extension of ToF rendering for dynamic scenes, with applications in simulating D-ToF cameras. D-ToF cameras use high-frequency modulation of illumination and exposure, and measure the Doppler frequency shift to compute the \jkimb{radial} velocity of dynamic objects. The time-varying scene geometry and high-frequency modulation functions used in such cameras make it challenging to accurately and efficiently simulate their measurements with existing ToF rendering algorithms.
We overcome these challenges in a twofold manner: To achieve accuracy, we derive path integral expressions for D-ToF measurements under global illumination and form unbiased Monte Carlo estimates of these integrals. To achieve efficiency, we develop a tailored time-path sampling technique that combines antithetic time sampling with correlated path sampling.
We show experimentally that our sampling technique achieves up to two orders of magnitude lower variance compared to naive time-path sampling. 
We provide an open-source simulator that serves as a digital twin for D-ToF imaging systems, allowing imaging researchers, for the first time, to investigate the impact of modulation functions, material properties, and global illumination on D-ToF imaging performance.
\end{abstract}

%% file: sections/01_introduction_v1.3.tex
\section{Introduction}
The last decade has witnessed a proliferation of time-of-flight (ToF) rendering algorithms, which simulate various ToF cameras in a physically accurate manner. 
These algorithms have facilitated several improvements in ToF-based imaging systems for applications such as depth 
 sensing~\cite{Marco:2017:DeepToF, Su:2018:Deep, Po:2022:Adaptive}, non-line-of-sight imaging~\cite{Tsai:2019:Volumetric, Iseringhausen:2020:Nonlineofsight, Pediredla:2019:SNLOS}, and imaging through scattering~\cite{Raghuram:2019:STORM}. 
ToF rendering algorithms have also facilitated sensor design~\cite{Zhang:2022:First} and large dataset generation for supervised learning~\cite{Chen:2020:Learned, Gutierrez-Barragan:2021:IToF2dToF}.
\blfootnote{ - Project page : \url{https://juhyeonkim95.github.io/project-pages/dopplertof/}}

Doppler time-of-flight (D-ToF) cameras are a class of ToF cameras that use the Doppler effect to estimate the radial velocity of moving objects~\cite{Heide:2015:Doppler, Hu:2022:Differential}.
In contrast to inter-frame methods~\cite{Whyte:2015:Resolving}, D-ToF cameras can instantly evaluate radial velocity~\cite{Heide:2015:Doppler}.
This capability makes them ideal for scenarios requiring high-speed operation, such as industrial robotics, automobiles, and drones. 
Despite the practical importance of D-ToF cameras, efficient physically accurate rendering algorithms for them do not exist, hindering research and engineering efforts.

In particular, prior ToF rendering algorithms~\cite{Jarabo:2014:Framework, Ament:2014:Refractive, Marco:2017:Transient, Marco:2019:Progressive, Pediredla:2019:Ellipsoidal, Liu:2022:Temporally} assume static scenes, and therefore cannot handle the dynamic scenes that D-ToF cameras typically image.
At first glance, it may appear that combining ToF and motion blur rendering algorithms could enable the simulation of D-ToF cameras. However, as we explain in Section~\ref{sec:DToFPathIntegral}, dynamic scenes break key assumptions underlying ToF rendering algorithms, making these algorithms inefficient or even incorrect.
% dynamic nature will break some important assumptions, such as path length importance function, utilized in their work, making them rather inefficient or inapplicable.
Moreover, due to the use of high-frequency modulation in D-ToF cameras, the naive time sampling of motion blur rendering algorithms results in significant noise, overwhelming the signal from subtle net-intensity differences between temporally adjacent light paths that D-ToF cameras measure.
% However, these combined algorithms converge poorly as D-ToF cameras use high-frequency illumination and sensor (also known as shutter) modulation functions. 
% The Doppler signal is proportional to the small net difference between the large positive and negative intensity values of various light paths. 
% Na\"ively combining ToF and motion blur rendering algorithms do not sample light paths that importance sample this net difference and rather sample paths based on their absolute contribution leading to high variance in the rendered image. 

In this paper, we introduce the first efficient, physically based rendering framework for D-ToF cameras. We start by deriving a ToF path integral for dynamic scenes (\cref{sec:DToFPathIntegral}) that generalizes both ToF and motion blur rendering. 
We then simulate D-ToF imaging systems by estimating this integral, which we can do in an unbiased and efficient manner using Monte Carlo integration with importance sampling.
However, importance sampling the D-ToF integrand is non-trivial: the integrand is the product of the path throughput and a sinusoidal function that depends on both time %
%
% \footnote{Time refers to the exposure time instant, and its integral is the total exposure duration.} 
%
and each path's time of flight.
% (ratio of optical path length and speed of light).
Therefore, naive time-path sampling will result in large variance due to the random cancellation of contributions from the positive and negative lobes of the integrand as shown in \cref{fig:teaser}.
% For the commonly used sinusoidal illumination and sensing modulation functions, 
% The D-ToF integrand is a product of path throughput and a function periodic with respect to the sum of exposure time instant and time-of-flight (ratio of optical path length and speed of light). 
% The frequency of the integrated is proportional to the difference in the frequencies of illumination and sensor modulation functions. 
% Therefore, naive time-path sampling will result in a lot of paths whose sum of contributions is zero, resulting in a large Monte Carlo variance. 
% Ideally, we should only sample paths that measure the net difference between the positive and negative lobes of the integrand. 
We overcome these challenges by introducing an efficient two-step time-path sampling strategy: time-domain antithetic sampling (\cref{section:4_time_domain_antithetic_sampling}), combined with correlated light path sampling (\cref{section:5_spatial_correlation}).
% For each sampled time step namely \textit{primal} sample, we also sample an \textit {antithetic} time step from the opposite lobe using either shifted or mirrored strategy, assuming that these paths have approximately equal path throughput and time-of-flights (Section~\ref{section:4_time_domain_antithetic_sampling}). %we sample at both these sampled time steps 
% Then, to ensure the spatial similarity between the primal and antithetic samples, we develop material-selective path correlation technique that is generally efficient for scenes with a mix of both diffuse and specular materials (Section~\ref{section:5_spatial_correlation}).

We implement both CPU and GPU versions of our method using Mitsuba 0.6%
%
% \footnote{
% \url{https://github.com/juhyeonkim95/MitsubaDopplerToF}
% }
%
~\cite{Jakob:2013:Mitsuba} and Mitsuba 3%
%
% \footnote{
% \url{https://github.com/juhyeonkim95/Mitsuba3DopplerToF}
% }
%
~\cite{Jakob:2022:Mitsuba3}, and demonstrate its effectiveness on scenes with varying geometries and material parameters (\cref{sec:experiments}). We show that our method can efficiently and accurately simulate different types of D-ToF cameras, including both homodyne and heterodyne at different frequency ranges.
%In particular, 
We replicate experimental results from prior D-ToF imaging work~\citep{Heide:2015:Doppler,Hu:2022:Differential}, and use our simulations to analyze failure cases (\cref{section:7_applications}). 
% We provide our implementation in the supplement and will make it publicly available upon acceptance. 
We expect that our open-source implementation will facilitate research toward improving D-ToF cameras and velocity estimation algorithms.

% To summarize, our main contributions are:
% \begin{itemize}
%     \item A path integral framework for D-ToF imaging.
%     \item An efficient sampling strategy using antithetic time sampling and path correlation, which results in orders of magnitude lower variance than commonly used sampling techniques such as uniform or stratified sampling.
%     \item An open-source implementation within Mitsuba 0.6
%     %
%     \footnote{
%     \url{https://github.com/juhyeonkim95/MitsubaDopplerToF}
%     }
%     %
%     and Mitsuba 3
%     %
%     \footnote{
%     \url{https://github.com/juhyeonkim95/Mitsuba3DopplerToF}
%     }
%     %
%     , which we share as a supplement.
%     % and will make publicly available. 
%     \item Evaluation of the proposed rendering framework on various scene geometries and material properties. 
%     \item Simulation experiments comparing various D-ToF techniques.
%     % and an algorithm that results in minor improvement compared to previous techniques. 
%     % \item We found a problem of velocity calculation in previous works and suggest possible solutions.
% \end{itemize}

%% file: sections/02_related_works_v1.1.tex
\section{Related Work}
\paragraph{Time-of-Flight Imaging}
Time-of-flight (ToF) cameras measure time-resolved flux, and include transient, time-gated, and continuous-wave time-of-flight (CW-ToF) cameras. 
Transient cameras~\cite{Velten:2011:Slow,OToole:2017:Reconstructing} use narrow pulsed laser sources and bin photons based on their time of travel. 
Time-gated cameras~\cite{Walia:2022:Gated2Gated, Pediredla:2019:SNLOS} use pulsed lasers and fast shutters to capture photons that travel a fixed time range. 
Continuous-wave time-of-flight cameras use amplitude-modulated illumination and exposure, measuring transients in the Fourier domain~\cite{Lin:2016:Frequencydomain,Peters:2015:Solving,OToole:2014:Temporal}.  
\jkimb{
All these ToF cameras have uses in applications such as depth sensing~\cite{Lange:2001:Solidstate, Foix:2011:Lockin, Gokturk:2004:Timeofflight, Gupta:2019:Asynchronous, Po:2022:Adaptive}, robot navigation~\cite{Prusak:2008:Pose, Yuan:2009:Laserbased}, non-line-of-sight (NLOS) imaging~\cite{Buttafava:2015:Nonlineofsight, Kadambi:2016:Occluded, Pediredla:2017:Reconstructing, OToole:2018:Confocal, Liu:2019:Nonlineofsight}, and imaging through scattering media~\cite{Naik:2014:Estimating, Satat:2016:All}. 
}
% All these ToF cameras have uses in applications such as depth sensing~\cite{gokturk2004time}, lifetime imaging~\cite{Lee:2019:Coding}, non-line-of-sight (NLOS) imaging~\cite{Buttafava:2015:Nonlineofsight}, and imaging through scattering media~\cite{Satat:2016:All}. 
%and material analysis~\cite{Su:2016:Material}. 
Some of these applications have been either enabled or enhanced due to advances in ToF rendering.

\paragraph{Time-of-Flight Rendering}
ToF rendering is the physically accurate simulation of measurements of ToF cameras. 
ToF rendering algorithms have mostly focused on synthesizing a sequence of images that show the evolution of light (transient rendering) or a single image for a specific time gate (time-gated rendering).
Both cases require handling the near-delta temporal manifold, which is a challenging problem.
Inspired by their success in steady-state rendering~\cite{Jensen:2001:Realistic, Jensen:1998:Efficient,Jarosz:2011:Comprehensive, Jarosz:2008:Beam}, several approaches have proposed using photon density estimation~\cite{Jarabo:2012:Femtophotography, Jarabo:2014:Framework, Ament:2014:Refractive} or photon beam methods~\cite{Marco:2017:Transient, Marco:2019:Progressive} for ToF rendering.
% Several works propose an approach based on photon density estimation~\cite{Jarabo:2012:Femtophotography, Jarabo:2014:Framework, Ament:2014:Refractive} inspired from photon mapping in steady-state rendering~\cite{Jensen:2001:Realistic, Jensen:1998:Efficient}, while some adapt photon beam strategy~\cite{Jarosz:2011:Comprehensive, Jarosz:2008:Beam} to ToF rendering domain~\cite{Marco:2017:Transient, Marco:2019:Progressive}.
% \cite{Jarabo:2018:Bidirectional}
\citet{Pediredla:2019:Ellipsoidal} adapted ideas from ToF participating media rendering~\citep{Jarabo:2014:Framework} and proposed ellipsoidal path connections to sample contributing paths for time-gated rendering.
\citet{Liu:2022:Temporally} extended this approach to general photon primitives.

Other works focus on accelerated ToF rendering for specific applications.
\Citet{Tsai:2019:Volumetric} and \citet{Iseringhausen:2020:Nonlineofsight} proposed simplified three-bounce rendering models for fast rendering in NLOS imaging.
% , while \citet{Royo:2023:Virtual} \wjarosz{subsequently removed the bounce limit}
\citet{Pan:2019:Transient} proposed a GPU-accelerated rasterization method using a transient version of instant radiosity.
Recently, \citet{Yi:2021:Differentiable}, \citet{Wu:2021:Differentiable}, and \citet{Plack:2023:Fast} developed differentiable ToF rendering algorithms to solve inverse imaging problems with analysis-by-synthesis.

In contrast to these prior works that largely focused on sampling the delta time-manifold, we focus on the unexplored problem of efficiently integrating a time-varying path space coupled with high-frequency illumination and sensor modulation signals.

\paragraph{Doppler Time-of-Flight Imaging}
Doppler time-of-flight (D-ToF) imaging is a CW-ToF technique that estimates the radial velocity of a moving object~\cite{Heide:2015:Doppler, Shrestha:2016:Computational}.
If we illuminate the scene with a high-frequency temporal signal, the moving objects cause a frequency shift to the observed signal due to the Doppler effect. 
\citet{Heide:2015:Doppler} achieve D-ToF imaging using two ToF imaging modes, homodyne and heterodyne, which differ in sensor modulation frequency. 
Homodyne mode uses sensor modulation with the same frequency as the illumination modulation (\qtyrange{10}{1000}{MHz}), whereas heterodyne mode uses a precisely shifted frequency---an integer multiple of the inverse exposure duration. 
A heterodyne image is proportional to the Doppler frequency shift, which in turn is proportional to the radial velocity. A homodyne image acts as a normalizing factor. 
Thus, the ratio of heterodyne and homodyne images provides the radial velocity. 
As a heterodyne mode image measures subtle frequency changes, it has low-intensity values, and thus a low signal-to-noise ratio (SNR). 
To overcome this problem, \citet{Hu:2022:Differential} proposed heterodyne mode imaging with arbitrary sensor frequencies that maximize the SNR.

Our algorithms can simulate the cameras proposed by \citet{Heide:2015:Doppler} and \citet{Hu:2022:Differential}, as well as their generalized variants that use arbitrary modulation waveforms. Whereas these papers use simple analytical models to analyze the D-ToF cameras they propose, we focus on how to efficiently and physically accurately simulate such cameras with rendering algorithms that account for both \jkimb{time-varying path throughput} and global illumination effects. 
% Additionally, our algorithms can simulate 

\paragraph{Motion Blur} D-ToF rendering is related to motion blur rendering techniques~\cite{Navarro:2011:Motion}, which also handle dynamic scenes that change within exposure. These techniques work by distributing samples along both pixel (or path) and time spaces. Example sampling techniques for this problem include uniform~\cite{Cook:1984:Distributed}, stratified~\cite{Mitchell:1996:Consequences}, adaptive~\cite{Whitted:1980:Improved}, and multidimensional adaptive~\cite{Hachisuka:2008:Multidimensional}, and frequency-aware ~\cite{Egan:2009:Frequency}. However, D-ToF involves high-frequency modulation functions with negative path contributions, which lead to extreme variance using existing techniques. 
We devise tailored time-path space sampling techniques to tackle this problem.
% \wjarosz{Even though motion blur rendering~\cite{Cook:1984:Distributed} can handle moving objects, D-ToF also involves high-frequency modulation functions and negative path contributions, which lead to extreme variance using existing techniques. We devise tailored time-path space sampling techniques to tackle this problem.}

\paragraph{Antithetic Sampling and Path Correlation}
% Finding a closely related path over the spatial or temporal domain is an important task in the computer graphics field.
% For example, motion vectors or optical flow are about temporal domain correlation.
% \apedired{editing this}
Antithetic sampling is a variance reduction technique that uses two correlated samples whose covariance is negative~\cite{Hammersley:1956:New}. 
Antithetic sampling significantly reduces variance if the integrand has regions that have negatively correlated parts. 

Antithetic sampling has found uses in differential and gradient-domain rendering~\cite{Bangaru:2020:Unbiased, Zhang:2021:Antithetic, Zeltner:2021:Monte, Kettunen:2015:Gradientdomain, Manzi:2016:Temporal}, leading to the development of \emph{shift mapping} techniques for sampling highly correlated antithetic pairs of paths.
%and calculate the difference between them.
% The two highly correlated paths are generated either in sampler domain or path domain. 
Path correlation can be achieved in the primary sample space---the same random numbers generate two correlated paths~\cite{Manzi:2016:Temporal, Zeltner:2021:Monte, Hua:2019:Survey}---or in the path space---the primal path generates the correlated path by deterministic shifting~\cite{Kettunen:2015:Gradientdomain, Zhang:2021:Antithetic}.
Prior work~\cite{Subr:2014:Error, Oztireli:2016:Integration,Singh:2019:Analysis} has additionally employed antithetic sampling in standard forward rendering for variance reduction.

Due to the use of illumination and sensor modulation, D-ToF rendering has a periodic integrand that lends itself to antithetic sampling. We show that antithetic sampling with appropriate path correlation reduces variance for a variety of modulation waveforms, including ones that do not have perfectly matching antithetic pairs.

%% file: sections/03_doppler_rendering_v3.5.tex
\section{Doppler Time-of-Flight Path Integral}
\label{sec:DToFPathIntegral}
We start by deriving a path integral expression for D-ToF rendering, and investigating its simplified form to devise an efficient sampling method.
\jkim{We summarize key notation in \cref{tab:notation}.}

\input{sections/03_doppler_rendering_notation_table.tex}

\subsection{ToF Path Integral for Dynamic Scenes}
\wjarosz{The \textit{ToF path integral}~\cite{Jarabo:2014:Framework, Pediredla:2019:Ellipsoidal},
\begin{equation}
    m(T) = \int_{0}^{T} \camF \int_{\mathcal{P}}f(\xbar) \lumFdop \dxbar \dt,
\end{equation}
models the measurements $m(T)$ captured by a ToF imaging system for a scene that remains static during an exposure $[0, T]$.}
Here, $\lumF$ and $\camF$ are the illumination and sensor modulation functions, \wjarosz{respectively, and $\xbar\coloneq\mathbf{x}_0\mathbf{x}_1\dots\mathbf{x}_{K}$ is a path consisting of $K + 1$ vertices, with $\mathbf{x}_0$ on the sensor and $\mathbf{x}_{K}$ on the light source}. \wjarosz{The path space $\mathcal{P}$ is the set of all light paths of all lengths $K > 0$, and $\dmu{\xbar}$ is the corresponding Lebesgue measure. The path throughput $f(\xbar)$ accounts} for visibility, geometric attenuation, and reflectance at all path vertices and edges. 
We \wjarosz{use $\xbarlen \coloneq \sum_{j=0}^{K-1} \norm{\mathbf{x}_j - \mathbf{x}_{j+1}}$ for the time of flight of a path,} where $\norm{\mathbf{x}_j - \mathbf{x}_{j+1}}$ is \wjarosz{the} time it takes light to travel from $\mathbf{x}_{j+1}$ to $\mathbf{x}_{j}$ considering the medium refractive index. 
% Assuming that all the media in the scene have unit refractive index and $c$ is the speed of light, $\xbarlen \coloneq \sum_{j=0}^{K-1} \norm{\mathbf{x}_j - \mathbf{x}_{j+1}}/c$. Finally, $\dmu{\xbar}$ is the path length measure. 

The assumption of a static scene allowed previous techniques to change the integration order in the ToF path integral:\wjarosz{
\begin{equation}
\label{eq:static_tof_path_integral}
    m(T) =\int_{\mathcal{P}} f(\xbar) W_T(\xbarlen)\, \dxbar,
\end{equation}
where $W_T(\xbarlen)\coloneq\int_{0}^{T}\!\! \camF \lumFdop \dt$ is the \textit{pathlength importance}~\citep{Pediredla:2019:Ellipsoidal}}.
\wjarosz{This form of the ToF path integral enabled the development of importance sampling strategies for} path length \cite{Pediredla:2019:Ellipsoidal}, \wjarosz{and even techniques} to analytically integrate the pathlength importance over \wjarosz{time}~\cite{Liu:2022:Temporally}. 
% \rev{In CW-ToF imaging, $W_T(\xbarlen)$ is the cross-correlation of the illumination and sensor modulation functions~\cite{Lin:2014:Fourier}.}

% For example, time-gated cameras could be simulated with $W_T(\xbarlen)$ of a narrow ranged rectangular or Gaussian function.

\begin{figure}[t]
\includegraphics[width=\linewidth]{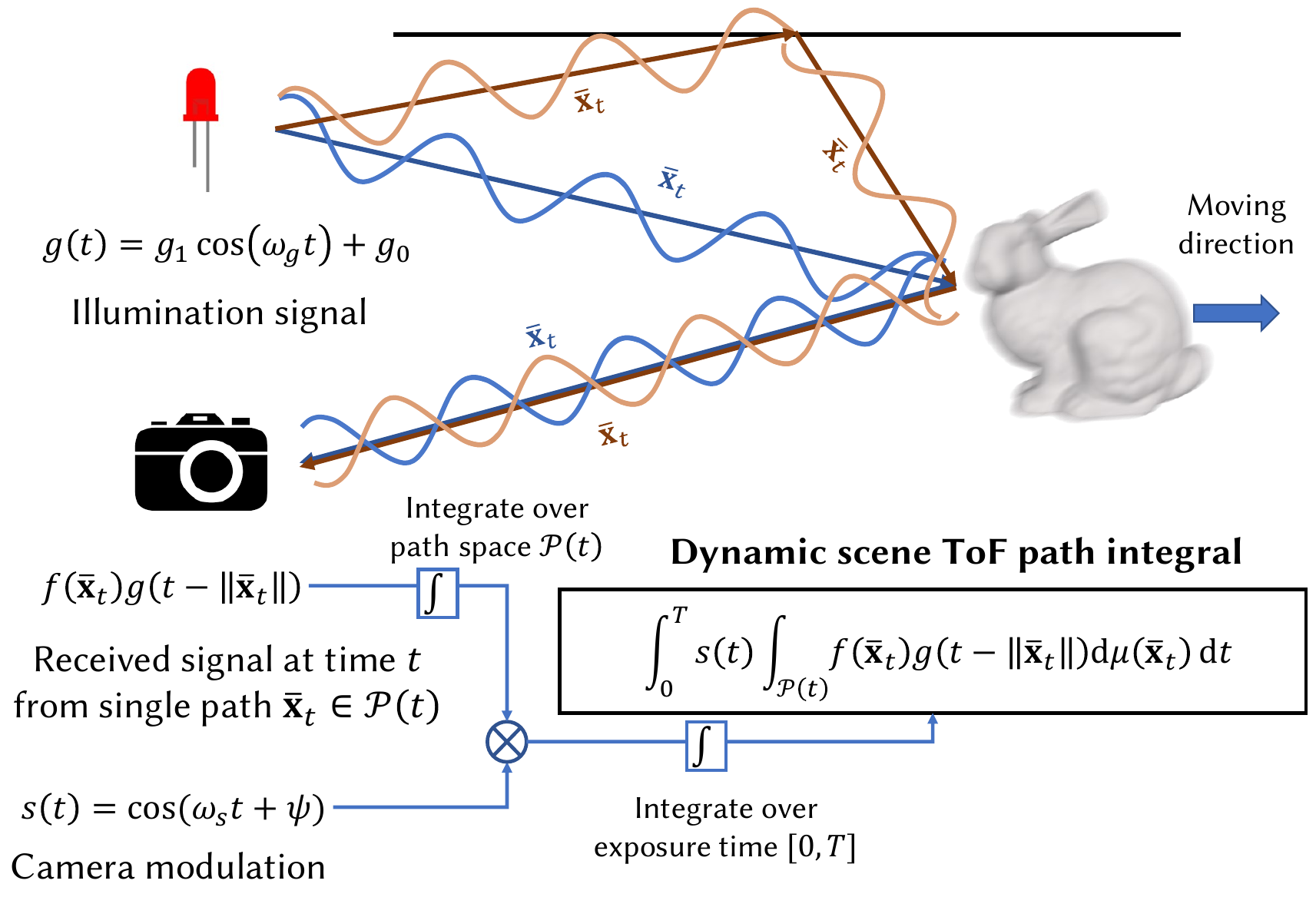}
\caption{Overview of ToF imaging and rendering of a dynamic scene. We show two paths $\xbarundert$ that arrive at the same image pixel. In D-ToF imaging, $\camF$ and $\lumF$ are high-frequency sinusoidal functions. The sinusoidal waves represent the illumination intensity at each point on the two paths.
% \igkiou{In figures: 1) Do not use the h argument, use t to bring them to the top of the page. Otherwise, make them wrapped figures. 2) Use linewidth and columnwidth for sizing, not textwidth.}
}
\label{figure:doppler_rendering_overview}
\end{figure}
% \jkim{Add scale of T? Such assumption is valid if T is in scale of ns, ps. However, in our case, T is in scale of ms.}
However, in general, the scene geometry is not static within exposure, and the path space $\mathcal{P}(t)$ varies over time $t$ as surfaces in the scene move. 
Then, the ToF path integral becomes 
\begin{equation}
\label{eq:dynamic_tof_path_integral}
    m(T) = \int_{0}^{T} \camF \int_{\mathcal{P}(t)}f(\xbarundert) \lumFdopundert \dxbarundert \dt. 
\end{equation}
Here, $\xbar_{t} \in \mathcal{P}(t)$ is a light path, similar to $\xbar$, but using scene geometry at time $t$. 
We call \cref{eq:dynamic_tof_path_integral} the \textit{dynamic ToF path integral}.
D-ToF imaging is a special case of \wjarosz{this} path integral \wjarosz{where the} modulation functions are \wjarosz{high-frequency} functions designed for radial velocity estimation~\cite{Heide:2015:Doppler}.
\wjarosz{\Cref{figure:doppler_rendering_overview} visualizes the terms of \cref{eq:dynamic_tof_path_integral} in a D-ToF imaging system}.
\wjarosz{\Cref{eq:dynamic_tof_path_integral} is, in fact, more general and can} reproduce not only D-ToF imaging, but also other dynamic scene-related \wjarosz{phenomena} such as motion blur.
However, \wjarosz{whereas lighting may change slightly during exposure in a typical motion blur setting, D-ToF imaging includes extremely high-frequency modulation functions, necessitating} new rendering approaches. 

% Doppler Time-of-Flight path integral is similar to above equation, but the only difference is that path space $\mathcal{P}$ is not constant over the time.

We ignore the subtle difference between the actual light path that arrives at the sensor at time $t$---and thus interact with the scene at times prior to $t$---and the path that is built using scene geometry at time $t$.
We show in the supplement that this approximation introduces a negligible bias that is proportional to the square of the ratio of the radial velocity $v$ to the speed of light $c$.
\subsection{Path Evolutions over Time}
% path trajectories?
\jkimb{The dynamic ToF path integral \eqref{eq:dynamic_tof_path_integral} does not assume any path mapping function over time, which means $\xbar_t$ are all independent.
However, imposing such a cross-time path correspondence makes the double integral more tractable.
}
%The dynamic ToF path integral \eqref{eq:dynamic_tof_path_integral} does not assume that scene paths at some time $t$ can be bijectively mapped to paths at some other time. However, imposing such a cross-time path correspondence makes the double integral more tractable.
% Assuming that the path trajectories are evolving with time imposes structure on the dynamic ToF path integral (Eq.~\ref{eq:dynamic_tof_path_integral}) and makes the double integral tractable. 
Therefore, we will \jkimb{introduce} a path mapping function $\mapping(t,\cdot): \mathcal{P}(0) \to \mathcal{P}(t)$ that is bijective for all $t\in [0, T]$ and maps a path $\xbar_0$ to a path at arbitrary time $t$. We write a \emph{path evolution} at time $t$ under this mapping as $\xbart \coloneq \mapping(t, \xbar_0)$, which implies $\xbart = \mapping(t, \xbarz)$. Replacing $\xbar_t$ with $\xbart$ in \cref{eq:dynamic_tof_path_integral}:
\begin{align}
\label{eq:dynamic_tof_path_integral_reordered}
% \begin{split}
    & \int_{0}^{T} \camF \int_{\mathcal{P}(t)}f(\xbart) \lumFdopt \dxbart \dt \nonumber \\
    % & \int_{0}^{T} \camF \int_{\mappingt(\mathcal{P}(0)) \cup \mathcal{P}_r(t) }f(\xbart) \lumFdopt \dxbart \dt \\
    &= \int_{0}^{T} \camF \int_{\mathcal{P}(0)}f(\xbart) J_{\mapping}(\xbarz)  \lumFdopt \dxbarz \dt \nonumber \\
    &= \int_{\mathcal{P}(0)} \int_{0}^{T}  \fmapping(\xbart)\camF \lumFdopt \dt \dxbarz, 
% \end{split}
\end{align}
where $J_{\mapping}(\xbarz) \coloneq \nicefrac{\dxbart}{\dxbarz}$ is the determinant of the Jacobian of the mapping, and  $\fmapping(\xbart) \coloneq f(\xbart) J_{\mapping}(\xbarz)$.
% \jkimb{$\mapping$ could be defined on both primary sample space or path space, but here we will consider
% }
% We define a \jkim{\textit{path evolution}} $\trajectory(\xbarz)\coloneq \{\xbart \mid t \in [0, T]\}$ as a set of corresponding paths over time.
%\rev{which is illustrated as a straight horizontal line in time-path domain (\cref{figure:overview})}.
%% IGKIOU: I moved the above comment to the figur ecaption.
\jkimb{This formulation is analogous to the material-form path space parameterization by~\citet{Zhang:2020:Path}.}
To justify the assumption that there exists such a bijective mapping $\mapping$, we provide one intuitive example: We consider a path $\xbarz$ with vertices attached to different points of the scene geometry at time $t = 0$. As the scene geometry evolves over time, these points will move to new locations, evolving at each time $t$ into a new path $\xbart$ in bijective correspondence with the original path $\xbarz$. 
\wjarosz{The mapping $\mapping$ induced by this correspondence is not the only possible one, and we discuss other mappings $\mapping$ we use for our rendering algorithm in \cref{section:5_spatial_correlation}}.
\subsection{Sampling Strategy for D-ToF Rendering}
% Now, going back to the Eq.~\ref{eq:dynamic_tof_path_integral_reordered} and let's examine the modulation term $\camF \lumFdopt$.
To start our investigation of efficient sampling techniques, we first examine the shape of the integrand in \cref{eq:dynamic_tof_path_integral_reordered} for the D-ToF case.
% Identifying the exact integrand for general case is not possible due to the complexity of changing geometry, but at least knowing the rough shape will be greatly helpful.
% To this end, we exploit the setting from ~\cite{Heide:2015:Doppler}.
% Note that their method is just an approximation for Eq.~\ref{eq:doppler_rendering_eq}.
Similar to the setup \citet{Heide:2015:Doppler} describe, we will consider \wjarosz{sensor and illumination} modulation functions that are high-frequency sinusoidal waves with frequencies $\camfreq$ and $\lumfreq$, respectively:
\begin{equation}
    \camF \coloneq \camFcos{t};\,\, \lumF \coloneq \lumFcos{t}.
\end{equation}
Here, $\psi$ is a programmable phase offset at the sensor, and $g_1, g_0$ are constant values.
We define the difference between the two frequencies as the \jkim{heterodyne frequency} $\omegadiff \coloneq \camfreq - \lumfreq$.
% Because we are using more general setting from ~\citet{Hu:2022:Differential}, let's say $\omegadiff$
\citet{Heide:2015:Doppler} used two imaging modes for radial velocity evaluation: \textit{heterodyne} mode with $\omegadiff = \nicefrac{2\pi}{T}$, and \textit{homodyne} mode with $\omegadiff = 0$.
% Heterodyne mode is proportional to radial velocity while homodyne mode functions as a normalization term, thus calculating ratio between two images gives the radial velocity.
% We can estimate radial velocity from ratio of homodyne and heterodyne images.
% To further simplify the notation, let's use $\omegaratio$ instead which is defined as $\omegadiff/\omegaref$.
In \wjarosz{subsequent} work, \citet{Hu:2022:Differential} proposed using a heterodyne mode where $\omegadiff$ can take any value within the range of $[0, \nicefrac{2\pi}{T}]$, to improve signal-to-noise ratio. 
We aim to reproduce this more general setting and thus assume that $\omegadiff\in[0, \nicefrac{2\pi}{T}]$.
% \jkim{set out target to more general setting from \citet{Hu:2022:Differential} and even expand for non-sinusoidal functions.} use \citet{Hu:2022:Differential} setting to derive efficient sampling algorithm, which works efficiently even for non-sinusoidal functions.

We define the path phase offset $\phi(\xbar) \coloneq -\lumfreq\xbarlen$, and use it to express the integrand in \cref{eq:dynamic_tof_path_integral_reordered} as:
\begin{equation}
\label{eq:multiplication_cos}
    \fmappingxbart \camFcos{t}  \left( \lumFcos{t+\phi(\xbart)} \right).
\end{equation}
Expanding the cosine terms in \cref{eq:multiplication_cos},
we get the following terms:
%the modulation terms become:
%
\begin{align}
\label{eq:modulation_term_full}
    &\frac{g_1}{2}\cosP{(\camfreq +\lumfreq) t + \psi + \phi(\xbart)} \nonumber \\
    &+ \frac{g_1}{2}\cosP{(\camfreq -\lumfreq) t + \psi - \phi(\xbart)}+ g_0 \camFcos{t}.
\end{align}
%
% \begin{align}
%     m(T) &= \int_{0}^{T} \camFcos{t} \\
%     & \int_{P(t)}f(\xbar) \left( \lumFcos{t+\phi} \right) \dxbar \dt
% \end{align}
% \begin{equation}
%     \int_{0}^{T} \int_{\mathcal{P}(t)}f(\xbart) \camFcos{t}  \left( \lumFcos{t+\phi(\xbart)} \right) \dxbart \dt
% \end{equation}
% Above multiplication generates cosine terms with frequencies $\camfreq + \lumfreq$, $\camfreq - \lumfreq$ and $\camfreq$ respectively.
As $\fmappingxbart$ and $\phi(\xbart)$ vary slowly relative to $\camfreq$, and $T \gg \nicefrac{1}{\camfreq}$, the high-frequency terms ($\camfreq + \lumfreq$ and $\camfreq$) approximately sum to zero when we integrate over $[0, T]$. 
Empirically, we found that ignoring the high-frequency terms reduces the variance by several orders of magnitude. 
This low-pass filtering operation introduces bias of order $\bigO(\omegadiff\nicefrac{v}{c})$ to the Doppler frequency shift, itself of order $\bigO(\lumfreq\nicefrac{v}{c}).$ 
%\jkim{where $v$ is the radial velocity and $c$ is the speed of light}. 
As $\omegadiff$ is typically around a few kHz and $\lumfreq$ ranges from \qtyrange{10}{1000}{MHz}, in practice the bias is less than 0.1\% and thus negligible. 
% However, this bias is negligible in practice since $\omegadiff$ is typically on the order of a few kHz, while $\lumfreq$ ranges from 10 to 1000 MHz.
Therefore, we will consider only the low-frequency ($\camfreq - \lumfreq = \omegadiff$) term of $\camF \lumFdopt$, \rev{which we will refer to as the \textit{modulation term} for simplicity,}
\begin{equation}
    \frac{g_1}{2}   \cosP{\omegadiff t - \phi(\xbart) +\psi }.
    % \corrmodul \coloneq \frac{g_1}{2}   \cosP{\omegadiff t - \phi(\xbart) +\psi } . 
    \label{eq:product_of_modulation_functions}
\end{equation}
%
% We will denote this low-pass filtered multiplication of modulation functions as $\corrmodul$.
We can do similar low-pass filtering for arbitrary periodic signals $\camF$, $\lumF$ using their Fourier series 
\jkimb{which results in correlation of two signals with frequency of $\omegadiff$}, as we show in the supplement.
% Using the Fourier series, we can derive that low-pass filtered of multiplication of modulation is equivalent to the correlation of two signals with a frequency of $\omegadiff$.
% We plot several example signals from ~\cite{gupta2018optimal} on Fig~\ref{figure:low_pass_filtering}.

\begin{figure} [t]
\includegraphics[width=\linewidth]{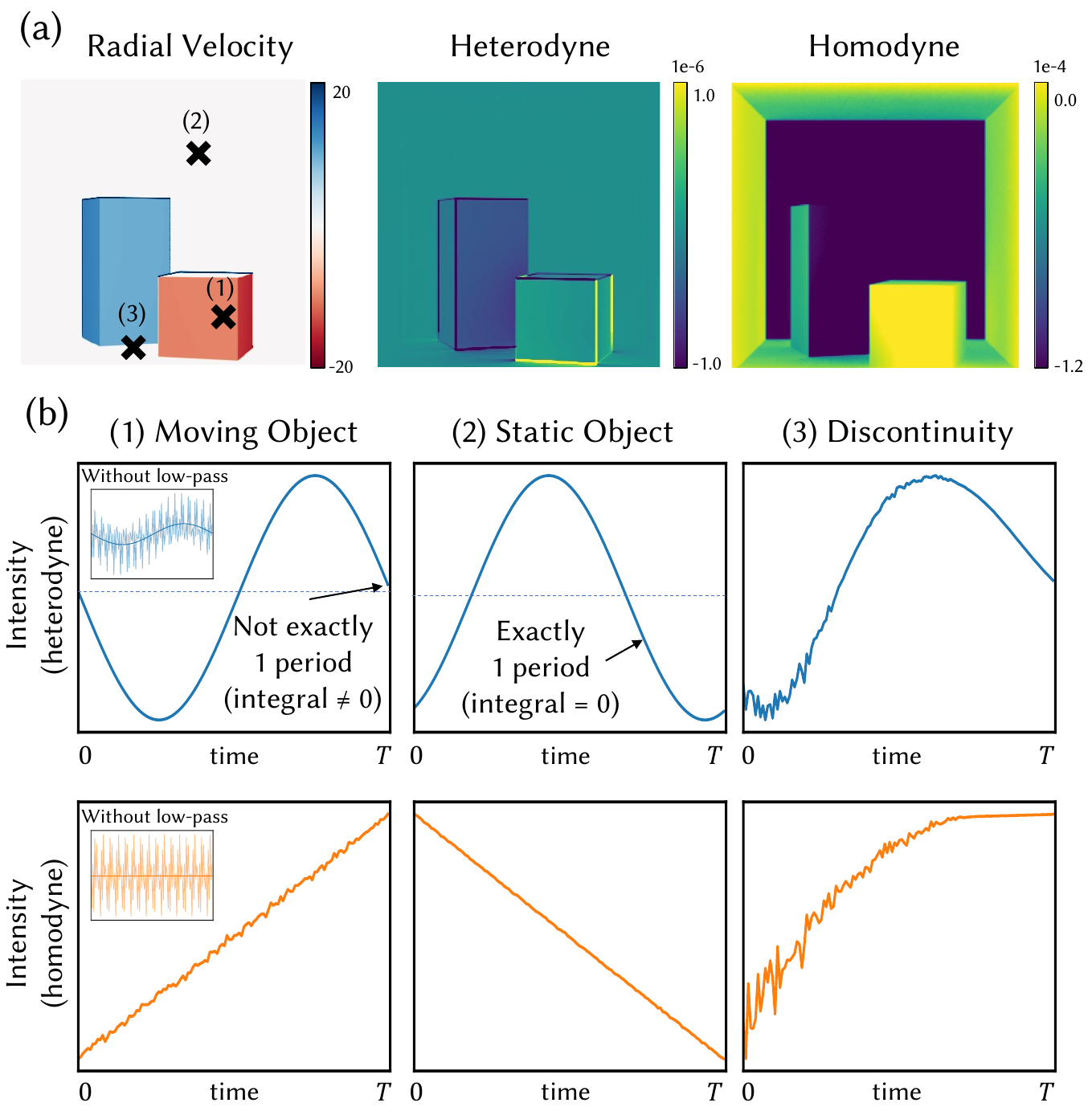}
\caption{Integrand of \cref{eq:doppler_rendering_path_integral_reordered} for a single-bounce path over time. Integrating the signal over the time-path space generates the homodyne and heterodyne images (first row). The unfiltered version is also plotted at the corner. 
% \apedired{change ``moving object" to ``dynamic" object in line with text. For static objects, the dashed line that represents zero should be higher; otherwise, it won't sum to zero}
}
\label{figure:doppler_over_time}
\end{figure}

% \begin{figure} [t]
% \includegraphics[width=\linewidth]{figures/3_doppler_tof_rendering/low_pass_filtering_v1.0.pdf}
% \caption{Low pass filtered modulation for some of the signals in ~\cite{gupta2018optimal}. 
% \TODO{remove this} Trapezoidal signal is also called Hamiltonian coding in their paper.}
% \label{figure:low_pass_filtering}
% \end{figure}
Substituting \cref{eq:product_of_modulation_functions} in \cref{eq:dynamic_tof_path_integral_reordered} gives a tractable D-ToF path integral:  
% \begin{equation}
% \label{eq:doppler_rendering_path_integral_reordered}
%      \int_{\mathcal{P}(0)} \int_{0}^{T} \fmapping(\xbart) \corrmodul \dt \dxbarz 
% \end{equation}
% \jkim{better to use L or keep cos?}
%
\begin{equation}
\label{eq:doppler_rendering_path_integral_reordered}
     \int_{\mathcal{P}(0)} \int_{0}^{T} \fmapping(\xbart) \frac{g_1}{2}   \cosP{\omegadiff t - \phi(\xbart) +\psi }  \dt \dxbarz .
\end{equation}
To understand the behavior of \cref{eq:doppler_rendering_path_integral_reordered}, in \cref{figure:doppler_over_time} we plot the integrand at three pixels assuming single-bounce paths $\xbart$. The illumination is a point light source collocated with the sensor. 
% To understand the behavior of the above Eq~\ref{eq:doppler_rendering_path_integral_reordered}, in Fig.~\ref{figure:doppler_over_time}, we plot the integrand
% %
% \footnote{We measured the pixel radiance and this is not exactly equivalent to the integrand of single $\xbart$. But the illumination is a point light and we only consider a single bounce, it is close to the shape of the integrand of a single path over time. The noise remains in Fig.~\ref{figure:doppler_over_time} after low-pass filtering is due to the variance inside the pixel.}
% %
% for three single-bounced paths $\xbart$ as a function of time. The illumination is a point light source that is collocated with the sensor. 
% (This is not exactly equal to the integrand, but the illumination is a point light and we only consider single bounce, it is close to the shape of the integrand of single path.)
% over the pixel for single-bounced paths as 
%we plot the integrand for three single-bounced paths $\xbart$ as a function of time. The illumination is a point light source that is collocated with the sensor. 
% Note that for the single-bounced path with delta light source, $\mapping$ is unique. 
In heterodyne mode, the integrand is a sinusoid
%of approximately one period for dynamic objects and
of exactly one period for static objects, making the integral zero in this case. 
For dynamic objects, $\phi$ is a function of time and contributes a Doppler frequency shift, making the integrand not a single-period sinusoid. 
This results in a non-zero integral and this non-zero value is important for computing the Doppler frequency shift, and thus the radial velocity. 
% Note that this is not exactly a one-period sinusoidal function because moving object changes $\xbartlen$ and $\fmappingxbart$.
% Note that even though $\fmapping(\xbart)$ is constant over the trajectory, the integrand is not exactly a one-period sinusoidal function because moving object changes $\xbartlen$, which causes Doppler frequency shift.
In homodyne mode, the integrand is close to a \jkimb{linear} function as $\omegadiff = 0$.
% For heterodyne mode, the signal is effectively a sinusoidal signal over one period, while for homodyne mode, it is close to a linear function.
% Note that this is not exactly a sinusoidal function because Moving objects make a small difference in $\xbartlan$
For paths near edges where scene discontinuities occur, we observe drastic changes in the integrand (third column in \cref{figure:doppler_over_time}). 
In this case, $\xbartlen$ and $\fmapping(\xbart)$ change significantly during exposure, and the integrand is neither a sinusoidal nor a linear function.
% This is not always true, because, for some paths, a discontinuity occurs (third column in Fig.~\ref{figure:doppler_over_time}) and also $\fmapping(\xbart)$ or $\xbartlen$ is not constant even discontinuity does not occur.
% Note that change in $\xbartlen$ due to a moving object causes a Doppler frequency shift.
% Multi-bounced paths exhibit analogous behavior.
%if $\mapping$ is strongly-correlated over time ($\fmapping(\xbart)$ or $\xbartlen$ is almost constant over time).

\begin{figure} [t]
\includegraphics[width=\linewidth]{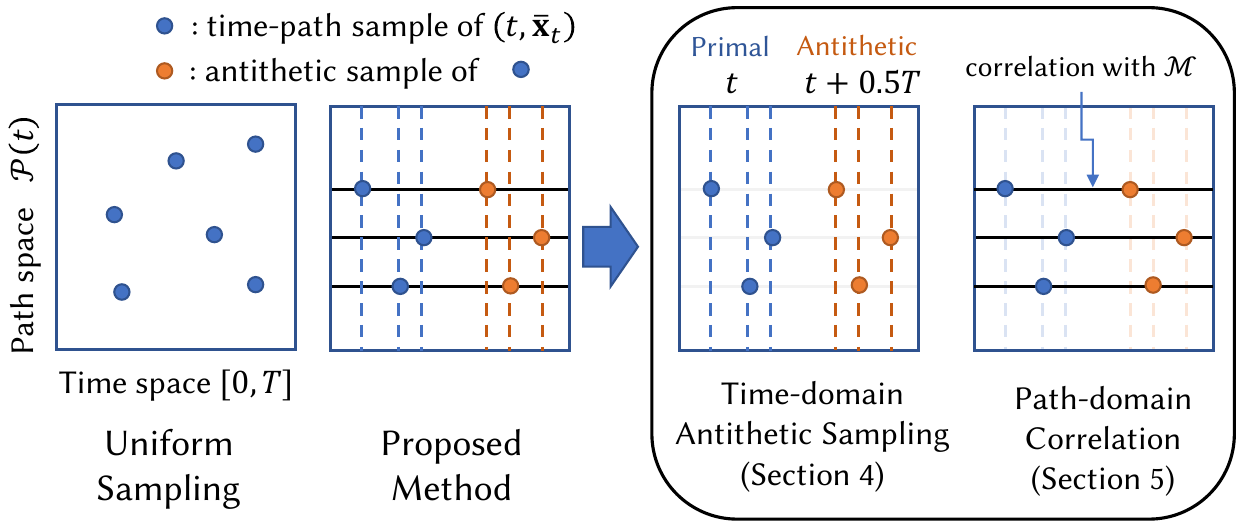}
\caption{Overview of our sampling strategy. We visualize path evolutions $\xbart$ for different paths $\xbarz$ as horizontal lines in the time-path space. We first sample time using antithetic sampling, then create correlated path samples as evolutions of different $\xbarz$ to the sampled antithetic time pairs.}
\label{figure:overview}
\end{figure}

Inspired by this pilot experiment, we estimate the double integral of \cref{eq:doppler_rendering_path_integral_reordered} by sampling its two domains, time and path, in two steps that we visualize in \cref{figure:overview}: 
\begin{enumerate}
    \item Assuming $\fmapping(\xbart)$ and $\xbartlen$ do not change significantly with time $t$ for some, still unknown, $\xbarz$ (i.e, assuming the behavior of the third column in \cref{figure:doppler_over_time}-(b) is rare), we find two \emph{antithetic} time samples for which \cref{eq:product_of_modulation_functions} has approximately the \jkimb{same deviation from ground truth but opposite sign.} %same absolute value but opposite sign.
%\footnote{We will refer this term as the modulation term for simplicity.}
%
    \item We use these time samples to create \emph{correlated} path samples, equal to the evolutions $\xbart$ of an underlying $\xbarz$ at the sampled times $t$, such that $\fmapping(\xbart)$ and $\xbartlen$ do not vary significantly with $t$, as we required in the first step. 
\end{enumerate}
We use these samples to form a Monte Carlo estimator of \cref{eq:doppler_rendering_path_integral_reordered}:
%
% \begin{equation}
%     \sum_{i=0}^{N_p} \sum_{j=0}^{N_t} \frac{\hat{f}(\xbar_i(t_j)) l(\xbar_i(t_j), t_j) }{p(\xbar_i(0), t_j)},
%     % \frac{f(t, \xbart)}{p(t, \xbart)} + \frac{f(t_a, \xbar_a)}{p_a(t_a, \xbar_a)}
% \end{equation}
\begin{equation}
    \!\!\langle m(T) \rangle \coloneq \frac{1}{N}\sum_{i=0}^{N_p} \sum_{j=0}^{N_t} \frac{\hat{f}(\xbar^i(t^j)) \frac{g_1}{2} \cosP{\omegadiff t^j - \phi(\xbar_i(t^j)) +\psi }  }{p(t^j) p(\xbar^i(0))},
    % m(T) \approx \frac{1}{N}\sum_{i=0}^{N_p} \sum_{j=0}^{N_t} \frac{\hat{f}(\xbar_i(t_j)) \frac{g_1}{2}  \left( \cosP{\omegadiff t_j - \phi(\xbar_i(t_j)) +\psi }  \right) }{p(\xbar_i(0), t_j)},
    % \frac{f(t, \xbart)}{p(t, \xbart)} + \frac{f(t_a, \xbar_a)}{p_a(t_a, \xbar_a)}
\end{equation}
where \jkim{$N_p$ is the number of paths $\xbarz$ used for evolutions, $N_t$ is the number of antithetic time samples for the evolution of each $\xbarz$ (which we set to 2 by default), $N \coloneq N_p \cdot N_t$ is the number of total path samples, and $p$ is the sampling probability density function (pdf).} We detail antithetic time sampling in \cref{section:4_time_domain_antithetic_sampling} and correlated path sampling in \cref{section:5_spatial_correlation}. 
It is challenging to reverse the sampling order---find antithetic samples in the path domain and align them in the time domain---as \wjarosz{this would} require finding an antithetic path with a specific length, which is expensive~ \citep{Pediredla:2019:Rendering}.
%\jkim{aligning them to have the same modulation function} 

% Because we are using time-domain antithetic sampling, $N_t=2$.
% Because we are using antithetic sampling for the time domain, note that $N_t=2$ and those two time samples have an antithetic relationship.

% Let's first focus on time domain sampling. 
% Ignore the geometric related term right now, and only consider time domain integration.
% Then, our goal is to find an efficient sampling strategy that minimizes variance of Eq. \ref{eq:doppler_rendering_eq_heide}.

% \begin{equation}
%     \int_{0}^{T}  \cosP{\omegadiff t + \phi(t) } \dt
% \end{equation}

%% file: sections/03_doppler_rendering_notation_table.tex
\begin{table}[t]
\caption{Definition of variables used throughout the paper.}
\resizebox{\columnwidth}{!}{%
\begin{tabular}{rl}
\toprule
\textbf{Notation}   & \textbf{Description}                                                                                      \\ \midrule

$\xbarundert$                & Path with scene geometry at time $t$.    \\
$\mathcal{P}(t)$        & Path space with scene geometry at time $t$.    \\
$\mapping(t, \cdot)$             & Path mapping function at time $t$. \\
$\xbart$                & Evolution of path $\xbarz\in\mathcal{P}(0)$ to time $t$ \jkimb{according to $\mapping$}.    \\
% $\trajectory(\xbarz)$                & \jkim{Set} of correlated paths $\{\xbart | t \in [0, T]\}.$    \\
$\xbarlen$                & Time of flight of $\xbar$.    \\
$\phi(\xbar)$                & Phase offset of path due to $\xbarlen$.    \\
$\camF, \lumF$                & Sensor and illumination modulation function.    \\
$\camfreq, \lumfreq$                & Sensor and illumination modulation function's frequency.    \\
$\psi$                & Programmable phase offset at sensor.    \\
% $\corrmodul$                & Low-pass filtered $\camF \lumFdopt$.  \\
$[0,T]$                & Sensor exposure time.    \\
% $\omegadiff, \omega_0, \omegaratio$                & $\camfreq - \lumfreq$, $2\pi /T$ ($\omegadiff$ at perfect heterodyne), $\omegadiff / \omega_0$    \\
$\omegadiff$                & Heterodyne frequency, $\camfreq - \lumfreq$. \\
$\omegaratio$                & Normalized heterodyne frequency, $\omegadiff / \omega_0$ ($\omega_0: 2\pi /T$). \\
$\Delta \omega$                & Observed Doppler frequency shift.   \\
$N_p$                & Number of sampled \jkim{path evolutions.}    \\
$N_t$                & Number of time samples along the evolution of a given \jkim{$\xbarz$}. \\
% $c$                & Speed of light.    \\
\bottomrule
\end{tabular}%
}
\label{tab:notation}
\end{table}

%% file: sections/04_time_domain_antithetic_sampling_v3.3.tex
\section{Time Domain Antithetic Sampling}
\label{section:4_time_domain_antithetic_sampling}
Assuming a given path evolution where $\fmappingxbart$ and $\xbartlen$ do not change significantly with $t$, we aim to find an efficient time-sampling strategy using $N_t$ samples for the modulation term time integral, %that minimizes the variance of
\begin{equation}
\label{eq:time_domain_integration}
    \int_{0}^{T}  \cosP{\omegadiff t - \phi(\xbart) + \psi } \dt, 
\end{equation}
where $\omegadiff\in[0, \nicefrac{2\pi}{T}]$. 
There is a trade-off between the \jkim{number of independent path evolutions} $N_p$ and the number of time samples $N_t$, given a fixed budget $N$ of total path samples. Increasing $N_t$ decreases the variance of the modulation term; but it also reduces $N_p$ which in turn increases the variance of the path throughput term.
% (\cref{figure:effect_of_correlation} (d)).
Empirically, we found that $N_t = 2$ results in lower overall variance, and thus fix it throughout the paper unless we state otherwise.

% result in more correlated paths that sample Doppler frequency shift term accurately but at the cost of \apedired{This actually feels out of place}

% Another thing that we have to point out here is the trade-off between the number of different trajectories $N_p$ and the number of time samples $N_t$. The product of these two values equals the total number of samples (spp) $N$ used in the estimation.
% % number of available time samples for the given trajectory $\mathcal{T}(\xbarz)$. 
% We can increase the accuracy of the estimator for Eq.~\ref{eq:time_domain_integration} by dedicating more time samples for certain trajectory.
% However, this comes at the expense of reducing the effective number of different trajectories that can be sampled.
% In extreme case such that $N_t = N$, we can minimize variance for the modulation term, but we can sample only one trajectory and this will significantly increase the total variance for Eq.~\ref{eq:doppler_rendering_path_integral_reordered}.
% We found that just using $N_t = 2$ is enough for many scenes, so we will assume that we are only allowed to use two time samples.
% Two samples may seem too small for the Monte Carlo integration, but there is a powerful technique called antithetic sampling that gives considerably low variance even using two samples.

\subsection{Antithetic Sampling for Modulation Term}
To further narrow down our problem, we simplify \cref{eq:time_domain_integration}.
We assume that $\xbartlen$ is near constant over a \jkim{path evolution}, so we can approximate $\phi(\xbart) \approx \phi(\xbarz)$.
We use this approximation \emph{only} to derive time sampling techniques and not for the actual evaluation. 
% Because of assumption on $\xbartlen$, we will use $\xbartlenz$.
% As $\xbartlen$ is not changing dramatically, we start with a Taylor series approximation $\phi(\xbart)=\phi(\xbar(0))+t \omegadelta$ where $\omegadelta$ is Doppler frequency shift) to derive our time-sampling technique. The measurement becomes 
% Because $\omegadiff \gg \omegadelta$, we can ignore the $\omegadelta$ term for deriving our sampling strategy.
We can represent $\phi(\xbarz)+\psi$ as a random variable $\theta$ of unknown, scene-dependent distribution.
Then, \cref{eq:time_domain_integration} \wjarosz{simplifies to}:
\begin{equation}
\label{eq:simplified_time_domain_integration}
    \int_{0}^{T}  \cosP{\omegadiff t + \theta} \dt.
\end{equation}
We know $\omegadiff$ before sampling, but not $\theta$. 
Our goal is to efficiently evaluate \cref{eq:simplified_time_domain_integration} for $\omegadiff \in [0, \nicefrac{2\pi}{T}]$ and unknown $\theta$.

% \begin{figure} [h]
% \includegraphics[width=0.45\textwidth]{figures/4_time_domain/time_domain_correlation_number.png}
% \caption{Number of samples used for time domain sampling.}
% \label{figure:time_domain_correlation_number}
% \end{figure}
%Instead of $\phi(\xbart) + \psi$, we will just denote it as a single phase offset term $\phi$ in this section.
%We assume that we know $\omega$ because it mainly depends on omega
%since path tracer incrementally traces rays and they shoul
%However, most of path tracer incrementally accumulates radiance as tracing the ray.
%This makes importance sampling hardly applicable to existing path tracing scheme.
% For example, if path consists of $\textbf{x}_1, \textbf{x}_2, ..., \textbf{x}_e$, $\phi(\textbf{x}_1, ... \textbf{x}_i)$ are all different, 

\begin{figure}
\includegraphics[width=\linewidth]{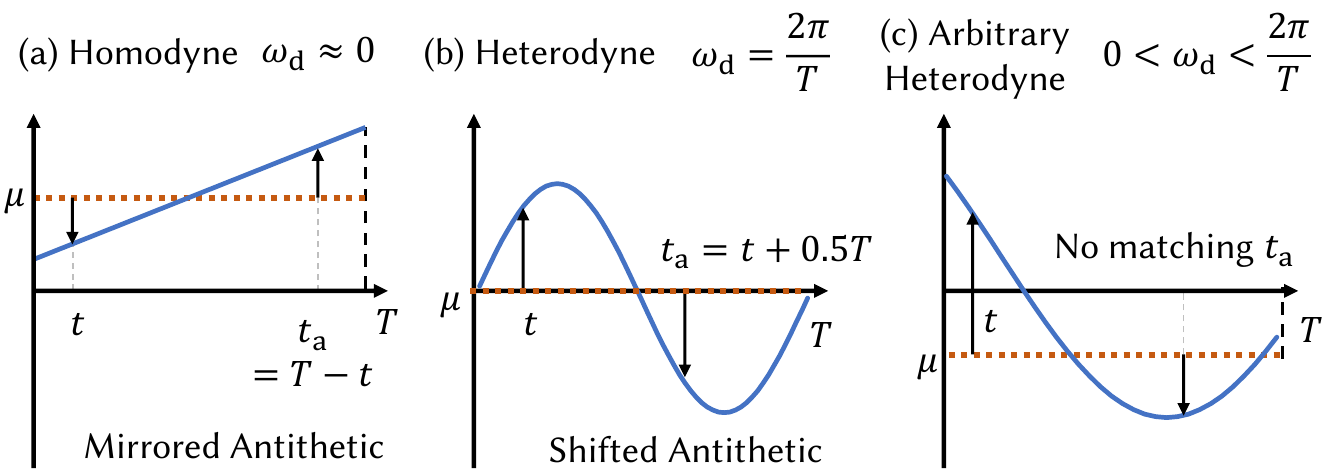}
\caption{Antithetic sampling for (a) homodyne, (b) heterodyne, and (c) arbitrary heterodyne mode. The ground truth integration value is $\mu$. In (a) and (b), we can find an antithetic pair with mirroring and shifting, respectively, that makes the primal and antithetic samples sum to $\mu$.
But in (c), we cannot find such perfectly matching antithetic sample.}
\label{figure:homodyne_heterodyne_antithetic_idea}
\end{figure}

For inspiration on how to sample time, we consider the heterodyne and homodyne cases (\cref{figure:homodyne_heterodyne_antithetic_idea}-(a,b)).
In these cases, we can easily find a zero-variance estimator using two-sample \textit{antithetic sampling}.
If $\omegadiff = \nicefrac{2\pi}{T}$ (perfect heterodyne),
given \textit{primal} sample $t$, we can select the \textit{antithetic} sample as $\ta=\mathrm{mod}(t+0.5 T, T)$. 
% with time shift $t+0.5 T$.
% If $t+0.5 T > T$, we can take mod of $T$.
Then $\cos(\nicefrac{2\pi}{T} t + \theta)$ is exactly a cosine function with period $[0, T]$, and antithetic sampling gives \wjarosz{exactly zero variance} regardless of $t$ and $\theta$.
We call this strategy \textit{\shifted antithetic sampling}.
%Inspired from OpenGL texture wrapping modes, we will call such antithetic sampling as \textit{shifted antithetic sampling}.
% Then, the sum of two samples is always zero regardless of $\phi$.
% \begin{equation}
%     \cos{\cP{\frac{2\pi}{T} t + \theta}} + \cos{\cP{\frac{2\pi}{T} \cP{t + 0.5T} + \theta}} = 0
% \end{equation}

% One may also come up with antithetic sampling with mirrored sign $-t - 0.5 T$.
% This also gives a zero variance if $\theta$ is zero, but it is not guaranteed for arbitrary $\theta$.
% Instead, such mirroring strategy is useful if $\omega$ becomes close to zero.

If $\omega_d \approx 0$ (homodyne), the integrand becomes close to a linear function.
Then, selecting a symmetric antithetic sample as $T - t$ will make the sum of primal and antithetic values have expected value $\cosP{\omega_d t + \theta}$ for $t \in [0, T]$. Thus we obtain a zero-variance estimator. We call this strategy \textit{mirrored antithetic sampling}.

However, our goal is to handle arbitrary $\omegadiff \in[0, \nicefrac{2\pi}{T}]$ as \citet{Hu:2022:Differential} proposed, so we cannot expect to find a perfectly matching antithetic sample in general (\cref{figure:homodyne_heterodyne_antithetic_idea}-(c)).
Instead, we aim to find an antithetic sample that minimizes the variance over $t$.
We use \shifted or mirrored antithetic sampling by setting the antithetic sample as $t + \ts$ or $-t + \ts$ for some constant $\ts$. 
(We omit the mod operation over $T$ for simpler notation.) For perfect heterodyne or homodyne operation, $\ts$ has an optimal value at $\ts=0.5T$ and $\ts=0$, respectively.
In the following subsections, we will discuss the relationship of our antithetic estimator to \textit{auto-correlation} and \textit{auto-convolution} in signal processing, and show that $\ts=0.5T$ and $\ts=0$ is also optimal for arbitrary heterodyne frequencies. 
% Then, our objective function is closely related to concept of \textit{auto-correlation} and \textit{auto-convolution} in signal processing theory.

\subsection{\Shifted Antithetic Sampling and Auto-correlation}
\label{section:4_shifted_antithetic_sampling}
\begin{figure}
\includegraphics[width=\linewidth]{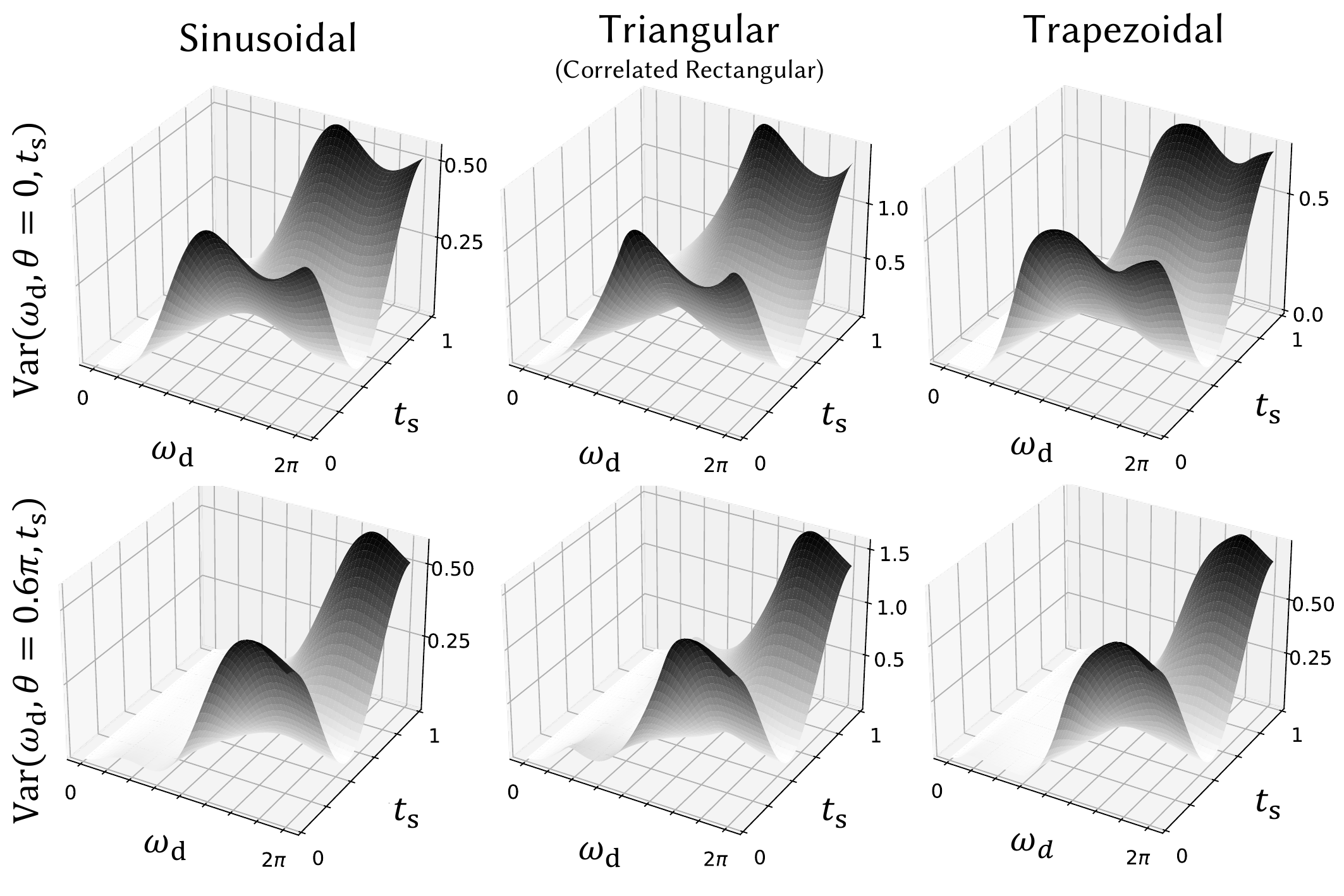}
\caption{$\mathrm{Var}(\omegadiff, \theta,\ts)$ of shifted antithetic sampling for various waveforms $x(t)$. 
They all have optimal values at $\ts=0.5T$ regardless of $\omegadiff$ and $\theta$.
Without loss of generality, we set $T=1$ and hence, $\omega_d \in [0, 2\pi]$.
We only show results for two cases, but the tendency was similar for other $\theta$s.}
\label{figure:autocorrelation_other_waveforms}
\end{figure}

We start with \shifted antithetic sampling ($\ta = t + \ts$), which is optimal for the heterodyne case.
% We show that minimizing the variance of the antithetic sampling is \wjarosz{equivalent to} minimizing the auto-correlation for \wjarosz{an} arbitrary integrand $x(t)$. 
% To do so, let's think of an arbitrary signal $x(t)$ and assume that we want to integrate its value from $0$ to $T$.
% \begin{equation}
%     \int_{0}^{T}  x(t) \dt
% \end{equation}
Using a primal sample $t$ drawn uniformly and an antithetic sample $t + \ts$, the variance of an antithetic estimator for an arbitrary integrand $x(t)$ is
\begin{equation}
    \mathrm{Var}(\ts) = \int_{0}^{T} \cP{\frac{x(t) + x\cP{\mathrm{mod}(t + \ts,T)}}{2} - \mu_x}^2 \dt,
\end{equation}
where $\mu_x$ is \wjarosz{the} mean of $x(t)$ in $[0,T]$.
\wjarosz{We see that} minimizing the variance is equivalent to minimizing the \textit{auto-correlation} : 
\begin{equation}
    R(\ts) = \int_{0}^{T} x(t) x\cP{\mathrm{mod}(t+\ts,T)} \dt.
\end{equation}
As auto-correlation is symmetric around $0.5T$, we can represent $R(\ts) = F(\ts) + F(T-\ts)$ for some $F$.
If $x(t) = \cosP{\omegadiff t + \theta}$, we can analytically calculate $F(\ts)$ as
\begin{equation}
    F(\ts) = \frac{1}{2\omegadiff} \cosP{\omegadiff T+2\theta} \sinP{\omegadiff \ts} + \frac{\ts}{2} \cosP{\omegadiff\cP{T-\ts}}.
\end{equation}
To better understand this expression, we express $R$ as a function of $\omegadiff, \theta, \ts$, rather than just $\ts$.
Then, we prove in the supplement that $R(\omegadiff, \theta, \ts)$ has a global minimum at $\ts = 0.5T$ regardless of the values of $\omegadiff$, $\theta$, if $\omegadiff \in [0, \nicefrac{2\pi}{T}]$.
% This optimal property also holds for some other special waveforms that can be mathematically proven. 
Interestingly, optimality at $\ts=0.5T$ also holds for other waveforms, such as triangular or trapezoidal, as \cref{figure:autocorrelation_other_waveforms} shows.
The fact that the optimal shift is independent of $\theta$ provides a significant benefit when we consider multi-bounce paths, which contain many subpaths with different $\theta$ values.
% In this case, we do not need to consider the time-of-flight dependent sampling strategy.
% , which was already pointed out as a problem of importance sampling in \cref{figure:importance_sampling_fails}.

% \begin{figure}
% \includegraphics[width=0.4\textwidth]{figures/temp_figure7.png}
% \caption{Autocorrelation and variance over frequency sweep ($a, \omega$) or offset sweep $a, \phi$}
% \label{figure:figure_antithetic_time_sinusoidal}
% \end{figure}

\subsection{Mirrored Antithetic Sampling and Auto-convolution}
\label{section:4_mirrored_antithetic_sampling}

For mirrored antithetic sampling $\ta = -t + \ts$, which is optimal for the homodyne case, we can repeat our analysis analogously to the previous subsection. 
%, except for the sign of the antithetic sample.
% Now let's consider antithetic sampling with sign-mirroring strategy ($\ta = -t + \ts$) which was used for homodyne case.
%We will skip the details because the process is the same except for the sign of the correlated sample.
The only difference is that the auto-correlation becomes \textit{auto-convolution} as the sign is inverted:
\begin{equation}
    C(\ts) = \int_{0}^{T} x(t) x\cP{\mathrm{mod}(-t+\ts,T)} \dt.
\end{equation}
Unfortunately, in this case, we cannot find a globally minimal $\ts$ that is independent of $\omegadiff$ and $\theta$ for $x(t) = \cosP{\omegadiff t + \theta}$. 
% So, which $\ts$ gives a minimal auto-convolution if $x(t) = \cosP{\omegadiff t + \theta}$?
% Unfortunately, in this case, we cannot find a globally minimal value independent of $\omegadiff$ and $\theta$.
Instead, we \jkimb{try to} find an optimal value for the expectation over $\theta$. 
% We use expectation over \jkim{uniform} $\theta$ as a fair alternative because $\theta$ depends on path length so calculating an optimal value for the expectation over $\theta$ can guarantee it to work well on the average \jkim{path}. 
As we do not know in advance the distribution of $\theta$, which depends on path time-of-flight, we assume it to be uniform over $[0, 2\pi]$.
%, which seems to be acceptable in practice. 
% \begin{equation}
%     \mathbb{E}_{\theta} [ R(\omegadiff, \theta, \ts) ] = \frac{1}{2\pi}\int_{0}^{2\pi} R(\omegadiff, \theta, \ts) \text{d}\theta
% \end{equation}
Then, for the sinusoidal wave, we can analytically calculate the expectation:
\begin{equation}
    \mathbb{E}_{\theta} [ C(\omegadiff, \theta, \ts) ] = \frac{\sinP{\omegadiff \ts} + \sinP{\omegadiff (T-\ts)}}{2\omegadiff},
\end{equation}
and this has a minimum at $\ts=0, T$ regardless of $\omegadiff \in [0, \nicefrac{2\pi}{T}]$.

% \begin{equation}
%     \mathbb{E}_{\phi} [ R(\omega, \phi, \ts) ] = \frac{\cP{\omega^2 + 4 \cosP{\omega} - 4}}{2\omega^2} + \frac{\sinP{\omega}}{4 \omega}
% \end{equation}

% \begin{figure}
% \includegraphics[width=0.5\textwidth]{figures/4_time_domain/sampling_method_comparison.png}
% \caption{Sampling method comparison.}
% \label{figure:sampling_method_comparison}
% \end{figure}

% \begin{figure}
% \includegraphics[width=0.4\textwidth]{figures/4_time_domain/sampling.png}
% \caption{Sampling method comparison.}
% \label{figure:f}
% \end{figure}

\subsection{Comparison with Uniform and Stratified Sampling}

\begin{figure}
\includegraphics[width=1.0\linewidth]{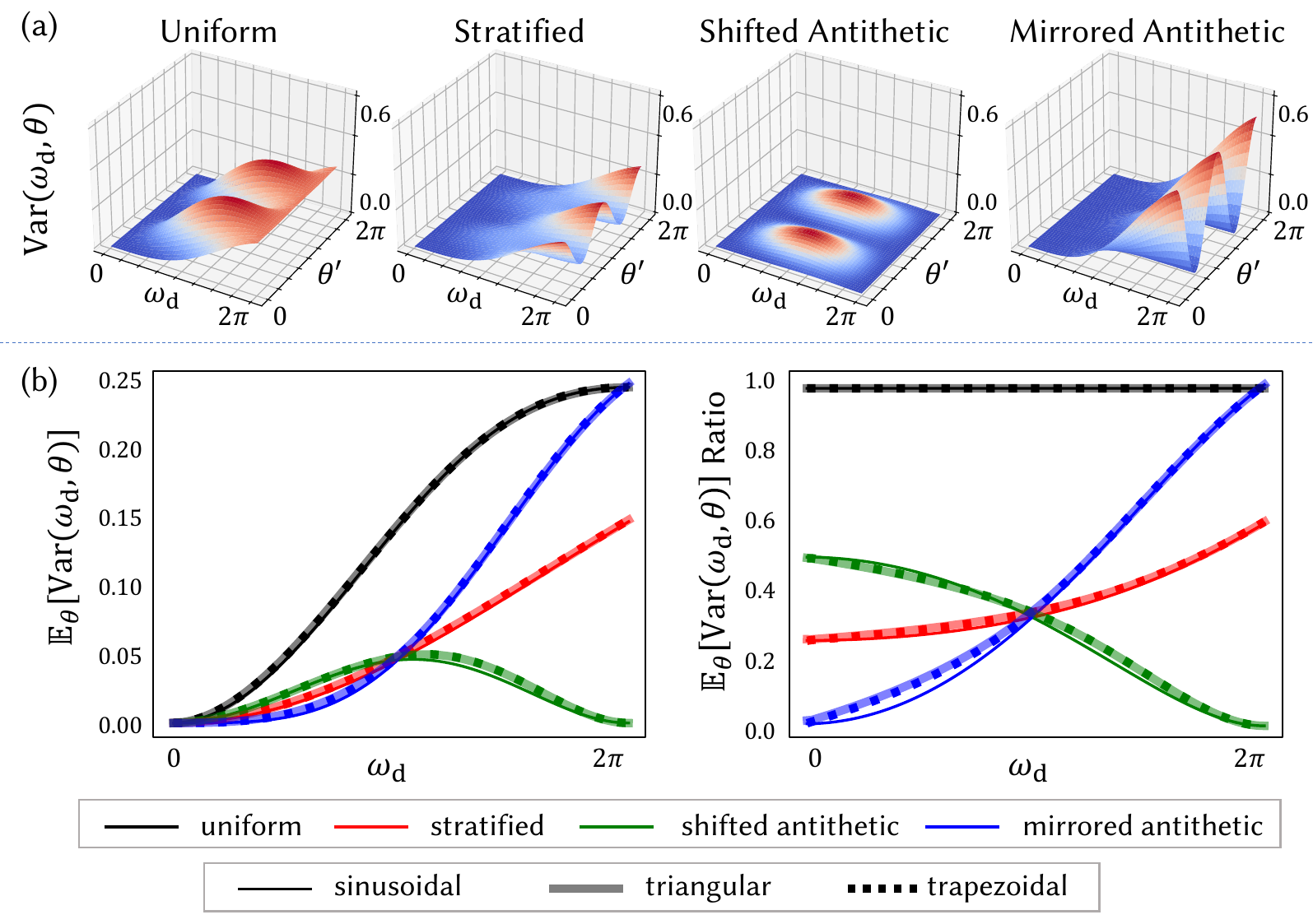}
% \vspace{0.2cm}
% \includegraphics[width=\linewidth]{figures/4_time_domain/time_sampling_method_2d_comparison_v1.5.pdf}
\caption{Comparison of time sampling methods with $N_t=2$. 
(a) We plot $\mathrm{Var}(\omegadiff, \theta)$ for \wjarosz{a} sinusoidal waveform using different sampling methods.
Like \citet{Hu:2022:Differential}, we use $\theta'=\theta+0.5\omegadiff T$ instead of $\theta$ for the x-axis to make the plot not skewed.
We set $T=1$.
(b) For different sampling methods and waveforms, we plot \jkim{(left)} $\mathbb{E}_{\theta}[\mathrm{Var}(\omegadiff, \theta)]$, and \jkim{(right)} its ratio to the variance from uniform sampling. 
%\jkim{Color indicates the sampling method, and line thickness and style indicate the waveform type.} 
Mirrored and shifted antithetic sampling are most effective for $\omegadiff \in [0, \nicefrac{\pi}{T}]$ and $\omegadiff \in [\nicefrac{\pi}{T}, \nicefrac{2\pi}{T}]$, respectively.}
\label{figure:time_sampling_method_2d_comparison}
\end{figure}

In \cref{figure:time_sampling_method_2d_comparison}, we compare the variance of antithetic sampling with other sampling methods---uniform sampling and stratified sampling, with $N_t = 2$.
For both antithetic sampling techniques, we use \wjarosz{our previously derived} optimal shift $\ts$. 
%Because shifting also affects the stratified sampling (in this case, shifting could be thought of as the starting point of the first stratum), we choose optimal shifting ($\ts=0$) to be fair. 
% For uniform sampling, shifting does not affect the performance, but it can effectively use twice more samples, so we halved the variance for a fair comparison. 
In \cref{figure:time_sampling_method_2d_comparison}-(a) we plot $\mathrm{Var}(\omegadiff, \theta)$ for different $\omegadiff, \theta$, \wjarosz{and in \cref{figure:time_sampling_method_2d_comparison}-(b) we plot the expectation of $\mathrm{Var}(\omegadiff, \theta)$ over $\theta$ and its ratio to the uniform sampling case}.
% We also plotted the variance ratio compared to uniform sampling.
Overall, antithetic sampling has \wjarosz{better} performance than both uniform and stratified sampling for different values of $\omegadiff$.
Mirrored antithetic sampling is most effective for $\omegadiff \in [0, \nicefrac{\pi}{T}]$, and \shifted antithetic for $\omegadiff \in [\nicefrac{\pi}{T}, \nicefrac{2\pi}{T}]$.
% But each antithetic sampling has an advantage at different frequency ranges.
% For $\omegadiff \in [0, \frac{\pi}{T}]$, mirrored antithetic sampling worked best, while for $\omegadiff \in [\frac{\pi}{T}, \frac{2\pi}{T}]$, \shifted antithetic sampling worked best.
% We also plotted the result for several different signals mentioned in \cref{figure:low_pass_filtering}.
% They show almost the same tendency with the sinusoidal case.
We also observe that the advantages of different antithetic sampling strategies persist across different waveforms, including triangular and trapezoidal (\cref{figure:time_sampling_method_2d_comparison}-(b)).

% \begin{equation*}
%     \mathbb{E}_{\phi} [ \text{Var}(\omega, \phi, a) ] 
% \end{equation*}

% \begin{equation}
%     \frac{\pi(\omega^2 + 2 \cosP{\omega} - 2)}{\omega^2}
% \end{equation}
% \begin{equation}
%     \frac{\pi(\omega^2 + 8 \cosP{\frac{\omega}{2}} - 8)}{2\omega^2}
% \end{equation}
% \begin{equation}
%     \frac{\pi(\omega^2 + 4 \cosP{\omega} -4)}{2\omega^2} + \frac{\pi}{2} \cosP{\frac{\omega}{2}}
% \end{equation}
% \begin{equation}
%     \frac{\pi(\omega^2 + 4 \cosP{\omega} -4)}{2\omega^2} + \frac{\pi}{2} \frac{\sinP{\omega}}{\omega}
% \end{equation}

% \begin{figure}
% \caption{Variance $\text{Var}(\omegadiff, \theta)$ comparison over $\omegadiff, \theta$ using different sampling methods. We used $T=1$.}
% \label{figure:time_sampling_method_3d_comparison}
% \end{figure}

% \begin{figure}
% \includegraphics[width=0.9\linewidth]{figures/4_time_domain/time_domain_correlation_stratified_v1.3.pdf}
% \caption{Stratified sampling methods. (a-c) is standard stratified sampling with different correlation strategies of (a) mirroring, (b) random, or (c) shifting, respectively.
% (d,e) is basically antithetic sampling, but with stratified primal sampling. If $N$ is large, (a, c) converge to (d, e).}
% \label{figure:time_domain_correlation_stratified}
% \end{figure}

% \begin{figure*}
% \includegraphics[width=\textwidth]{figures/4_time_domain/time_domain_multiple_sampling_concept.png}
% \caption{Expected variance $\expectedvariance$ comparison using different sampling methods. We plotted both variance and variance ratio to uniform sampling.}
% \label{figure:time_domain_multiple_sampling_concept}
% \end{figure*}

\begin{figure} [t]
\includegraphics[width=0.95\linewidth]{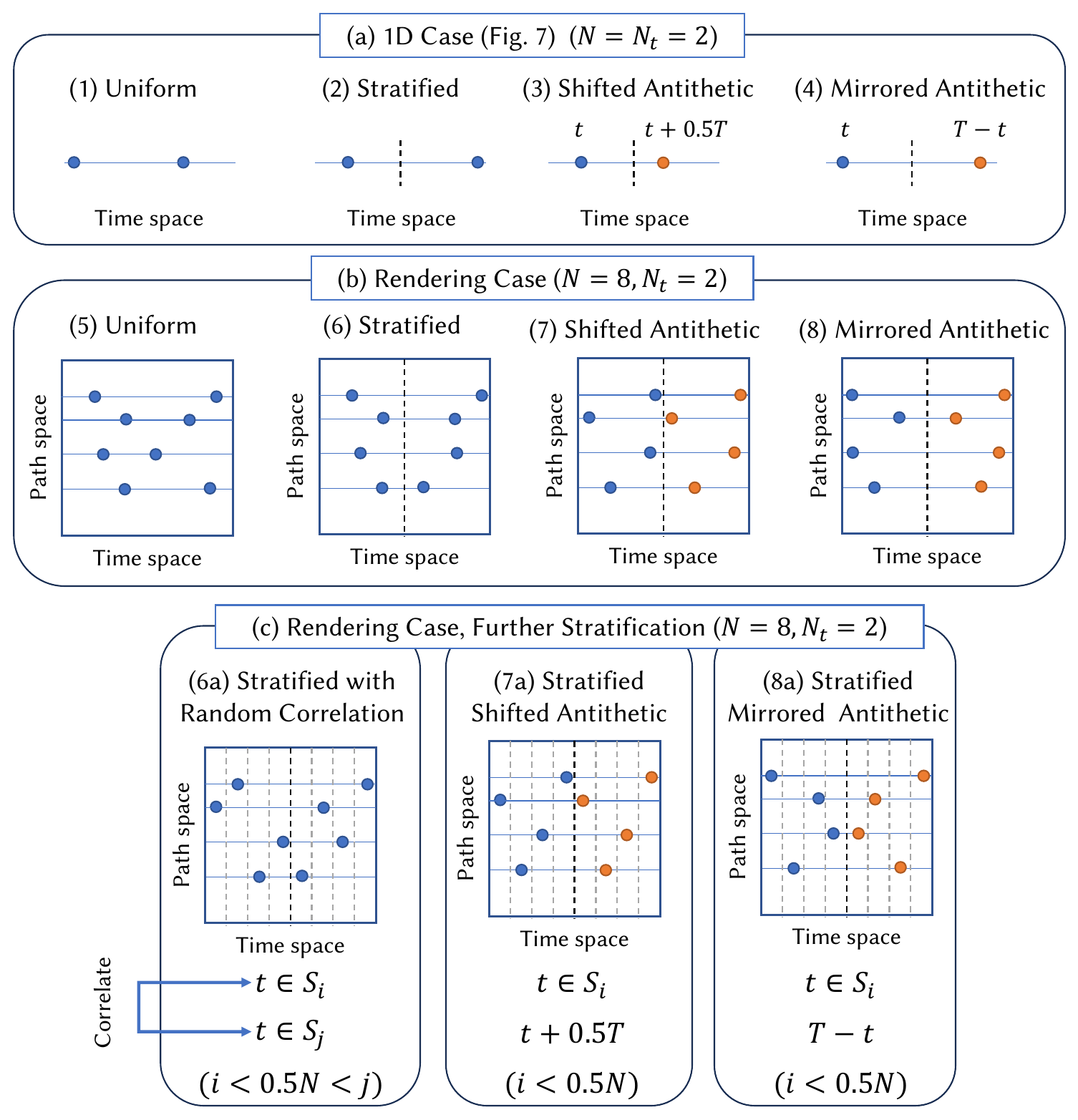}
\caption{\jkim{Comparison of time-domain stratification techniques. We visualize only cases where the primal samples are in $[0, 0.5T]$. Horizontal lines are path evolutions, as in \cref{figure:overview}. The 1D case in \cref{figure:time_sampling_method_2d_comparison} considers only the time domain and corresponds to (a) with $N=N_t=2$.
The rendering case jointly considers both time and path domains.
The direct expansion of the 1D to the rendering case is (b), with $N_t=2$.
In (c) we further improve (b) with $N_t=2$, by stratifying the time domain.
$S_i$ is a stratum of $[\nicefrac{i}{N}T, \nicefrac{(i+1)}{N}T]$.
% Finally (d) describes sampling without path correlation.
}}
\label{figure:time_domain_correlation_stratified}
\end{figure}

\begin{figure}  [t]
\includegraphics[width=0.89\linewidth]{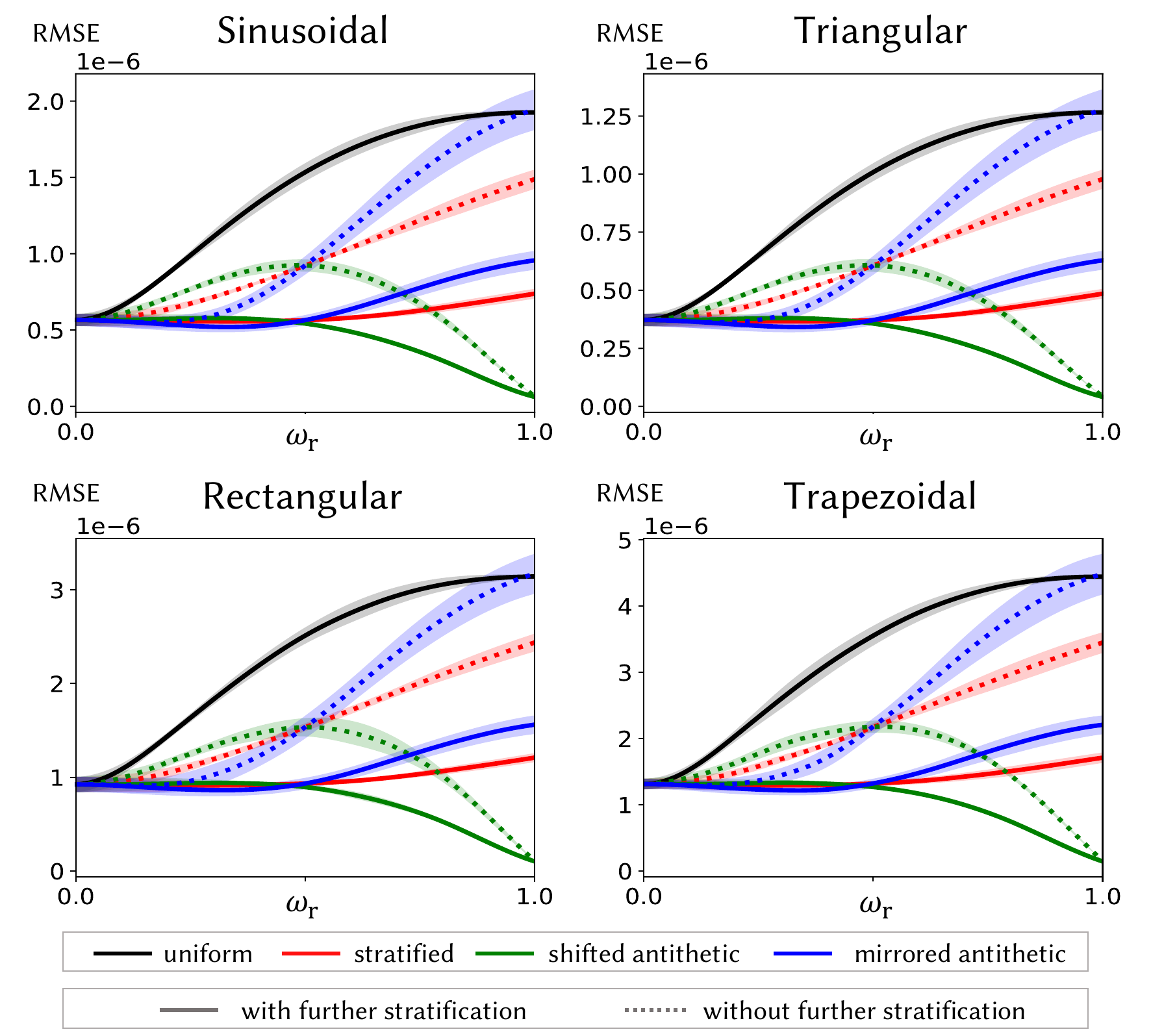}
\caption{RMSE versus $\omegaratio$ (normalized $\omegadiff$) for various sampling configurations and waveforms for the \textsc{Cornell-Box} scene. 
% \rev{Color indicates the time-sampling method while line style indicates further stratification, which namely correspond to \cref{figure:time_domain_correlation_stratified}-(b), (c).
The relative performance when not using further stratification (dotted line) is similar to the 1D case in \cref{figure:time_sampling_method_2d_comparison}.
Enforcing further stratification (solid line) improves it in all cases.}
\label{figure:result_time_sampling_methods_subregion_plot}
\end{figure}

\subsection{Further Stratification in the Time Domain}
\label{section:4_subregion_stratified_sampling}
\rev{Until now, we have only considered the simple 1D case where $N=N_t=2$, as in \cref{figure:time_domain_correlation_stratified}-(a).
For rendering, however, we will need to consider sampling over the path domain while additionally increasing $N$ (for example, $N=8$ in \cref{figure:time_domain_correlation_stratified}-(b) and (c)).
% We found that correlating more than two paths is harmful in terms of path diversity, we will only consider $N_t=2$ here.
% Extending the 1D case naively will work, but since stratification is dictated by $N_t$, it will not improve as we increase $N$ but keep $N_t = 2$; increasing $N_t$, on the other hand, is also undesirable because it reduces path diversity
The direct extension of the 1D case with $N_t=2$ would correspond to \cref{figure:time_domain_correlation_stratified}-(b).
Unfortunately, this extension does not stratify the collection of all $N$ samples across the time domain, as the stratification is still fixed by $N_t=2$.
We could improve stratification by increasing $N_t>2$, but doing so reduces path diversity for a fixed sample budget $N=N_t \cdot N_p$.
}

% \revw{Use this command for new revision.}
\rev{
Instead, we propose to combine antithetic sampling with \textit{further stratification} over the time domain (\cref{figure:time_domain_correlation_stratified}-(c)).
%In other words, instead of uniform sampling, we use stratified sampling for the primal samples which also makes the antithetic samples stratify well over the time domain.(\cref{figure:time_domain_correlation_stratified}-(7a), (8a)).
% For better visualization, we limit the range of primal samples to $[0, 0.5T]$ in \cref{figure:time_domain_correlation_stratified}, but note that it is actually $[0, 0.5T]$, and 
For $N_t = 2$, our approach distributes the primal samples within the left half of the time domain, and places the antithetic samples at either a shifted (7a) or mirrored (8a) location in the right half of the domain, providing full stratification with $N$ strata (in the actual implementation, we may swap the primal and antithetic samples).
In order to evaluate the impact of antithetic sampling, we also consider in \cref{figure:time_domain_correlation_stratified}-(6a) a fully stratified extension of \cref{figure:time_domain_correlation_stratified}-(6), but which does not exploit any antithetic properties. Instead, this approach enforces $N$ strata in time and correlates the paths of random strata with each other (stratum $i$ is correlated with a random stratum $j$ \jkim{on the other half}).}

\rev{In \cref{figure:result_time_sampling_methods_subregion_plot}, we compare the effect of using further stratification in the \textsc{Cornell-box} scene.
(We describe the experiment details in \cref{sec:experiments}.)
When we do not use further stratification, the relative performance of different techniques resembles the 1D case with $N_t=2$ in \cref{figure:time_sampling_method_2d_comparison}.
Applying further stratification keeps relative performance similar, but reduces overall variance for all time-domain sampling techniques.
Thus, to ensure optimal performance, we use in our experiments further stratification in the time domain for all sampling methods except uniform sampling.}

\subsection{Analytic Approximation}
\label{section:4_integration_based}
\jkim{
So far, we have discussed Monte Carlo integration techniques to estimate \cref{eq:doppler_rendering_path_integral_reordered}. 
We can alternatively derive an analytic approximation to this equation, which results in a fast but biased estimate.
The first-order Taylor series approximations of $\fmappingxbart$ and $\xbartlen$ are:
\begin{equation}
    \fmappingxbarz + t \frac{\partial}{\partial t} \fmappingxbart, \xbarzlen + t \frac{\partial}{\partial t} \xbartlen.
\end{equation}
We evaluate the derivatives with the finite-difference method, using values at $t=0, T$. With this approximation, we can integrate \cref{eq:doppler_rendering_path_integral_reordered} analytically.
}
\rev{
This analytic approximation is similar to the method of \citet{Heide:2015:Doppler}, except that we use first-order approximations for both $\xbartlen$ and  $\fmappingxbart$, whereas \citeauthor{Heide:2015:Doppler} use zeroth-order approximation for $\fmappingxbart$.
We found that our analytic approximation, while still biased, significantly improves rendering performance compared to \citeauthor{Heide:2015:Doppler}'s approximation (\cref{figure:result_different_spatial_correlation}).
}

%% file: sections/05_spatial_correlation_v3.6.tex
\section{Path Correlation by Temporal Shift Mapping}
\label{section:5_spatial_correlation}
We have discussed an efficient antithetic time domain sampling assuming a \rev{path mapping} $\mapping$ such that $\fmappingxbart$ and $\xbartlen$ remain approximately constant over time $t$. In this section, we will focus on finding such a mapping $\mapping$.
As a $\mapping$ that results in near-constant $\fmappingxbart$ also results in near-constant $\xbartlen$, we will focus on only $\fmappingxbart$. 
We will consider mappings $\mapping$ that we can represent as a composition of \jkimb{bijective} path vertex mapping functions $\raymapping_i(t, \cdot) : \mathbb{R}^{3} \to \mathbb{R}^3$ for each vertex $\mathbf{x}_i(0)$ until path depth $K$.
%
%\footnote{
%We abuse the notation here. For random replay, $\raymapping_i$ should be defined on random numbers required to sample $\mathrm{x}_i$.
% But exploiting theoretical bijectivity between random numbers and paths
%To be accurate, $\raymapping_i$ is conditional on previous path $\mathbf{x}_0(0)\mathbf{x}_1(0)\dots\mathbf{x}_{i-1}(0)$, but we omit this for simpler notation.
%}
%
Therefore:
\begin{equation}
    \mapping (t, \xbarz) \coloneq \raymapping_0(t, \mathbf{x}_0(0))\raymapping_1(t, \mathbf{x}_1(0))\dots\raymapping_K(t, \mathbf{x}_K(0)).
\end{equation}
% \begin{equation}
%     \mapping (t, \rbar) \coloneq \raymapping_0(t, \mathbf{r}_0)\raymapping_1(t, \mathbf{r}_1 | \mathbf{r}_0)\dots\raymapping_K(t, \mathbf{r}_K | \mathbf{r}_{K-1}\dots\mathbf{r}_{0}).
% \end{equation}
%
% Then, $f_{\mapping}(\xbart) = \prod_{i=0}^{N} f_{\raymapping_i}(\xbart_i)$.
We assume that \jkimb{pixel coordinate of} the camera ray $\raymapping_0(t, \mathbf{x}_0(0))\rightarrow\raymapping_1(t, \mathbf{x}_1(0))$ is unaltered by $\mapping$. 
% Here, we enforce one condition that camera ray $\raymapping_0(t, \mathbf{x}_0(0))\rightarrow\raymapping_1(t, \mathbf{x}_1(0))$ (\apedired{there should be a better way to say this}) should always intersect with same coordinate in camera's image plane.
% Such constraint makes camera sensor importance $W_e$ remain the same and prevents inter-pixel correlation which can be tricky to implement.
Such \wjarosz{a} constraint makes sensor importance remain the same, decreasing variance significantly. 
%\jkim{is this true?}
\jkim{Prior work has studied finding good mappings $\mapping$ in the context of \emph{shift mappings} between adjacent pixels~\cite{Lehtinen:2013:Gradientdomain, Kettunen:2015:Gradientdomain, Manzi:2016:Temporal, Hua:2019:Survey}.
\jkimb{
Our $\mapping$ is a time-domain shift, not an image-domain shift as most of the works, so we call it \textit{temporal shift mapping} following \citet{Manzi:2016:Temporal}.
However, our $\mapping$ can be defined on arbitrary shift values, not a unit frame shift as \citet{Manzi:2016:Temporal}.
}
% \jkimb{Note that our primal-antithetic path relation is equivalent to the base-offset path in these works, but shift-mapped in the temporal domain, so we call $\mapping$ as \textit{temporal shift mapping.}
% Also note that our $\mapping$ is not limited to single }
We will adopt some simple shift mapping strategies, as we discuss in the following subsections.}
\jkimb{Previously, we defined the reference path space at $t=0$, but we now equivalently use the primal time sample as a reference for convenience}.

% for delta  prevents inter-pixel correlation which can be tricky to implement.
% \begin{equation}
%     \fmappingxbart = \prod_{i=0}^N f_{\raymapping_i}(\mathbf{x}_i(t))
% \end{equation}
%We will show a few examples of $\raymapping$ in the following subsections.
% Because we are only considering two paths, primal and antithetic, we will simplify some notations.
% Let's denote primal path as just $\xbar$ and antithetic path as $\xbar_\mathrm{a}$.
% Also instead of $\mapping$ which works for sequence of the paths, we will use simpler version of it, antithetic mapping function $\mapping_a(\xbar) = \xbar_\mathrm{a}$ and corresponding $\raymapping_a$. 
% set the reference point of $\mapping$ not $\xbarz$ but $\xbart$ which is a primal path.
% Also, to make the notation simpler, let's denote primal path as just $\xbar$ and antithetic path as $\xbar_\mathrm{a}$.

\subsection{Temporal Random Replay}
\label{ssec:samplercorrelation}
\jkimb{
In temporal random replay $\raymapping_\mathrm{s}$, we use the same sequence of random numbers to generate both primal and antithetic paths~\cite{Manzi:2016:Temporal, Hua:2019:Survey}.
Temporal random replay is a mapping on \textit{primary sample space} (PSS)~\cite{Kelemen:2002:Simple}.
Therefore, to be accurate, $\raymapping_i$ should be defined on random numbers required to sample $\mathrm{x}_i$.
However, exploiting the theoretical bijectivity between random numbers and paths \cite{Bitterli:2018:Reversible}, we will keep using the path vertex representation with a bit of abuse.
We found that temporal random replay works well when path vertex locality is preserved---the transformed points still intersect the same object, with roughly the same normal, texture coordinates, etc.---but fails otherwise---near silhouette edges or high-frequency textures.
}
% Assuming that the underlying path generator uses BSDF sampling, the path correlation occurs in terms of ray direction. 
%\jkimb{Random replay, which is also known as \textit{random replay} \cite{Hua:2019:Survey}, is a mapping on primary sample space (PSS).
% Bijectivity between PSS and path space has been shown in \citet{Bitterli:2018:Reversible}.
%}
% \jkim{Random replay is similar to \textit{random replay} \citep{Hua:2019:Survey}, except we apply it to two samples in the same pixel but at different times.}
% Note that random replay is conceptually equivalent to the  \textit{attached} strategy by \citet{zeltner2021monte} or half-way vector correlation by ~\citet{Kettunen:2015:Gradientdomain}. 
% (\apedired{This sentence is not clear. Also, the fact that random replay works better for specular points, this case specular wall is not established properly.})
% \wjarosz{}
%quality deteriorates near silhouette edges}. 
% \jkim{As random replay does not preserve locality, it is also vulnerable to a high-frequency textured mesh.}

\subsection{Temporal Path Reconnection}
\label{ssec:positioncorrelation}
\jkimb{
In temporal path reconnection $\raymapping_\mathrm{p}$, we reconnect antithetic path with transformed primal path vertex along with the geometry it resides on. 
Temporal path reconnection is conceptually equivalent to \textit{path reconnection} in ~\citet{Kettunen:2015:Gradientdomain}, but transforms the target vertex according to time.
Temporal path reconnection has the advantage of preserving locality at the transformed point.
However, it does not consider BSDF importance sampling, thus it is less advantageous for making $\fmappingxbart$ constant over $t$. 
This is especially problematic for specular materials where a small change in ray direction causes significant BSDF change.
}

\jkimb{For shorter notation, we will omit the word \textit{temporal} from both temporal random replay and temporal path reconnection from now.}
\begin{figure}[t]
\includegraphics[width=\linewidth]{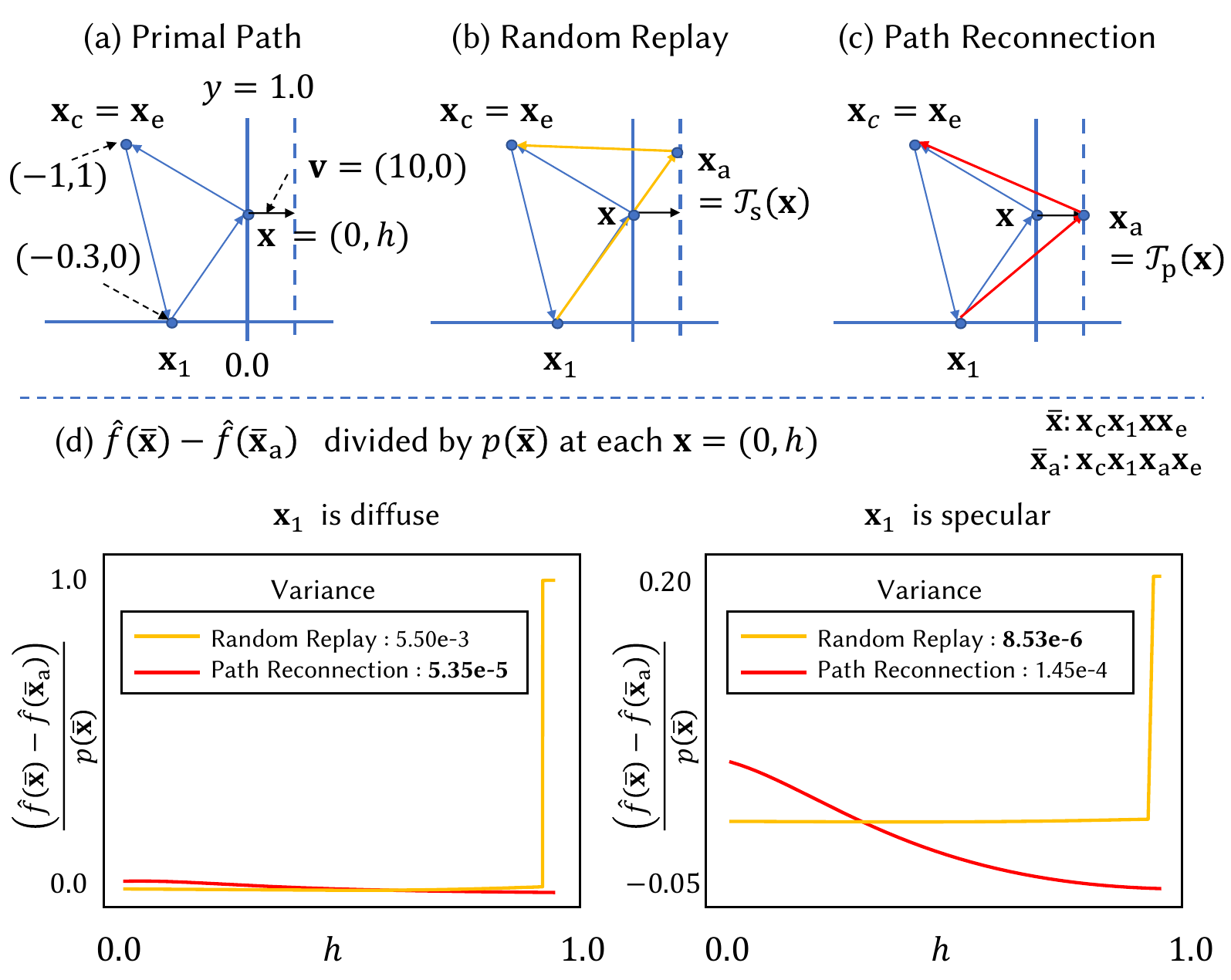}
\caption{We exploit two temporal shift mapping techniques to generate \wjarosz{an} antithetic path for a given (a) primal path. \wjarosz{Random replay (b) uses the same random numbers as the primal path to generate the antithetic path}. \wjarosz{Path reconnection (c) keeps vertices attached to the moving geometry to preserve locality}. We show (d) the difference of $\hat{f}$ between the primal and the antithetic path, divided by the sampling pdf, for vertical wall points. Random replay has better correlation everywhere except near \wjarosz{the} edge, which results in significant variance for \wjarosz{the} diffuse case, but not \wjarosz{the} specular case. 
% \apedired{Please add y-axis label which is currently the title (Just say the math equation) and state that constant is better. There is redundancy in notation. The x-axis for d should be h.}
}
\label{figure:sampler_position_correlation_concept}
\end{figure}

\begin{figure}[t]
\includegraphics[width=0.85\linewidth]{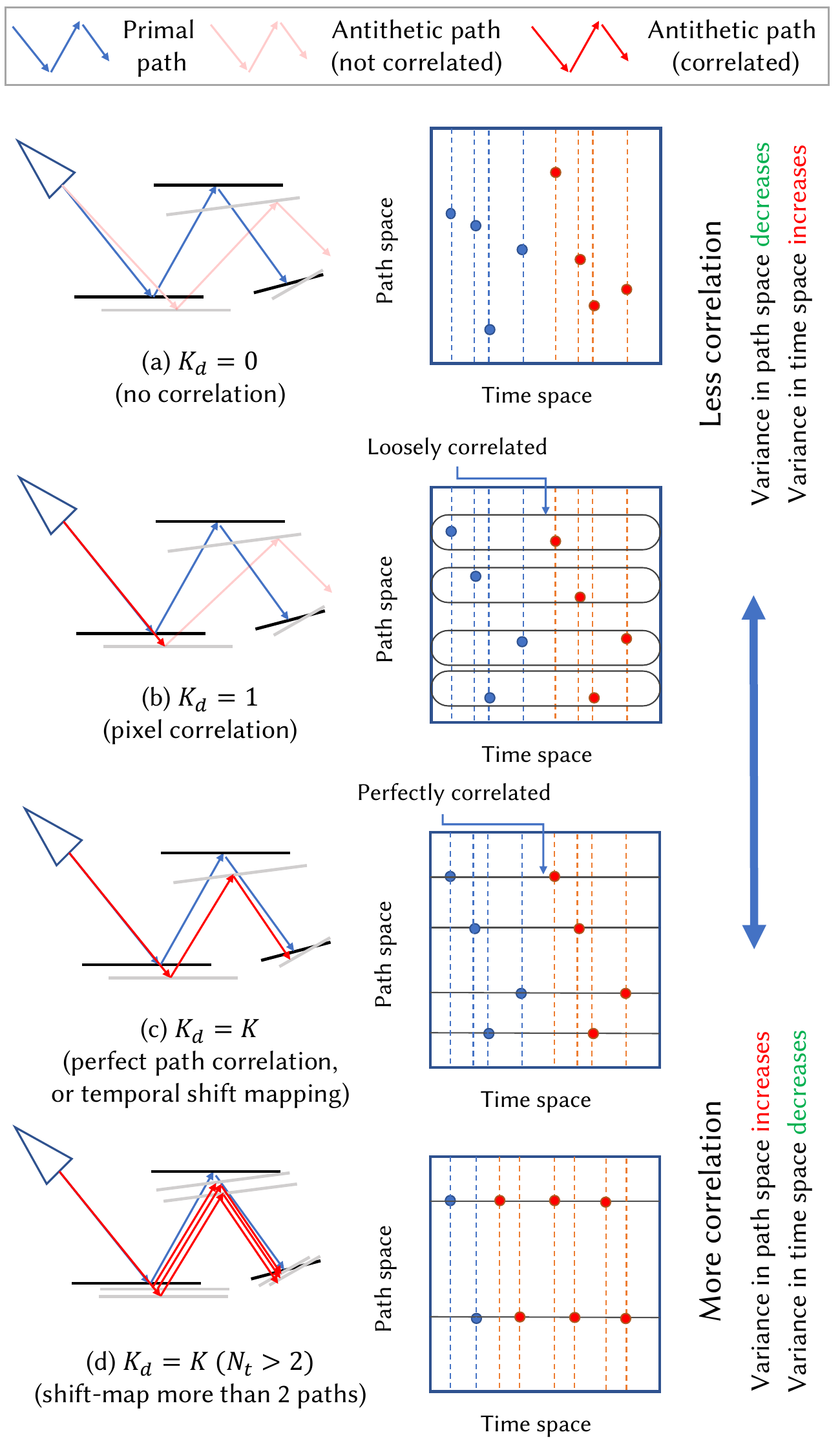}
\caption{Effects of varying path correlation strength. By increasing the shift mapping depth \jkimb{or mapping more paths}, and thus making paths more correlated, we can decrease variance of the time sub-integral, but increase variance of the path sub-integral. The optimal shift mapping depth depends on the scene configuration and heterodyne frequency.}
\label{figure:effect_of_correlation}
\end{figure}

\subsection{\jkim{Analysis of Random Replay and Path Reconnection}}
To better understand the two types of \jkimb{temporal shift} mappings, we consider the simple 2D scene in \cref{figure:sampler_position_correlation_concept}. 
We denote \wjarosz{a} primal path $\xbar$. and \wjarosz{its} antithetic path  $\xbar_\mathrm{a}$. 
% Because we only consider primal and antithetic samples, we denote each as $\xbar$ and $\xbar_\mathrm{a}$ for simpler notation.
The \wjarosz{floor} is fixed and the \wjarosz{wall that runs vertically between $y\in[0,1]$ moves in the positive $x$ direction with speed of $10$}. 
The point light source ($\mathbf{x}_\mathrm{e}$) and sensor ($\mathbf{x}_\mathrm{c}$) are collocated. 
We use BSDF sampling to sample the primal path at $t=0$, which has the form $\xbar=\mathbf{x}_\mathrm{c}\mathbf{x}_1\mathbf{x}\mathbf{x}_\mathrm{e}$ with vertices $\mathbf{x}_\mathrm{c} = (-1, 1), \mathbf{x}_1 =(-0.3,0), \mathbf{x} = (0, h)$, \wjarosz{and thus only depends on $h\in[0,1]$}. 
% \jkim{
% }
% \jkim{Because the light source is point light and we assume to preserve the camera ray,}
To map $\xbar$ to its antithetic path $\xbar_\mathrm{a}$ \jkimb{at $\ta=0.5T$, where $T=0.001$}, our only degree of freedom is changing the intersection point $\mathbf{x}_\mathrm{a}$ on the moved wall at $\ta$, which we can do using either $\raymapping_\mathrm{s}$ or $\raymapping_\mathrm{p}$. 
% Because path only depends on $\mathbf{x}$, we will represent path just with $\mathbf{x}$.
% If we use random replay, it gives the same direction as the primal path, which makes $\mathbf{x}_1, \mathbf{x}, \mathbf{x}_\mathrm{a}$ locate on the same line.
To compare the two mappings, in \cref{figure:sampler_position_correlation_concept} we plot the difference of $\hat{f}(\xbar) - \hat{f}(\xbar_\mathrm{a})$ divided by the primal path pdf $p(\xbar)$, which is an unbiased estimate for \cref{eq:doppler_rendering_path_integral_reordered} with perfectly matching modulation term. 
We observe that random replay gives near-zero difference for most of the $h$ values.
However, for primal points near the edge of the wall, random replay gives an invalid antithetic point, and $\hat{f}$ drastically changes. 
This results in a large spike near the edge if $\mathbf{x}_1$ is diffuse.
On the other hand, path reconnection does not suffer from this discontinuity problem.
For specular materials, random replay still suffers from the discontinuity, but is better than path reconnection. This is because path reconnection causes a dramatic change in BSDF, which makes $\fmapping$ differ significantly between primal and antithetic paths.
% is much larger which will be discussed in the following subsection.
% As random replay does not preserve locality, it is also vulnerable to a high-frequency textured mesh.
% the error is significantly lower in this case. 
% much alleviated because sampling pdf for near-edge regions becomes small.

% Therefore, $f_{\raymapping}$ remains same which is a good news for antithetic sampling.
% The disadvantage of random replay is that locality is not guaranteed to be preserved.
% For example in Figure~\ref{figure:sampler_position_correlation_concept}, if primal point locate near the edge of the wall, random replay gives invalid antithetic point and $f_{\raymapping}$ does not match at all.
% You could easily check this by difference spike near the edge.
% Because the locality is not preserved, random replay is also vulnerable to a high frequency textured mesh.

\paragraph{Primal and Antithetic Multiple Importance Sampling}
As any path could be sampled as either primal or antithetic, we use multiple importance sampling (MIS)~\citep{Veach:1995:Optimally} between primal and antithetic samples. MIS is known to reduce variance in path reconnection~\cite{Kettunen:2015:Gradientdomain}.

\jkimb{
\paragraph{Ensuring Unbiasedness}
MIS between primal and antithetic samples also helps ensure unbiasedness---it effectively rejects a shift mapping if \jkimb{bijectivity fails because} the antithetic path has zero throughput (e.g., due to a missing ray or path blocking) and thus could not have been sampled as the primal path.
In this case, the integration falls back from \cref{eq:dynamic_tof_path_integral_reordered} to  \cref{eq:dynamic_tof_path_integral}, but is still unbiased.
One thing we need to be careful about is to ensure the primal sample covers whole time domain so that there exists no unsampled area even though shift mapping fails.
This is why we randomly swapped the stratum of primal and antithetic samples in \cref{section:4_subregion_stratified_sampling}.
}

% \paragraph{MIS}
% First, we can do MIS between primal and antithetic path as path could be sampled as either a primal or antithetic paths.
% Second, we can do MIS between multiple antithetic mapping functions. But this is 
% \paragraph{Handling invalid antithetic sample}
% \jkim{Finding antithetic path may fail in some cases like using random replay for near edge in \cref{figure:sampler_position_correlation_concept}.
% In this case, we do not average with the primal path, but just reject the antithetic path to ensure unbiasedness.
% Detail could be found in supplementary material.}
% \jkim{MIS also has advantage to guarentee the unbiasedness by handling the case that antithetic path becomes invalid.}
% This is also important to make the renderer unbiased.
% To elaborate, if antithetic path becomes invalid, primal path should not be averaged with antithetic path because antithetic path cannot be sampled as a primal path.
%it refers two thing - (1) the antithetic path cannot be sampled as a primal path and (2) the primal path could not be sampled an antithetic path.

\paragraph{Implementation}
Random replay is preferable to path reconnection in terms of implementation complexity. We can implement random replay \jkimb{implicitly} by simply repeating the path tracing process with the same sampler and random seed, but replacing the primal with the antithetic time sample. 
On the other hand, to implement path reconnection, we need to either trace primal and antithetic paths at the same time, or store the primal path and update its vertices to form the antithetic path.%
\footnote{Random replay can be also implemented this way, but the implicit method is preferable.}
Both approaches have considerable implementation overhead.

\subsection{\jkimb{Adaptive Temporal Shift Mapping}}
\label{ssec:selective}
\wjarosz{As neither random replay nor path reconnection is universally better, we follow \citet{Kettunen:2015:Gradientdomain} and combine the two mapping strategies adaptively based on the vertex material:}
% To this end, exploiting the idea from \citet{Kettunen:2015:Gradientdomain}, we use a selective correlation method that depends on material properties.
\jkimb{If current, next vertex of primal path ($\mathbf{x}_i, \mathbf{x}_{i+1}$), and current vertex of antithetic path ($\mathbf{x}_{\mathrm{a},i}$) are identified as diffuse material, we use path reconnection, otherwise, we use random replay.}
% If the current vertex is identified as a specular material, we use random replay, otherwise, we use path reconnection.
\jkimb{We found this adaptive approach to be effective for most scenes.}
% For example in \cref{figure:sampler_position_correlation_concept}, even $\mathbf{x}_1$ is diffuse, if point is close to the wall($\mathbf{x}_1 \rightarrow (0, 0)$), using random replay is much better because path reconnection causes large cosine term change.
An alternative approach would be to use MIS between the two shift mapping strategies. Unfortunately, we found this approach to be prohibitively expensive, as it requires storing $2^K$ paths created with all possible per-vertex combinations of random replay versus path reconnection.
% \wjarosz{Whereas for MIS between BSDF sampling and next-event estimation in standard path tracing we need to store only the main path that BSDF sampling produces, we would need to store} $2^K$ paths created with all possible combinations of sampler and path reconnection. 
% Even though we only consider path reconnection path and random replay only path, we need to trace at least 5 paths - (1) primal, (2) correlation position only, (3) primal whose random replay is (2), (3) random replay only and (5) primal whose path reconnection is (2).
Furthermore, we show in the supplement that such an MIS approach is not guaranteed to perform well under antithetic sampling.
% \apedired{Juhyeon: Do you mean supplementary? We need to discuss this paragraph. }.

% Meanwhile, we also have to consider path occlusion.
% If antithetic path is blocked, we reject it instead of tracing ray from blocked point.
% antithetic path may be blocked by other geometry
% We also need to discuss MIS of primal and antithetic path.
% This is important because 
% One other thing that we have to notice is that if point is close to the wall($\mathbf{x}_1 \rightarrow (0, 0)$), using random replay is much better regardless of the material.
% This is due to drastic cosine term change.

\subsection{\wjarosz{Depth-limited} \jkimb{Temporal Shift Mapping}}
\label{ssec:partialcorrelation}
% Finally, we also need to discuss partial correlation.
In practice, instead of mapping the entire path until vertex $K$, we can limit mapping only to the first $K_d < K$ vertices,
% \TODO{cite \cite{Loubet:2019:Reparameterizing}}
%
\begin{equation}
    \mapping (t, \xbarz) \coloneq \raymapping_0(t, \mathbf{x}_0(0))\raymapping_1(t, \mathbf{x}_1(0))\dots\raymapping_{K_d}(t, \mathbf{x}_{K_d}(0)).
\end{equation}
%
% We correlate for the first few path segments and do not correlate after a certain depth.
% In this case, $\mappingt$ is not smooth on path space, so we cannot use Jacobian.
% Instead, we can evaluate covariance.
% Assuming that we already sampled in time domain with $t$ and antithetic sample of $t_a$, and also path $\xbar(t)$, our goal is to sample  $\xbar(t_a)$ that minimizes variance of following equation.
% \begin{equation}
%     \int_{\mathcal{P}(t)}f(\barbf{x}(t), t) \dmu{\barbf{x}(t)} + 
%     \int_{\mathcal{P}(t_a)}f(\barbf{x}(t_a), t_a) \dmu{\barbf{x}(t_a)}
% \end{equation}
% Let's say that we are using path sampling pdf $p$ and $p_a$ respectively for primal and antithetic samples.
% % For primal path sampling, it is equal to traditional sampling strategies such as BSDF sampling or direct illumination sampling method.
% To make the notation easier, let's define $\barbf{x}(t)$ as $\barbf{x}$ and $\barbf{x}(t_a)$ as $\barbf{x}_a$.
% Then, the variance of the above equation is
% \begin{equation}
%     \variance{\frac{f(\barbf{x})}{p(\barbf{x})}} + \variance{\frac{f(\barbf{x}_a)}{p_a(\barbf{x}_a)}} +
%     2\covariance{\frac{f(\barbf{x})}{p(\barbf{x})}, \frac{f(\barbf{x}_a)}{p_a(\barbf{x}_a)}}
% \end{equation}
% If $\barbf{x}$ and $\barbf{x}_a$ are not correlated (independent), then the covariance term becomes close to zero.
After $K_d\textsuperscript{th}$ vertex, we continue the path using independent path tracing.%
\footnote{Our choice to stop shift mapping at vertex $K_d$ means that our rendering algorithm uses a hybrid between the path integrals of \cref{eq:dynamic_tof_path_integral} and \cref{eq:dynamic_tof_path_integral_reordered}, \jkimb{with partial Jacobian}.}
\jkimb{Note that this is no longer deterministic shift mapping, but correlation with some stochasticity.
}
This \emph{depth-limited} approach makes the primal and antithetic paths less correlated, and thus makes antithetic time sampling less effective.
%$\fmappingxbart$ not constant over time $t$, thus makes 
However, it is advantageous in terms of increasing diversity in path space. \Cref{figure:effect_of_correlation} shows this trade-off between path correlation and path diversity. %The more we correlate path samples, we get the advantage in time-space to evaluate the modulation term, but we lose path space diversity.
% We also plotted two cases where either path-space variance dominates or time-space variance dominates in \cref{figure:path_time_variance_dominate}.
% If the variance in time-space dominates, it is better to use more correlation, but if the variance in path space dominates, it is better to use less correlation.
Empirically, we found that perfect path correlation, (or temporal shift mapping), works best at near-heterodyne modes, whereas limited correlation works best at near-homodyne modes. 
% The result will be explained in detail in the later part.

% \begin{figure} 
% \includegraphics[width=0.5\textwidth]{figures/5_spatial_correlation/path_time_variance_dominate_v1.0.pdf}
% \caption{\TODO{remove this} MAE of using multiple sampling methods.}
% \label{figure:path_time_variance_dominate}
% \end{figure}

%% file: sections/06_results_v3.5.tex
\section{Experiments}
\label{sec:experiments}

\begin{figure*} [t]
\includegraphics[width=0.95\linewidth]{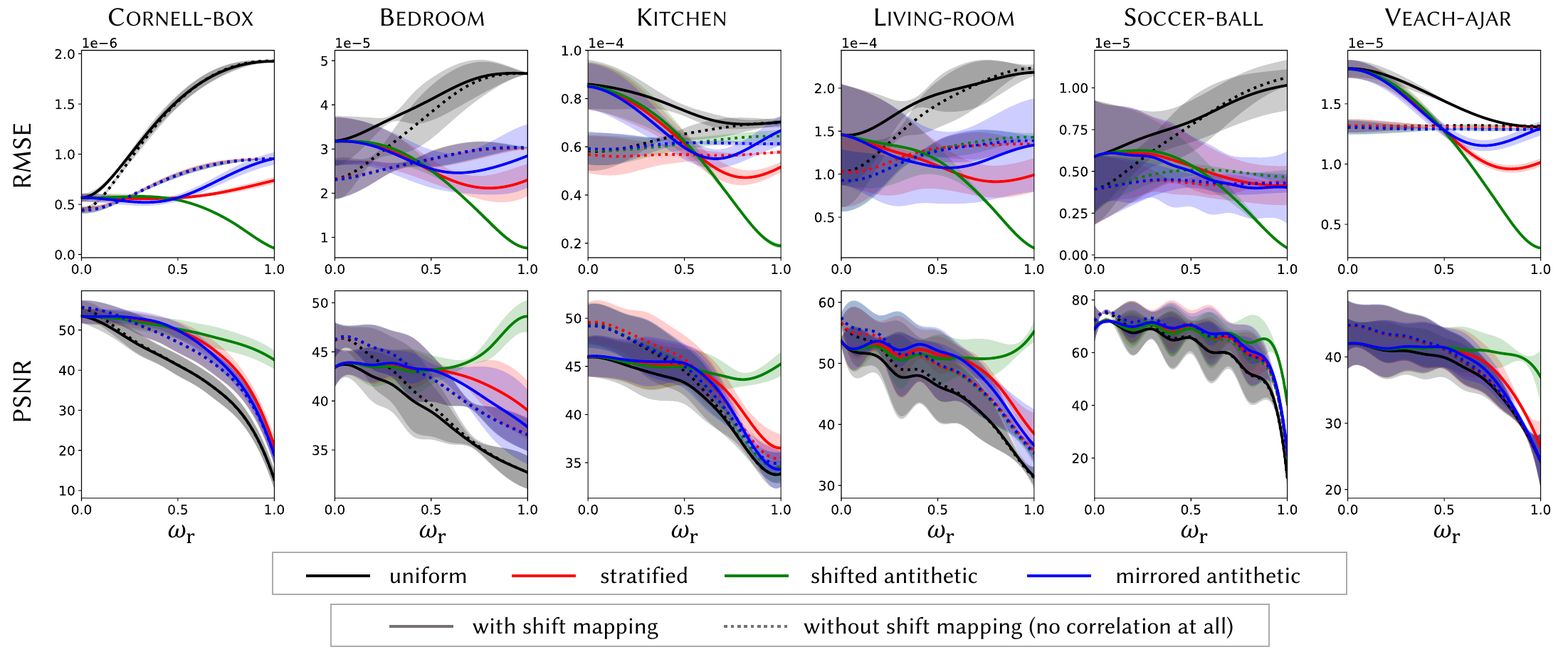}
\caption{RMSE and PSNR of different sampling methods for $\omegaratio \in [0, 1]$ ($x$-axis for all of the plots). We plotted mean and variance for uniformly spaced values of $\psi$.
% \jkim{Note that color indicates time-sampling method while line style refers whether using \jkimb{\temporalmapping}, which namely corresponds to \cref{figure:time_domain_correlation_stratified}-(c) with and without \jkimb{\temporalmapping}.
Shifted antithetic sampling with \jkimb{\temporalmapping} had the best performance for $\omegaratio \in [0.5, 1.0]$.
For $\omegaratio \in [0.0, 0.5]$, among techniques with \jkimb{\temporalmapping}, mirrored antithetic sampling performed the best; but \jkimb{\temporalmapping} did not always improve performance.
}
% As expected, mirrored antithetic sampling converges faster for $\omegaratio \in [0, 0.5]$, and shifted antithetic sampling converges faster for $\omegaratio \in [0.5, 1]$. 
% \apedired{Juhyeon: Please increase the font size for numbers. Please add x-axis labels.}
% \apedired{Please add x-axis labels and change the first sentence appropriately. }
\label{figure:result_error_plot}
\end{figure*}

In this section, we evaluate the time-path sampling techniques we proposed in \cref{section:4_time_domain_antithetic_sampling,section:5_spatial_correlation} for various scene geometries, modulation functions, and modulation frequencies. 
% We thoroughly demonstrate the effectiveness of our method under various configurations. 
% In this section, we focus on rendering and do not address velocity reconstruction, which will be discussed in Section~\ref{section:7_applications}.
% \url{https://github.com/juhyeonkim95/MitsubaDopplerToF}
% \url{https://github.com/juhyeonkim95/Mitsuba3DopplerToF}
\paragraph{Implementation and Experimental Settings}
We implement CPU and GPU versions of our algorithm using Mitsuba 0.6
%\url{https://anonymous.4open.science/r/MitsubaDopplerToF-EAC7/README.md}
~\cite{Jakob:2013:Mitsuba} and Mitsuba 3
%\url{https://anonymous.4open.science/r/Mitsuba3DopplerToF-4DE7/README.md}
~\cite{Jakob:2022:Mitsuba3}.
\wjarosz{We use six scenes for evaluation, with a collocated point light source and camera.} Across all scenes, we set $\omega_g = \qty{30}{MHz}$ and $T=\qty{1.5}{ms}$, same as ~\citet{Heide:2015:Doppler}.
We use \wjarosz{$1024$ samples per pixel (spp)} for all experiments.
\wjarosz{We also set the maximum bounce depth to $4$, or $8$ if the scene contains refractive objects.}
Unless we state otherwise, we use \wjarosz{\jkimb{$N_t=2$ and random replay} with full depth by default}.
To simplify notation, we use a normalized version $\omegaratio \in [0, 1]$ instead of $\omegadiff \in [0, \nicefrac{2\pi}{T}]$ in this section.

\begin{figure} [t]
\includegraphics[width=0.9\linewidth]{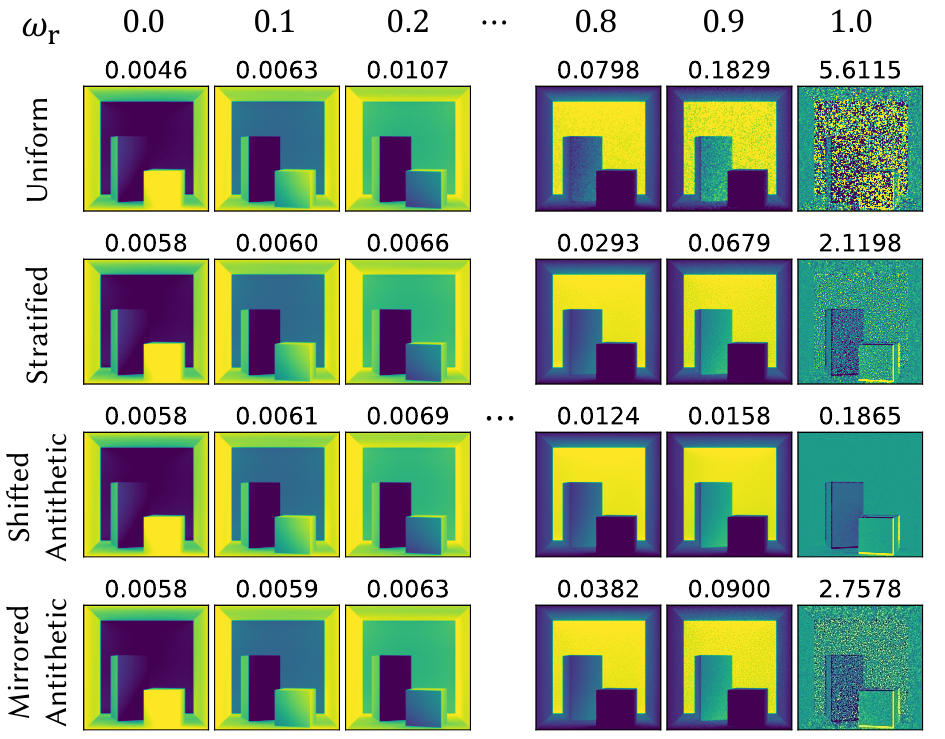}
\caption{\jkim{Qualitative results in \textsc{Cornell-Box} for different values of $\omegaratio$ at $\psi=0$. 
% Each image at certain $\omegaratio$ has same image value range.
Above each image we denote relative RMSE. We use \jkimb{\temporalmapping} for all methods except uniform sampling. Performance differences are very noticeable at $\omegaratio=1.0$, but become essentially indistinguishable for $\omegaratio<0.8$.}}
\label{figure:images_over_frequency}
\end{figure}

\begin{figure*} [t]
\includegraphics[width=\linewidth]{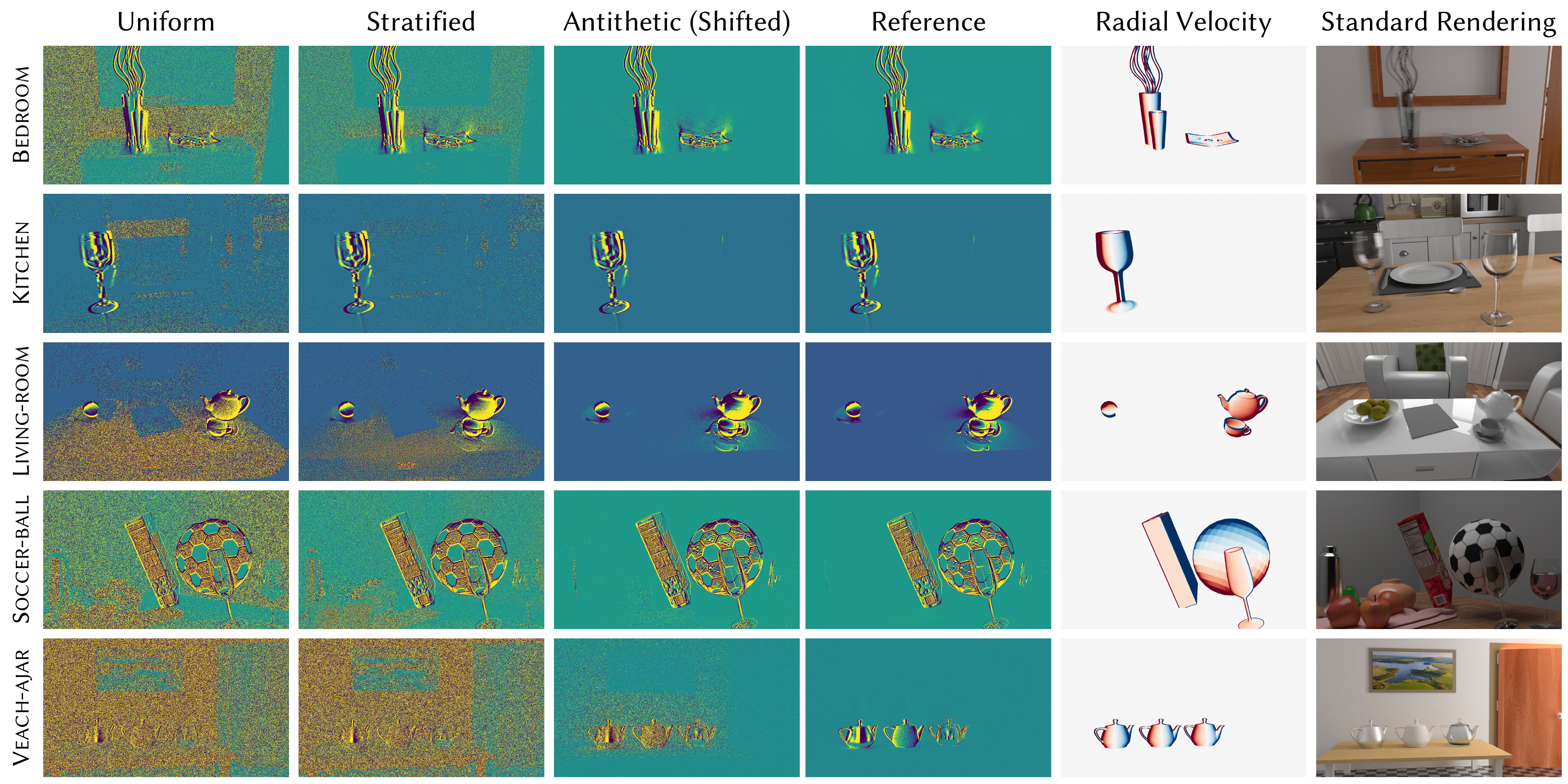}
\caption{Qualitative results using \wjarosz{different} time sampling methods for perfect heterodyne mode ($\omegaratio=1.0$) with $\psi=0$.
\jkim{We used \jkimb{\temporalmapping} in all results, as it performs the best at $\omegaratio=1.0$ (\cref{figure:result_error_plot}).
We omit mirrored antithetic sampling for easier visualization. For each scene, we show D-ToF renderings, standard intensity renderings, and the ground truth radial velocity maps computed from depth differences at $t=0,T$.
}
Our proposed shifted antithetic sampling method has significantly lower variance than the uniform or stratified sampling methods.}
\label{figure:result_time_sampling_methods}
\end{figure*}

\subsection{Effectiveness of Antithetic Sampling}
\label{section:6_comparison_of_time_sampling}
To demonstrate the effectiveness of our proposed antithetic sampling methods \jkim{with \jkimb{path correlation}}, we compare them against standard sampling techniques \jkim{with and without \jkimb{\temporalmapping}} for various scenes \jkim{under} different heterodyne frequencies $\omegaratio$ and sensor phase offsets $\psi$ (\jkim{$11\times11$ configurations of $\omegaratio \in [0, 1]$, $\psi \in [0, 2\pi]$ with uniformly-spaced intervals}).
\jkim{In each method except uniform sampling, we use further stratification as we described in \cref{section:4_subregion_stratified_sampling}.}
% To verify this, we use $11\times11$ configurations of $\omegaratio \in [0, 1]$, $\psi \in [0, 2\pi]$ with equal intervals.

% We demonstrate the effectiveness of our proposed antithetic sampling method (Section~\ref{section:4_time_domain_antithetic_sampling}) across arbitrary heterodyne frequencies $\omegaratio$ and sensor phase offsets $\psi$.
% To verify this, we use $11\times11$ configurations of $\omegaratio \in [0, 1]$, $\psi \in [0, 2\pi]$ with equal intervals.
\jkim{
\Cref{figure:result_error_plot} shows averages and standard deviations of RMSE and PSNR computed across $\psi$ values.
For $\omegaratio \in [0.5, 1.0]$, shifted antithetic sampling with \jkimb{\temporalmapping} clearly shows the best performance, especially at perfect heterodyne mode ($\omegaratio = 1$), where it is around two orders of magnitude better in terms of squared error (variance) \jkimb{compared to the worst case of uniform sampling}.}

\jkim{On the other hand, for $\omegaratio \in [0.0, 0.5]$, there is no clear winner. 
If we consider only sampling methods with shift mapping (solid line), their relative performance is similar to the 1D case in \cref{figure:time_sampling_method_2d_comparison}, where mirrored antithetic sampling works best for $\omegaratio \in [0.0, 0.5]$, shifted antithetic sampling works best for $\omegaratio \in [0.5, 1.0]$, and stratified sampling falls somewhere in between.
However, \jkimb{such perfect path correlation} turned out to be not always helpful for $\omegaratio \in [0.0, 0.5]$. 
%which makes mirrored antithetic sampling with \jkimb{\temporalmapping} not always the best one for $\omegaratio \in [0.0, 0.5]$.
}

\jkim{This difference in performance is due to the trade-off we explained in \cref{figure:effect_of_correlation} between variance in path space and time space.
For the homodyne case ($\omegaratio = 0$), the modulation term is near constant, so the variance in path space is the main bottleneck and we should dedicate more samples for path diversity, which makes \jkimb{\temporalmapping} harmful.
As $\omegaratio$ increases, the variance in time space due to the modulation term increases and eventually outweighs the variance in path space. Then, we should favor better time sampling with \jkimb{\temporalmapping} at the cost of reduced path diversity. The exact value of $\omegaratio$ where this change occurs is scene-dependent.
% Variance in time-space clearly dominates variance in path space for $\omegaratio \in [0.5, 1.0]$, but for $\omegaratio \in [0.0, 0.5]$, it depends on the scene.
For example, for a simple scene like the \textsc{Cornell-box}, path-space variance is relatively low, so improving time sampling using mirrored antithetic with shift mapping starts to become helpful from $\omegaratio\approx0.2$.
\jkimb{This value is $0.3$ for the \textsc{Living-room}, while does not appear for other scenes.}
\jkimb{We have more discussion on this correlation-diversity trade-off in \cref{sec:effect_of_path_correlation_strength}.}
% but this does not hold for other scenes.
% However, for other scenes, the range that 
}

\jkim{
Fortunately, when we consider PSNR, performance differences between methods are relatively small for $\omegaratio \in [0.0, 0.5]$
% In other words, the error compared to ToF image intensity itself is relatively low for $\omegaratio \in [0.0, 0.5]$ which makes the difference between the sampling methods less noticeable 
(\cref{figure:images_over_frequency}).
However, for $\omegaratio \in [0.5, 1.0]$, the differences become more pronounced, especially at $\omegaratio=1.0$. 
\jkimb{This implies that as $\omegaratio$ becomes larger, it becomes more important to use carefully designed sampling techniques, in which case the proposed method can be particularly helpful.}
The qualitative evaluation for perfect heterodyne mode in \cref{figure:result_time_sampling_methods} shows clear visual differences between sampling methods.
}

\paragraph{\jkim{Using Other Waveforms}}
\rev{In \cref{figure:result_time_sampling_methods_subregion_plot}, we show that our method performs well with other widely used non-sinusoidal modulation waveforms \jkimb{(rectangular, triangular and trapezoidal)}.
Overall, relative performances remain consistent with the sinusoidal case.
}
% \Cref{figure:result_time_sampling_methods_subregion_plot} also demonstrates that the proposed method performs well with other modulation waveforms.
%Furthermore, we found that antithetic sampling also performs well with other modulation waveforms, as demonstrated in \cref{figure:result_time_sampling_methods_subregion_plot}. 

% We only plotted full heterodyne mode, that the difference was most dramatic.
% In addition to ToF images, we also show radial velocity which is defined by depth map difference at $t=0$ and $t=T$ divided by $T$.
% Standard rendering is also included to show material properties.

% Overall, the result was consistent with our analysis in 1D case.
% If $\omegaratio \in [0.5, 1.0]$, \shifted antithetic sampling performed best,
% while for $\omegaratio \in [0, 0.5]$, mirrored antithetic sampling with mirroring was best.
% However, the effectiveness of using antithetic sampling is most noticeable for $\omegaratio$ close to 1.
% Without using antithetic sampling for $\omegaratio=1.0$ generates noise almost 10 times of original image intensity, which makes the result disastrously noisy.
% But using antithetic sampling reduced this error to acceptable range.
% On the other hand, the advantage of using mirrored antithetic sampling for $\omegaratio \in [0.0, 0.5]$ was not dramatic.
% We enlarged the part of the graph to show it more clearly.

% Finally, we also showed the result for using different modulation waves in \cref{figure:result_other_signals}.
% They almost showed identical tendency with sinusoidal case.
\begin{figure} [t]
\includegraphics[width=0.95\linewidth]{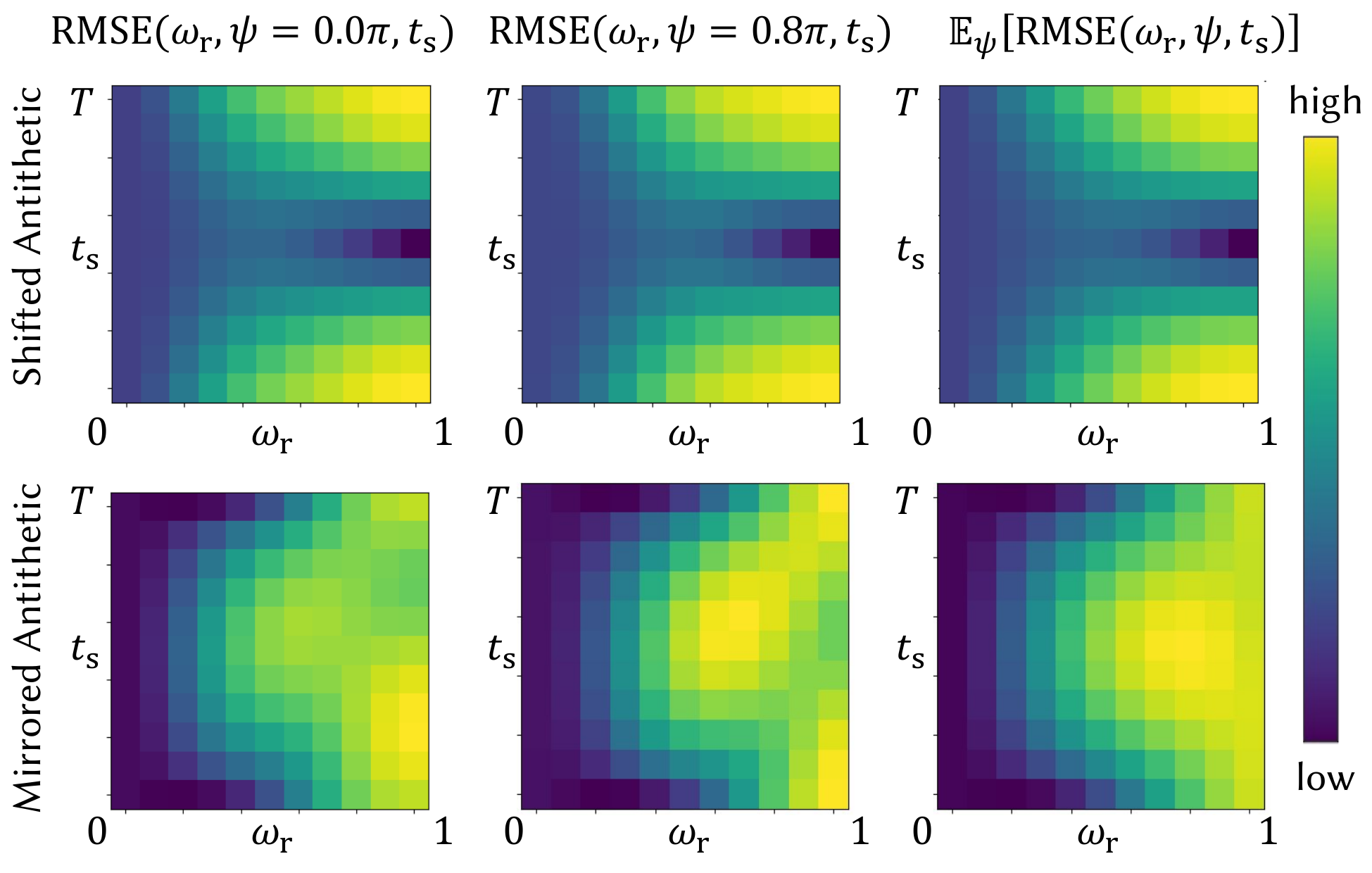}
\caption{\jkim{RMSE by using different antithetic shifts ($\ts$) for \textsc{Cornell-Box}. For \shifted antithetic, the optimal value occurs at $0.5T$ independent of $\omegaratio, \psi$. 
For mirrored antithetic, the optimal value depends on $\psi$. However, the optimal expected RMSE over uniform $\psi$ occurs at $\ts=0$, independent of $\omegaratio$.}}
\label{figure:result_different_shift}
\end{figure}

% \begin{figure} 
% \includegraphics[width=\linewidth]{figures/6_result_render/result_other_signals_v1.0.pdf}
% \caption{MAE for other modulation signals for \textsc{Cornell-Box}. The tendencies are almost identical to sinusoidal case. \TODO{remove this?}}
% \label{figure:result_other_signals}
% \end{figure}

\paragraph{Verification of Optimal Antithetic Shift}
To verify our claim for the optimal shift $\ts$ for antithetic sampling (\cref{section:4_shifted_antithetic_sampling,section:4_mirrored_antithetic_sampling}), in \cref{figure:result_different_shift} we show experiments using varying shift $\ts$ in $[0, T]$. The first two columns show results for specific values of $\psi$, and the last column shows averages across all values of $\psi$. Shifted antithetic sampling performs the best results when $\ts = 0.5T$ for all combinations of $\omegaratio$ and $\psi$.
For mirrored antithetic sampling, the optimal shift value varies depending on $\psi$. However, when we consider the averaged results across all values of $\psi$, optimal performance occurs at $\ts=0$ and $\ts=T$. These observations are consistent with our theory.
% The result is shown in \cref{figure:result_different_shift}.
% For the first two columns, we plot the result with specific $\psi$, while we average over $\psi$ for the last column.
% As expected, shifted antithetic sampling showed the best result at $\ts=0.5T$ for every $\omegaratio, \psi$.
% But for mirrored antithetic sampling, it showed different optimal value depends on $\phi$.
% However, when we average them assuming that $\phi$ has an uniform distribution, it showed the minimum value at $\ts=0, T$ as expected.

\subsection{Effect of Path Correlation Strength}
\label{sec:effect_of_path_correlation_strength}
We investigate the effect of varying degrees of correlation between sampled paths, as illustrated in \cref{figure:effect_of_correlation}.
% In this part, we repeated Section~\ref{section:6_comparison_of_time_sampling} with different amount of \jkimb{\temporalmapping} as described in \cref{figure:effect_of_correlation}.
% Compared to the default full \jkimb{\temporalmapping} of two paths, we increased the correlation intensity by increasing the number of correlated paths or decreased the correlation intensity by reducing the number of correlated rays.
\jkimb{First, we examine the impact of increasing correlation by using more than two shift-mapped paths ($N_t > 2$) (\cref{figure:effect_of_correlation}-(d)).}
For stratified sampling, we increase $N_t$ from $2$ to $1024$ while keeping the number of time strata fixed at $1024$.
%We utilize random correlation as we explain in \cref{section:4_subregion_stratified_sampling} for a fair comparison.
\wjarosz{For} antithetic sampling, we only consider the shifted method, as extending the mirrored method to $N_t > 2$ is not straightforward.
There \wjarosz{are} different ways to implement antithetic sampling with multiple samples. \wjarosz{We use a periodic approach analogous to uniform jittered sampling~\citep{Pauly:2000:Metropolis,Ramamoorthi:2012:Theory}:
%which is empirically observed to be optimal.
% In Section~\ref{section:4_subregion_stratified_sampling}, we listed several ways to implement antithetic sampling with multiple samples, but we use periodic sampling which is empirically observed to be optimal. (see supplementary for details). 
\wjarosz{instead} of a single antithetic shift with $0.5T$, we use $\left[\nicefrac{T}{N_t}, \nicefrac{2T}{N_t}, \ldots, \nicefrac{(N_t-1)T}{N_t} \right]$ where $N_t \in [2, 1024]$, and correlate these paths with shift mapping.}

% \jkimb{\temporalmapping} more than two paths corresponding to 
% is expected to be not that helpful because we already have efficient primal and antithetic pairs.
% However, an increased number of antithetic samples would be helpful for arbitrary heterodyne mode that does not have perfectly matching primal and antithetic pairs.
% To this end, we tested periodic sampling, which is a multiple sampled version of \shifted antithetic sampling.
% Instead of \shifted antithetic shift with $0.5T$, we use $\left[\frac{T}{N}, \frac{2T}{N}, ..., \frac{(N-1)T}{N} \right]$ where $N \in [2, 1024]$ and correlated these $N$ paths.
\begin{figure} [t]
\includegraphics[width=\linewidth]{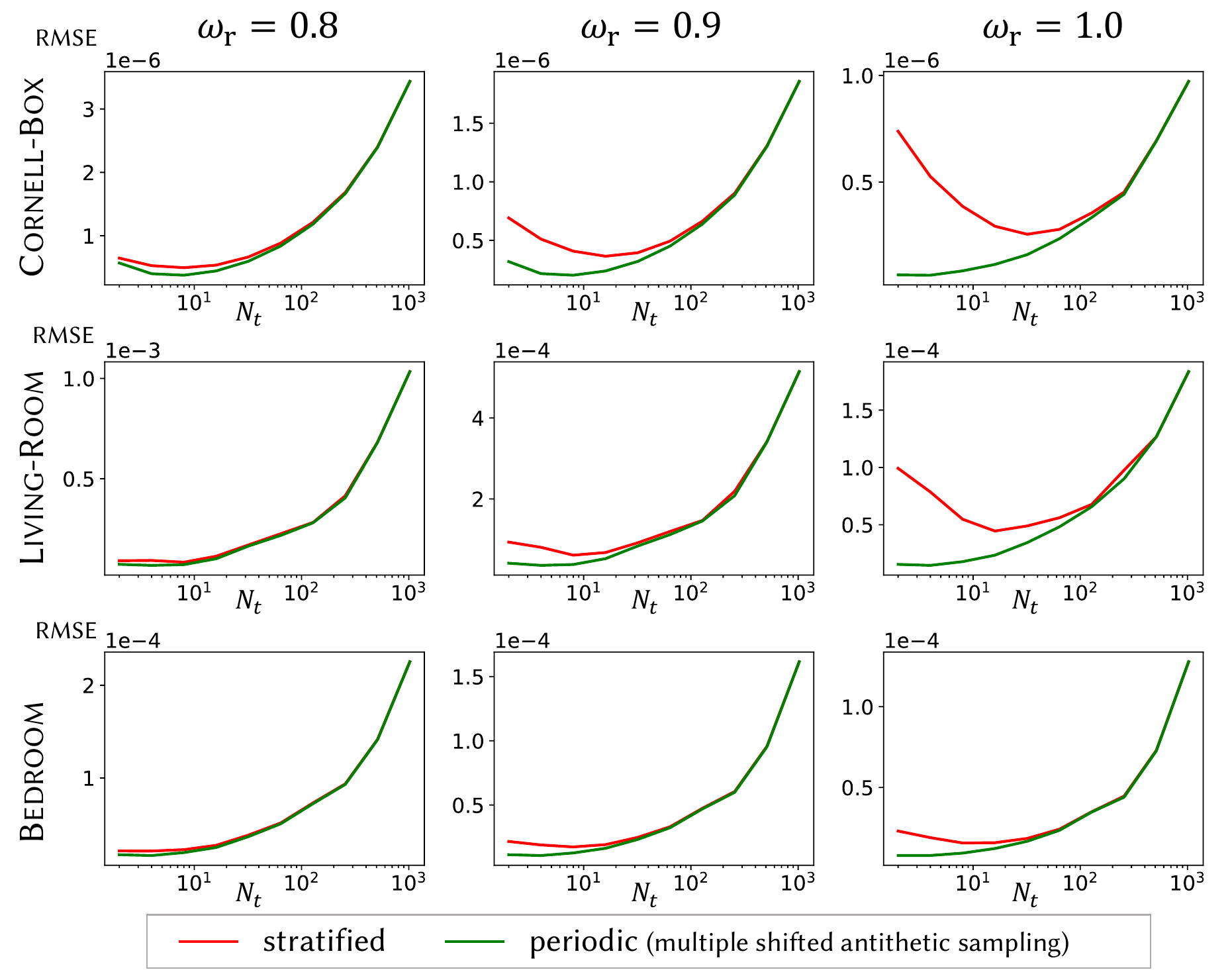}
\caption{RSME versus number of correlated paths $N_t$ \wjarosz{for the \textsc{Cornell-Box}, \textsc{Living-Room}, and \textsc{Bedroom} scenes. Even though the minimum RMSE for our method (green) sometimes occurs at $N_t>2$, using $N_t=2$ is close to the optimum and consistently outperforms stratified sampling (red).}}
\label{figure:result_multiple_path_correlation_plot}
\end{figure}

From \cref{figure:result_multiple_path_correlation_plot}, we observe that increasing $N_t$ \wjarosz{above $2$ generally} degrades performance,
% below $\omegaratio = 0.8$,
but can provide some improvement at high $\omegaratio$ values for stratified sampling. 
This improvement is because the variance of the modulation term dominates performance, thus dedicating more samples \wjarosz{to} the time domain is helpful. Too many correlated paths\wjarosz{, however,} result in \wjarosz{higher} variance due to limited path diversity.
Modulation variance increases with larger $\omegaratio$, so stratified sampling \wjarosz{achieves its minimum RMSE} for $N_t > 2$ as $\omegaratio \rightarrow 1$.
Periodic sampling shows a similar trend, but it \wjarosz{reaches its lowest RMSE at a much lower $N_t$ (usually $N_t =2$ or $N_t = 4$)}. Furthermore, performance at that point is better than stratified sampling.
The performance improvement is largest for perfect heterodyne mode, where periodic sampling with just two samples already gives near zero-variance estimation, thus making it better to allocate the remaining samples towards increasing path diversity. 
% Meanwhile, we could also observe some scene-dependent tendencies.
\wjarosz{Within the above general trends, there are some notable scene-dependent differences.}
\wjarosz{As scene complexity increases (\textsc{Cornell-Box} < \textsc{Living-Room} < \textsc{Bedroom}), the penalty from reduced path diversity increases, and using more correlated paths becomes more harmful}.
In summary, antithetic sampling with just $N_t = 2$ correlated paths was best for many cases and especially for perfect heterodyne mode.

% As we discussed before, increasing number of correlated path gives advantage for making an antithetic sampling in time domain more useful, but affects negatively in terms of path diversity.
% For most of frequencies, the disadvantage of latter one overwhelms the advantage of former one.
% At perfect heterodyne mode, using more correlated path appears to be somewhat helpful.
% But we could find the advantage only for stratified sampling and it even started to increase if $N$ becomes too large.
% For periodic sampling, increased number of correlated path was not useful at all.
% This infers that just using two antithetic sample is enough and it is better to dedicate more samples for path diversity.

% KKKKKKKKKKKKKKKK

\begin{figure}  [t]
\includegraphics[width=\linewidth]{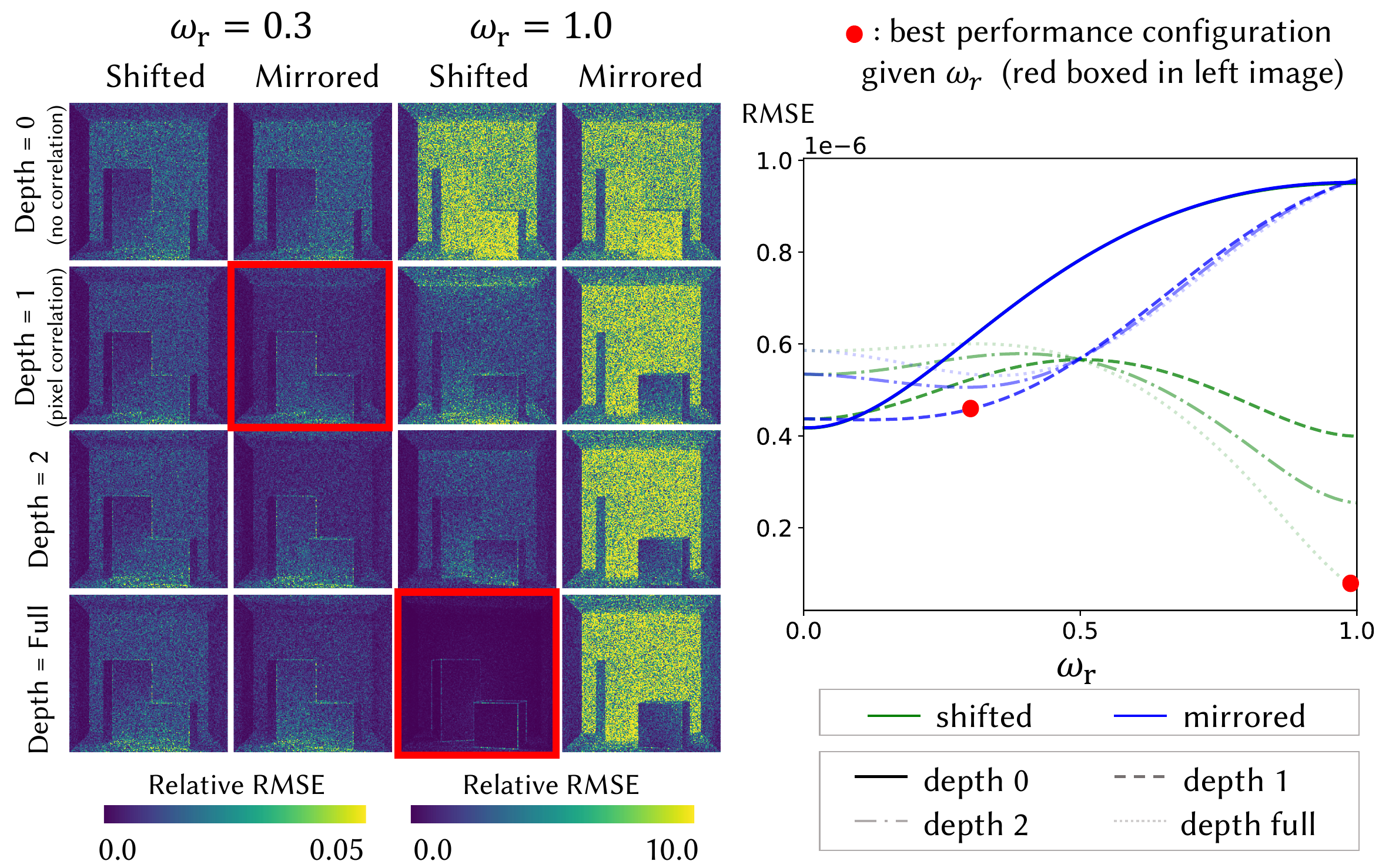}
\caption{\wjarosz{The} effect of using different \wjarosz{maximum} \jkimb{\temporalmapping} depths. Large correlation depths work best for higher $\omegaratio$, and worst for lower $\omegaratio$.}
\label{figure:result_path_correlation_depth}
\end{figure}

Second, we examine the impact of varying the maximum \jkimb{\temporalmapping} depth $K_d$ (\cref{figure:effect_of_correlation}-(a-c)).
% show the result of using different \jkimb{\temporalmapping} depths.
% The purpose of this experiment is somewhat similar to the previous section, to evaluate the effect of correlation.
% The difference is that the previous experiment is about inter-\jkimb{\temporalmapping}, this experiment is about intra-\jkimb{\temporalmapping}.
% For this, \wjarosz{we fix the number of correlated paths to $N_t=2$}. 
Maximum shift mapping depth $K_d = 0$ implies using no \jkimb{correlation at all}, $1$ implies camera ray (pixel) correlation, and so on until full path correlation (shift mapping). 
\Cref{figure:result_path_correlation_depth} shows qualitative results.
% As we discussed before, increasing correlation gives the advantage for making an antithetic sampling useful, but affects negatively in terms of path diversity.
% \jkim{rewrote this part}
For $\omegaratio\in[0.5,1.0]$, using larger $K_d$ gives better results, but the opposite is generally true for $\omegaratio\in[0.0,0.5]$. 
% The best performance for each region is by shifted antithetic sampling with full correlation, and mirrored antithetic sampling with pixel correlation, respectively.
This change occurs because the \wjarosz{variance of the modulation term} becomes less dominant as $\omegaratio \rightarrow 0$.
\rev{Interestingly,} completely abandoning variance from the modulation term ($K_d = 0$) gives the worst performance \jkim{except over a small region near $\omegaratio=0$}. 
\jkimb{This indicates that there should be the optimal balance between path correlation and diversity for $\omegaratio\in[0.0,0.5]$, but this seems to be highly scene-dependent, making it hard to find an absolute rule.}

\begin{figure} [b]
\includegraphics[width=0.9\linewidth]{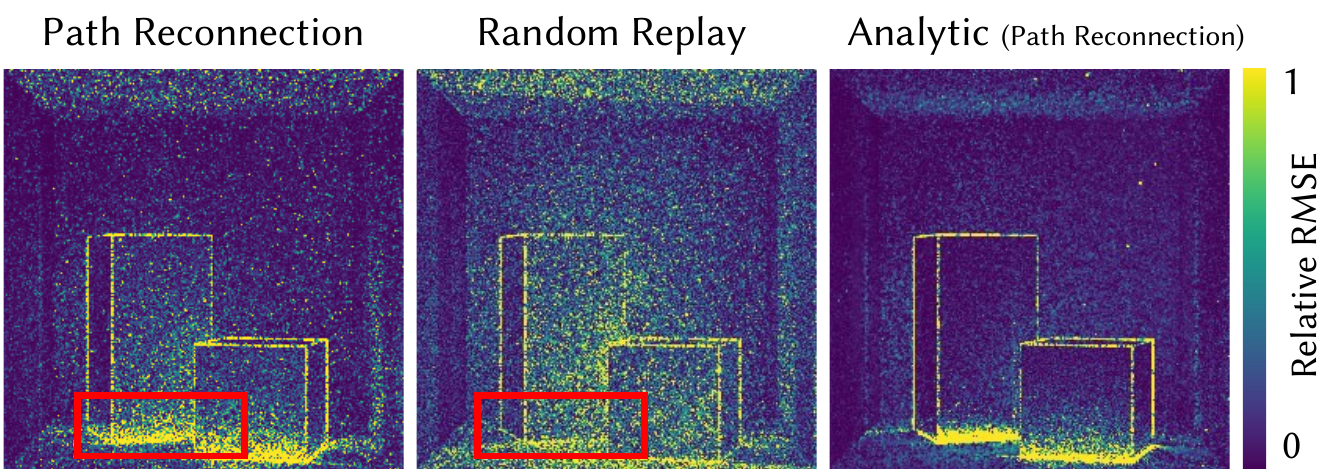}
\caption{Comparison of different \jkimb{\temporalmapping} methods. 
The scene is made up of diffuse materials, so \jkimb{path reconnection} works better than \jkimb{random replay}. 
However, it fails near the region highlighted in red. There, the lengths of interreflected rays are short, and thus \jkimb{path reconnection} causes a large change in $\fmappingxbart$, which increases variance.}
\label{figure:result_path_correlation_methods_detail}
\end{figure}

\begin{figure*} [t]
\includegraphics[width=\linewidth]{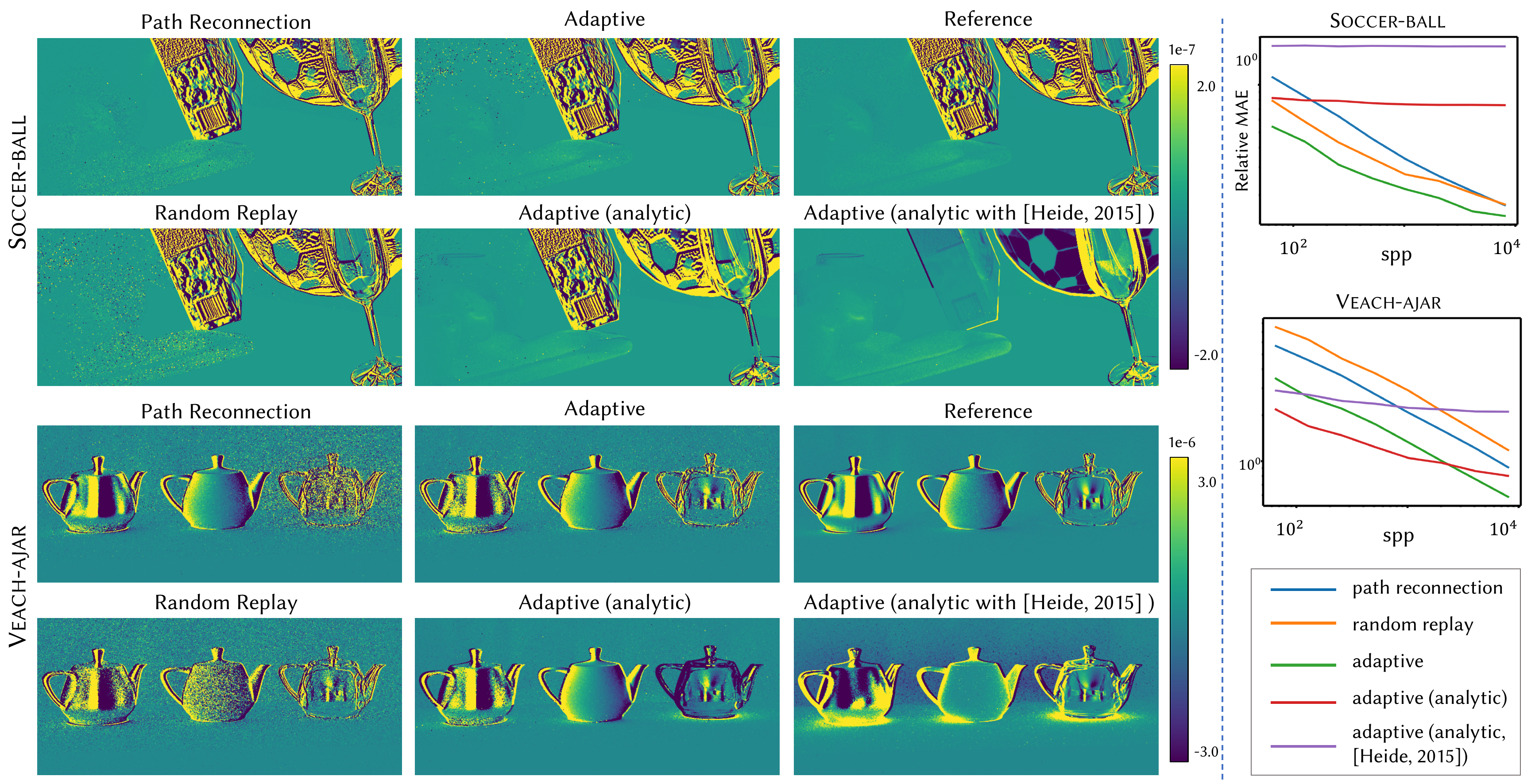}
\caption{Qualitative result and relative MAE using different \jkimb{\temporalmapping} strategies. We use MAE instead of RMSE because RMSE overestimated the failure case depicted in \cref{figure:result_path_correlation_methods_detail}. Among unbiased techniques, adaptive correlation works best in most cases. The integration-based method (first-order approximation) shows bias similar to \citet{Heide:2015:Doppler}, but has significantly lower error than their zero-order approximation of integrand.}
\label{figure:result_different_spatial_correlation}
\end{figure*}

\subsection{\jkimb{Shift} Mapping Strategy Comparison}
In \cref{figure:result_different_spatial_correlation}, we compare the different \jkimb{shift} mapping strategies we introduced in \cref{section:5_spatial_correlation}.
% Because implementation of \jkimb{\temporalmapping} tends to be biased, we also plotted error by spps. \apedired{What is biased?}
% Overall, there was no consistent superiority between positional and \jkimb{random replay}.
In general, if specular materials are dominant in the scene, \jkimb{random replay} works better; conversely, \jkimb{path reconnection} works better for scenes with mostly diffuse materials. 
However, exceptions can arise: In \cref{figure:result_path_correlation_methods_detail}, even though the scene is diffuse, there is a region where \jkimb{path reconnection} fails.
In this region, the length of interreflected rays is small, and thus geometric attenuation varies significantly for \jkimb{path reconnection}. 
% This is because the inter-reflection ray's distance is small in that region, and the discrepancy of $\fmappingxbart$ for primal and antithetic samples becomes extreme if we use \jkimb{path reconnection}.
Adaptive strategy generally works better than either \jkimb{path reconnection} or \jkimb{random replay}.
% However again, it also is affected by the specific scene property.
% Further study could be done to find a consistently best-performing correlation.
For all three methods, MAE (\cref{figure:result_different_spatial_correlation}, right column plots) decreases at the rate of $1/\sqrt{\text{spp}}$, showing that they are unbiased.
% \jkim{should we make another section for this? analytic is not that related to \jkimb{\temporalmapping} strategy.}

\subsection{Analytic Approximation}

In \cref{figure:result_different_spatial_correlation}, we include results to show the speed-up and bias of our analytic approximation. This approximation results in a less noisy image compared to Monte Carlo methods but shows noticeable bias in both the images and the MAE plot over spp. 
Our \wjarosz{first-order} approximation still gives better results compared to the \wjarosz{zeroth-order} approximation from \citet{Heide:2015:Doppler}, which demonstrates the importance of considering non-constant $\fmappingxbart$ over exposure time.
% This would be further discussed in next section.

\begin{figure} [t]
\includegraphics[width=\linewidth]{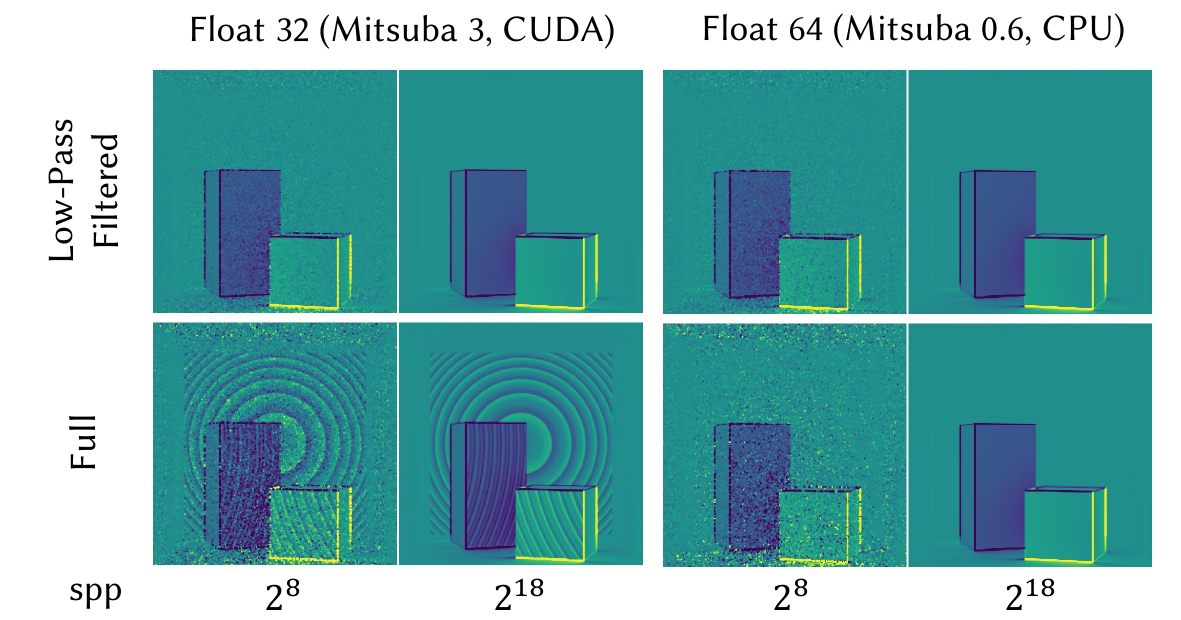}
\caption{
% Comparison with different precision level and error convergence.
We compare renderings at \texttt{float32} and \texttt{float64} precision, for the low-pass filtered and full D-ToF integral. The full rendering at \texttt{float32} has ringing artifacts, which we can resolve by increasing precision or removing high frequencies, at the cost of longer runtimes and bias, respectively.
}
\label{figure:precision_comparison}
\end{figure}

\subsection{High Frequency Terms and Precision Issue}
\label{section:high_frequency_term_precision}
As D-ToF rendering handles very small numerical values, such as $\nicefrac{v}{c}$, it is vulnerable to numerical precision issues.
We run our algorithm in both \qty{32}{bit} (Mitsuba 3, CUDA) and \qty{64}{bit} (Mitsuba 0.6, CPU) floating-point precision, and compare rendered results in \cref{figure:precision_comparison}. 
We also compare rendered results \jkimb{without ignoring the high-frequency terms in \cref{eq:modulation_term_full}}, which are most sensitive to such precision issues.
% The result is plotted in \cref{figure:precision_comparison}.

In the \jkimb{full signal rendering without low-pass filtering}, we notice strong ringing artifacts when using \qty{32}{bit} precision, which persists as we increase the number of samples. We found that scaling up the scene helps resolve these artifacts---generally, a $10-100\times$ scaling was effective, but we could not increase the scale arbitrarily due to intersection precision problems.
\jkimb{For \qty{64}{bit} precision, such artifacts do not occur, but noise is increased universally.
Low-pass filtering the modulation term as \cref{eq:product_of_modulation_functions} resolves ringing artifact and noise with the cost of negligible bias ($ < 1\%$) for both precisions.
}
%These artifacts disappear also when using \qty{64}{bit} precision (at the cost of longer runtimes), as well as when ignoring high-frequency terms (at the cost of bias).
% There still was a small difference between \wjarosz{the float32 and float64 low-pass} filtered images, but it was quite small ($ < 1\%$).
% To summarize, if the user wants to render images in an acceptable error range with faster speed, the CUDA float32 version with low-pass filtered signal will be a good choice as it is about $10$--$100 \times$ faster than the CPU version.
% On the other hand, if high image precision is important, then simulation with double precision would be a possible option.

\subsection{Area Light Sources}
\begin{figure}[t]
\includegraphics[width=0.85\linewidth]{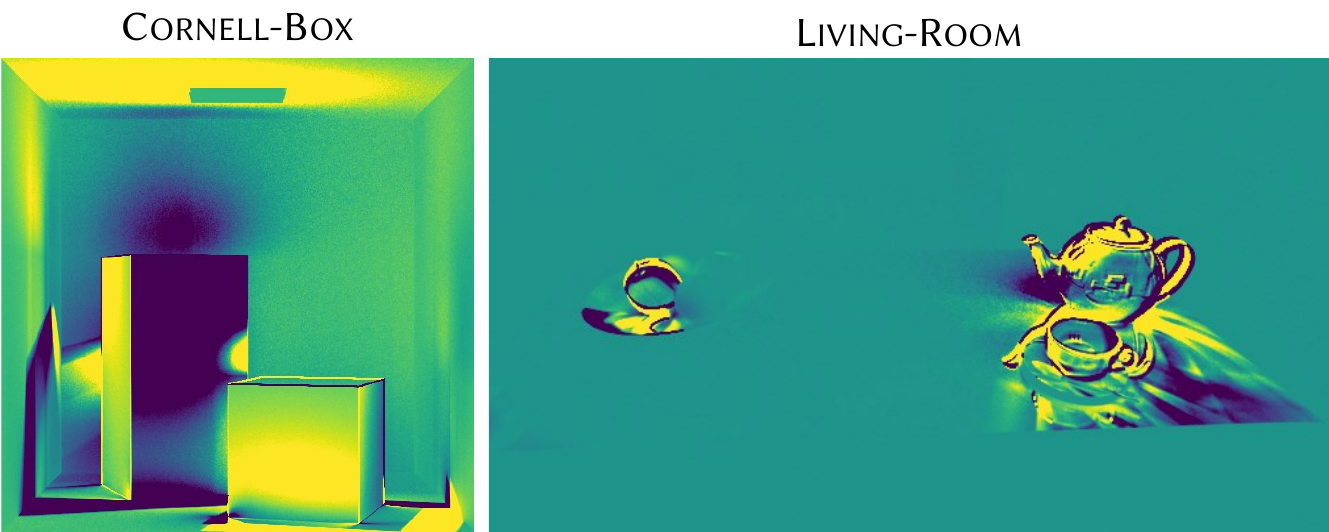}%
\caption{An example scene using area light source demonstrating that the renderer can handle a variety of light sources along with material properties. }
\label{figure:result_area}
\end{figure}

To demonstrate that our algorithm can handle a variety of light sources and materials, in \cref{figure:result_area} we render \wjarosz{the} \textsc{Cornell-box} and \textsc{Living-room} scenes with area light sources. 
The figure illustrates the complex interplay of diverse light paths.
Overall, it took more time ($16$--$64\times$) to converge compared to a point light.

%% file: sections/07_applications_v3.2.tex
\section{Application to Radial Velocity Estimation}
\label{section:7_applications}
In this section, we reproduce the radial velocity estimation algorithms from previous D-ToF imaging systems. 
We identify conditions under which these algorithms fail and explain the failure cases. With the help of our D-ToF rendering engine, we identify new imaging system parameters that, if realized experimentally, can make the  velocity estimation more accurate. 

\begin{figure} [b]
\includegraphics[width=\linewidth]{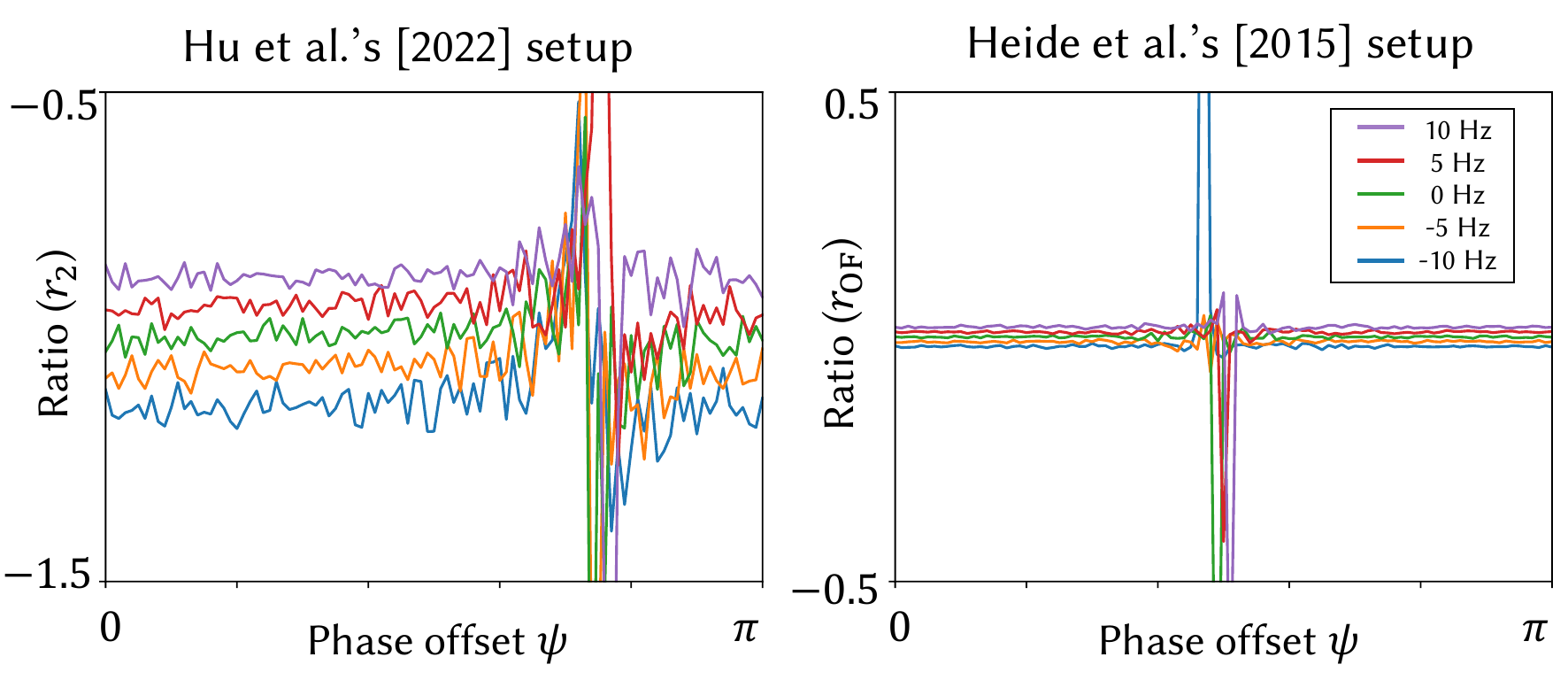}
\caption{We reproduce the hardware results from \citet{Hu:2022:Differential} with our renderer. This figure illustrates that \citeauthor{Hu:2022:Differential}'s technique can \wjarosz{separate Doppler frequencies better} than \citeauthor{Heide:2015:Doppler}'s~[\citeyear{Heide:2015:Doppler}] technique.}
\label{figure:hu_2022_comparison}
\end{figure}

% \subsection{Velocity Estimator}
% Doppler ToF imaging is mainly used for velocity estimation in practice.
\subsection{Reproducing Existing D-ToF Imaging System}
We first use simulation to reproduce the systems and algorithms of \citet{Heide:2015:Doppler} and \citet{Hu:2022:Differential} in virtual experiments, using the same experimental setup as \citeauthor{Hu:2022:Differential} ($\lumfreq = \qty{74}{MHz}$, $T=\qty{2}{ms}$) \jkim{with collocated sensor and point light source}. 
% , we measured the ratio between some ToF images.
For \citeauthor{Heide:2015:Doppler}, \wjarosz{we use the ratio of homodyne ($\omegaratio = 0.0$) to heterodyne ($\omegaratio = 1.0$) measurements} to estimate radial velocity; for \citeauthor{Hu:2022:Differential} \wjarosz{we use the ratio of $\omegaratio = 0.6625$ (optimal for sinusoidal) measurements} with four different $\psi$ offsets. 
We tested five different $\omegadelta$s, $\qtylist{-10; -5; 0; 5; 10}{Hz}$ over $\psi \in [0, \pi]$.
\Cref{figure:hu_2022_comparison} shows the simulated result, which is similar to Fig.~8 from~\citet{Hu:2022:Differential}. 
The measurement ratio of \citeauthor{Hu:2022:Differential}'s ($r_2$) shows a larger margin compared to \citeauthor{Heide:2015:Doppler}'s ratio ($r_\mathrm{OF}$), which implies that \citeauthor{Hu:2022:Differential}'s technique can better separate different Doppler frequencies. 
% This infers \citet{Hu:2022:Differential}'s method is more robust against Poisson noise.
For more details, we refer to \citet{Hu:2022:Differential}.
\jkim{We note that we did not model the sensor noise profile. The noise in \cref{figure:hu_2022_comparison} is Monte Carlo noise, not simulated sensor shot noise.}

For both methods, \wjarosz{however,} we found that velocity \wjarosz{estimation} works well only for certain offsets (\cref{figure:hu_2022_comparison_velocity}).
For others, the reconstructed velocity direction can even be reversed.
\wjarosz{One interpretation is that such deviations are simply variance, but next we provide a better interpretation in terms of the dynamic ToF path integral}. 
% \citet{Hu:2022:Differential} termed such deviation as variance, but using the rendering equations, we could give a better interpretation. 

\subsection{Limitations of Velocity Estimation Algorithms}
% In the rest of the section, 
Existing velocity estimation algorithms for D-ToF imaging systems~\citep{Heide:2015:Doppler,Hu:2022:Differential} assume that the object normal is parallel to the viewing direction, the light sources are directional, and global illumination is absent. When these assumptions fail, the velocity estimation algorithms become inaccurate. In this section, we use our rendering algorithm to investigate each factor in turn. 
% First consider a simple case of single-bounced ray.

% We will next discuss the limitations of velocity calculation algorithms proposed in \citet{Heide:2015:Doppler} and \citet{Hu:2022:Differential} and some methods to improve it.
% We will first consider single-bounced light rays only. 
% To make the  easier, we will consider single bounced case only with point light source.
% Because of some oversimplified physical rules in these works, velocity turned out to be valid in only limited cases during our simulation on general scenes.

\begin{figure} [t]
\includegraphics[width=0.95\linewidth]{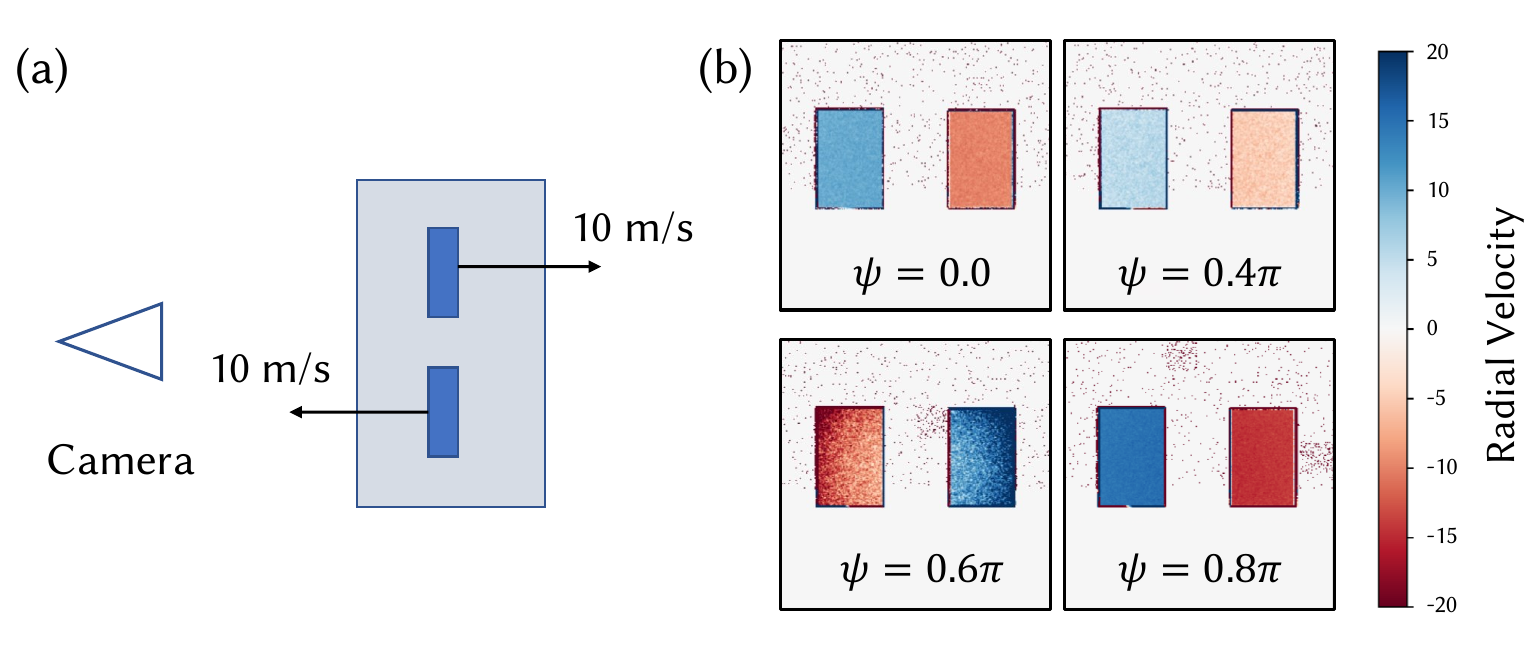}
\caption{We observe that for the same scene configuration shown in (a), the reconstructed radial velocity in (b) varies significantly for different phase offset $\psi$ values. Only the left top corner ($\psi=0$) gives the correct velocity. 
\jkimb{We only show result of \citet{Heide:2015:Doppler}, but similar for \citet{Hu:2022:Differential}.}
%This happens due to singularity at certain offset values.
\apedired{show the $\psi$ values as insets}}
\label{figure:hu_2022_comparison_velocity}
\end{figure}

\begin{figure} [b]
\includegraphics[width=0.95\linewidth]{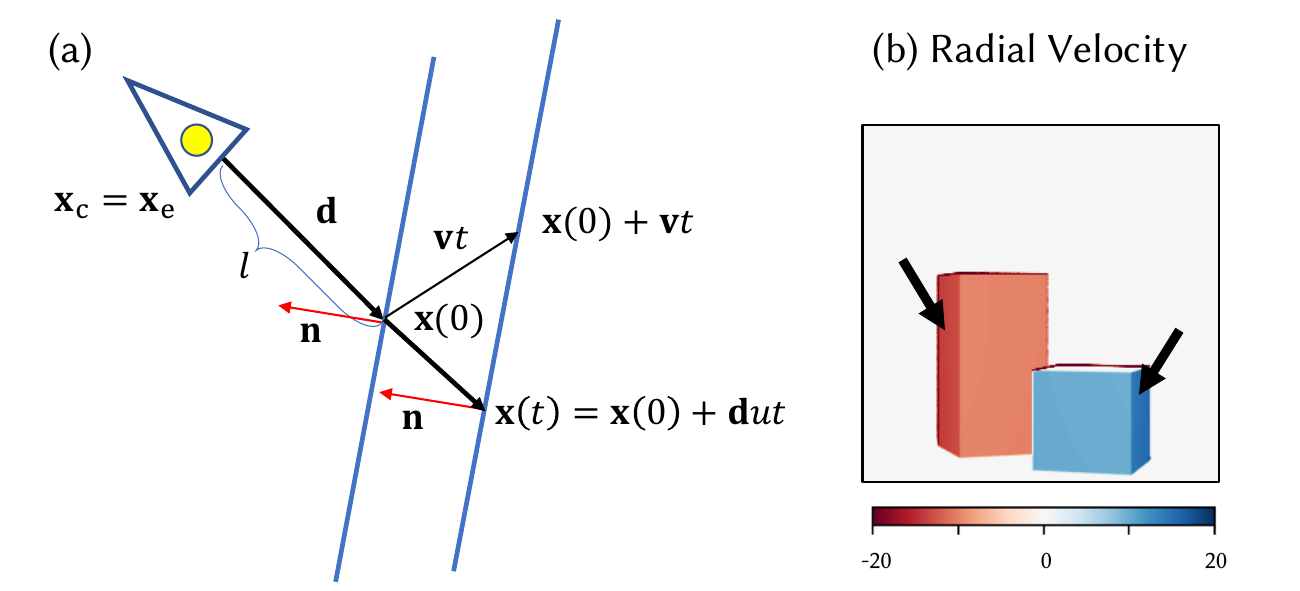}
\caption{Effective radial velocity for a planar patch. The Doppler imaging algorithms assume that the surface normal of the object aligns with the viewing direction. However, for a general planar patch shown in (a), the effective radial speed $u = {(\mathbf{n} \cdot \mathbf{v})}/{(\mathbf{n} \cdot \mathbf{d})}$. This results in large radial velocity estimates for objects viewed at grazing angle as shown in (b). 
% \apedired{Juhyeon: Please add ground truth image as right column figure here. }
}
\label{figure:velocity_calculation_concept}
\end{figure}

\paragraph{Effective Radial Velocity}
% First, we have to notice that what we can effectively calculate from Doppler imaging is slightly different from the general meaning of radial velocity.
First, we clarify the exact definition of the velocity that Doppler imaging estimates. 
We consider \cref{figure:velocity_calculation_concept}-(a), where the camera $\mathbf{x}_{\mathrm{c}}$ and point light source $\mathbf{x}_{\mathrm{e}}$ are collocated, and locality is preserved over single-bounce paths $\mathbf{x}_{\mathrm{c}}\mathbf{x}(t)\mathbf{x}_{\mathrm{e}}$. Then:
\begin{equation}
    \mathbf{x}(t) = \mathbf{x}(0) + \mathbf{d} u t \, \text{ and } (\mathbf{x}(0) + \mathbf{v}t - \mathbf{x}(t)) \cdot \mathbf{n} = 0,
\end{equation}
\jkim{where $\mathbf{d}$ is the ray direction, $\mathbf{n}$ is the surface normal, $\mathbf{v}$ is the object velocity, and $u$ is the \emph{effective radial speed} in direction $\mathbf{d}$ which equals}
% \begin{equation}
%     (\mathbf{x}(0) + \mathbf{v}t - \mathbf{x}(t)) \cdot \mathbf{n} = 0
% \end{equation}
% Solving these two equations, we can calculate $u$ as
%
\begin{equation}
    u = \frac{(\mathbf{n} \cdot \mathbf{v})}{(\mathbf{n} \cdot \mathbf{d})}.
\end{equation}
This is different from the actual \emph{radial speed}, $(\mathbf{v} \cdot \mathbf{d})$. We refer to the velocity vectors associated with these speeds as the effective radial velocity and radial velocity, respectively. Therefore, the effective radial speed is equal to the radial speed only if the object normal aligns with the viewing direction. At grazing angles (\cref{figure:velocity_calculation_concept} (b)), the effective radial speed previous techniques report could be larger than the object's radial speed (side of the box in $\textsc{Cornell-box}$). 

\begin{figure} [t]
\includegraphics[width=\linewidth]{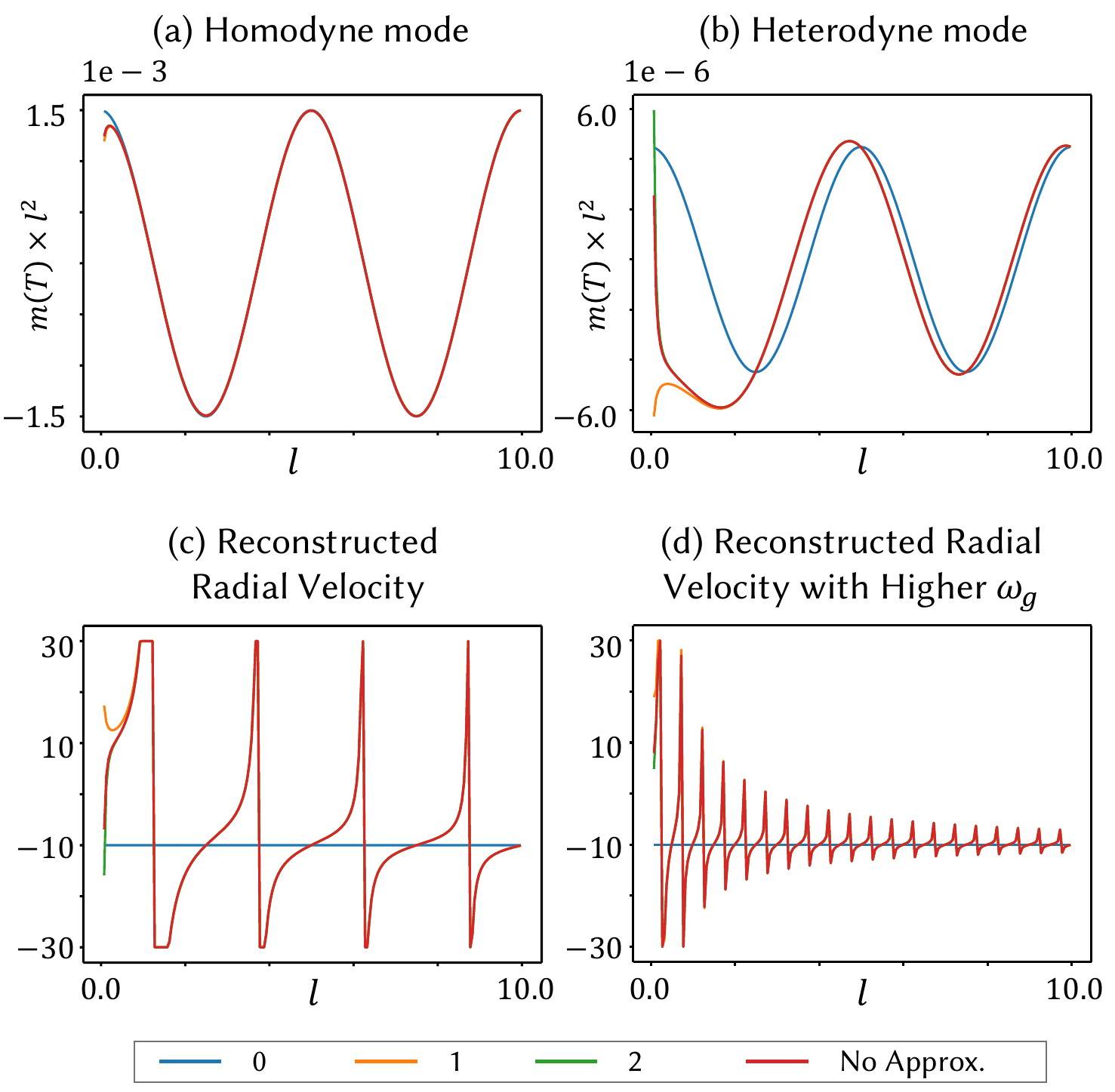}
\caption{
Effect of geometric attenuation term. \jkim{We multiplied $m(T)$ by $l^2$ to aid visualization.} We show the integrated radiance for homodyne and heterodyne mode for 0, 1, 2-order Taylor series approximations of the geometric attenuation term. Previous velocity computation techniques derived velocity expressions assuming 0-order approximation for the forward model. These algorithms fail when the approximation is not valid. 
% Integrated radiance for a simple 1-d case / cornell and calculated velocity from homodyne and heterodyne image.
}
\label{figure:velocity_calculation_demo_plot}
\end{figure}

\paragraph{\jkimb{Time-varying Geometric Attenuation}}
We proceed to analyze $\widehat{f}$ in \cref{eq:doppler_rendering_path_integral_reordered} for single-bounce paths in \cref{figure:velocity_calculation_concept}-(a).%with a point light source with unit intensity.
The distance toward the object at $t=0$ is $l$ and the object is moving away from the camera with effective radial speed $u$.
Then we can write $\widehat{f}$ in \cref{eq:doppler_rendering_path_integral_reordered} as
% \begin{equation}
%     m(T) = \int_{0}^{T} \camFcos{t} \left(g_1 \cos{\left(\omega_g \left(t+\frac{2(l+vt)}{c}\right) \right)} + g_0 \right) \frac{1}{(l+ut)^2} \text{d}t
% \end{equation}
%
\begin{equation}
    \fmappingxbart=\frac{f_r(\mathbf{x}(t), -\mathbf{d}, -\mathbf{d})(\mathbf{n} \cdot -\mathbf{d})}{\left(l + ut\right)^2}
\end{equation}
%
% \begin{equation}
%     m(T) = \int_{0}^{T} \cosP{\cP{\camfreq - \lumfreq + \lumfreq\frac{2u}{c}}t + \lumfreq \frac{2l}{c} } \frac{f_r(\mathbf{x}(t), -\mathbf{d}, -\mathbf{d})}{\left(l + ut\right)^2} \text{d}t, 
% \end{equation}
where $f_r$ is the BRDF.
\jkim{We assume $f_r$ and $\mathbf{n}$ are constant over $t$, so that only the geometric attenuation term $\left(l + ut\right)^{-2}$ affects $\widehat{f}$.}
Both \citet{Heide:2015:Doppler} and \citet{Hu:2022:Differential} assume a constant geometric attenuation term and integrate cosine modulation terms over time, which is accurate if either the distance $l$ is large, the light source is directional, or the cosine frequency is small (homodyne). 
In terms of Taylor series expansion, this is a zeroth-order approximation of $\left(l + ut\right)^{-2}$. 
On the other hand, our approximate analytic expression uses first-order approximation. 
To validate our approximation, we calculated $m(T)$ with different approximation orders.

\Cref{figure:velocity_calculation_demo_plot}-(a,b) show analytically integrated $m(T)$ for homodyne and heterodyne modes. 
The red line is the ground truth value, while lines with labels $0, 1, 2$ represent different order Taylor-series approximations.
For homodyne mode, even the zeroth-order Taylor approximation is accurate and hence, CW-ToF cameras that use homodyne for depth measurement do not suffer from this problem.  
However, there is a lot of deviation in the heterodyne mode, especially when we use zeroth-order approximation. 
The first-order approximation also fails at close distances, but overall, it is significantly more accurate than the zeroth-order, which explains why our analytic integration outperforms \citet{Heide:2015:Doppler}.

\Cref{figure:velocity_calculation_demo_plot}-(c) shows the result for radial velocity calculation.
% Using zeroth-order approximation gives the exact true value ($-10 m/s$), but notice that this is not from physically true values.
% The physically accurate $m(T)$ gives significantly deviated velocity with many discontinuities.
% This is the reason why existing technique fails for certain offsets in \cref{figure:hu_2022_comparison_velocity}.
The physically accurate $m(T)$ gives significantly deviated velocity with many discontinuities, implying that velocity computation techniques fail \jkimb{when geometric attenuation (or in general, path throughput $\fmapping$) varies over exposure time.}
%
%
%for point light sources. 
This is why the existing ratio-based techniques fail for some $\psi$s in \cref{figure:hu_2022_comparison_velocity}.

\begin{figure} [b]
\includegraphics[width=\linewidth]{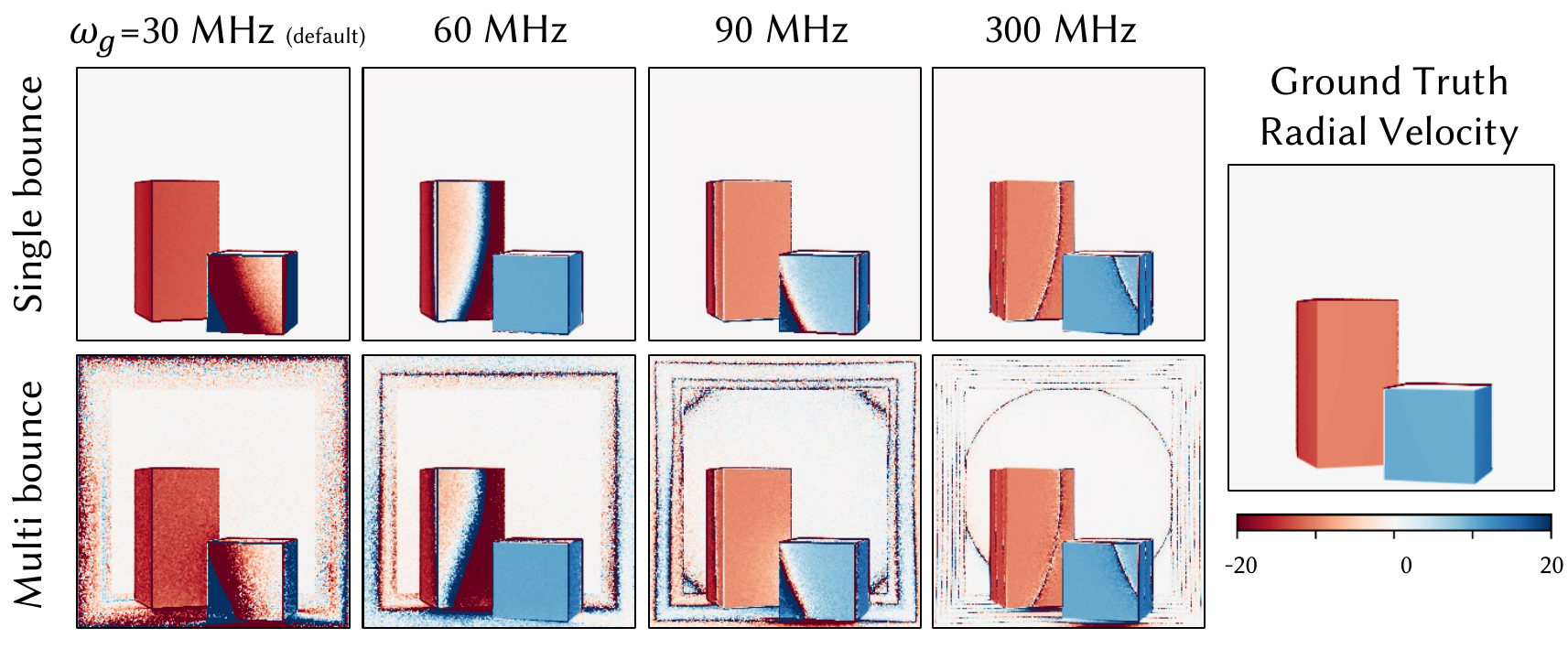}
\caption{
In the presence of global illumination (multi-bounce), velocity estimation algorithms fail. We observed, however, that using higher modulation frequency illumination diminishes the global illumination effects.}
\label{figure:velocity_calculation_high_freq}
\end{figure}

\paragraph{Global Illumination}
Another scenario where the velocity estimation fails is in the presence of strong global illumination. 
In \cref{figure:velocity_calculation_high_freq}, we show how global illumination results in inaccurate velocity estimates. 
\jkim{
Quantifying the impact of global illumination (usually termed \textit{multi-path interference} (MPI) in ToF literature) is crucial for ToF applications, as evinced by the extensive prior work on suppressing MPI in CW-ToF imaging~\cite{Whyte:2015:Resolving, Kadambi:2013:Coded}.
}

\subsection{Improving Radial Velocity Calculation}
% \apedired{needs more work}
% We have found some methods to improve the quality of velocity calculation.
To mitigate the challenges imposed by time-varying geometric attenuation and global illumination, we propose a few solutions that require operating the D-ToF camera at high frequency, which is feasible with recent advances in CW-ToF cameras~\cite{Baek:2023:Centimeterwave}. 

\paragraph{Mitigating Effects due to Time-varying Geometric Attenuation}
One way to mitigate the effects of time-varying geometric attenuation on velocity estimation is to increase the working distance or use a directional light, both of which make geometric attenuation (approximately) constant.
Another approach is to increase the modulation frequency $\lumfreq$. 
From \cref{figure:velocity_calculation_demo_plot} (d), we observe that increasing this frequency makes the velocity estimation error small over a larger working distance range.
% It also turned out to be helpful in rendering cases (\cref{figure:velocity_calculation_high_freq}, \cref{figure:bunny_translate_rotate_scale}).
% One drawback of increasing frequency is that the speed range which has a linear relation between homodyne and heterodyne becomes smaller.
% However, Considering Doppler shift is $\frac{v\omega_g}{c}$, it could be easily accepted.
% Another problem of using higher frequency is that discontinuity tends to appear more, but this will be less noticeable compared to low-frequency cases.
% Also if we use high frequency, we found that the multibounce effect could is alleviated (\cref{figure:velocity_calculation_high_freq}, \cref{figure:bunny_translate_rotate_scale}) because modulations of multi-bounced paths tend to be canceled out~\cite{Nayar:2006:Fast}.
However, increasing modulation frequency makes homodyne mode \wjarosz{become zero more} frequently, which causes numerical precision problems resulting in noticeable ringing artifacts.
\jkim{We can resolve these problems by using two different phases which have zero homodyne values at different positions (\cref{figure:more_scenes}) and combining them as in the \emph{six-sample method} in \citet{Hu:2022:Differential}.}
%Also, although we did not focus on distance estimation, increasing modulation frequency reduces maximum detectable distance.

\begin{figure} [t]
\includegraphics[width=0.95\linewidth]{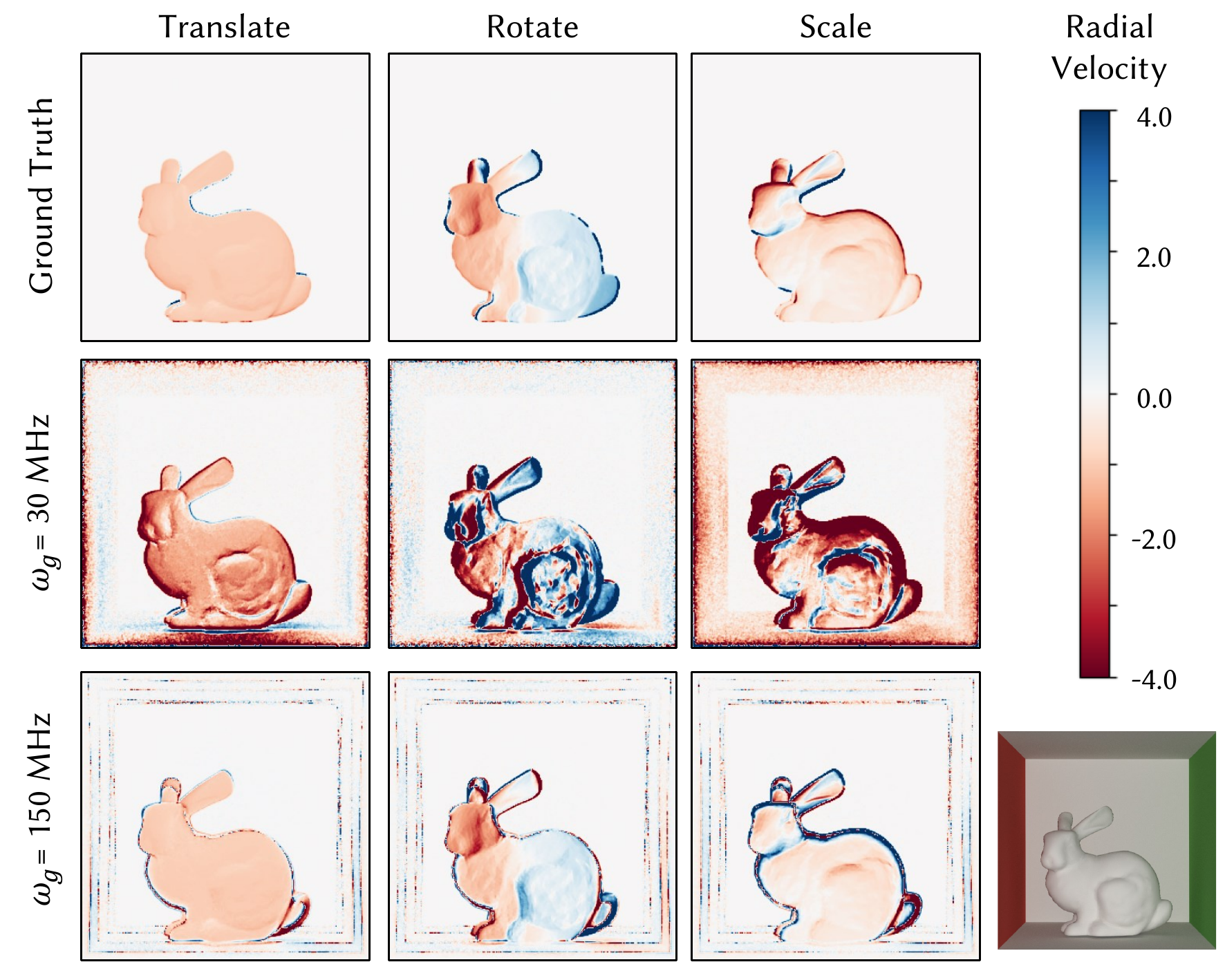}
\caption{For various object motions (translation, rotation, scaling), we observed that global illumination significantly affects velocity computation at low illumination frequency but not at high illumination frequency.}
\label{figure:bunny_translate_rotate_scale}
\end{figure}

\begin{figure} [t]
\includegraphics[width=\linewidth]{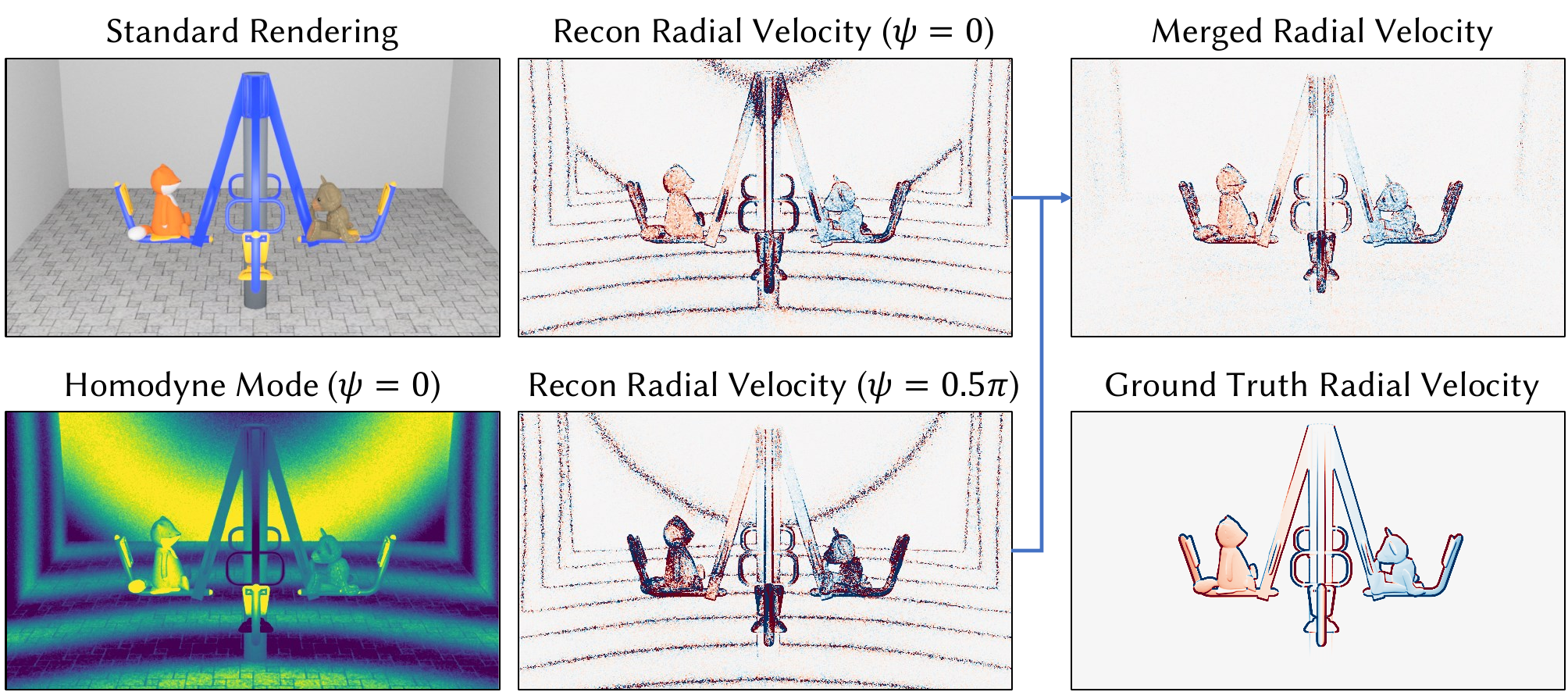}
\caption{
\jkim{Homodyne mode has zero values at certain positions which causes ringing artifact in velocity reconstruction. This could be improved by using two different phases and mixing them.
The rendering time for this scene with \qty{22}{\kilo\nothing} triangles and $N=1024$ is $\qty{4}{sec}$ per frame.
}
%Our rendering algorithm and velocity reconstruction techniques scale for complex scene geometry. The rendering time for this scene with 22k triangles and $N$=1024 is $\unit[4]{sec}$ per frame. We included more videos in supplementary with animation.
}
\label{figure:more_scenes}
\end{figure}

\paragraph{Mitigating Global Illumination}
We found that using higher frequencies also mitigates global illumination effects. 
In \cref{figure:velocity_calculation_high_freq}, we show that increasing the modulation frequency reduces the inaccuracy of velocity calculation due to global illumination.
\Cref{figure:bunny_translate_rotate_scale} shows another example under various scene motions.
\jkim{This behavior is because contributions from multi-bounce paths tend to cancel out as modulation frequency increases.}
The same effect is already known in structured light~\cite{Nayar:2006:Fast} and CW-ToF~\cite{Gupta:2015:Phasor} imaging systems and it is exciting to see that it also holds true for D-ToF imaging systems.

% This phenomenon is known in structured light~\cite{Nayar:2006:Fast} and CW-ToF~\cite{Gupta:2015:Phasor} imaging systems, and it is exciting to see that it holds true for D-ToF cameras as well. 
% In \cref{figure:bunny_translate_rotate_scale}, we observe the same phenomenon for various scene motions (translation, rotation, scaling).
% \jkim{In \cref{figure:more_scenes}, we ran our rendering algorithm and velocity reconstruction techniques on more complex scene geometries and the observations hold true independent of scene complexity. We provided more videos in the supplementary.}

\jkim{Finally, we run our rendering algorithm and velocity estimation techniques on more complex scene geometries under several frames and animate the results (\cref{fig:teaser}, \cref{figure:more_scenes}).
Our algorithm reliably and efficiently reproduces both D-ToF images and radial velocity.
We provide more videos in the supplement.}

% First and second row of \cref{figure:geom_atten_improve} shows the improvement of each method respectively.

% \begin{figure}
% \includegraphics[width=0.5\textwidth]{figures/7_applications/img_1d_case.png}
% \includegraphics[width=0.5\textwidth]{figures/7_applications/img_render_case.png}
% \caption{Integrated radiance for a simple 1-d case / cornell and calculated velocity from homodyne and heterodyne image.}
% \label{figure:geom_atten}
% \end{figure}

% \begin{figure}
% \includegraphics[width=0.5\textwidth]{figures/7_applications/img_improve_increase_freq_1d.png}
% \caption{Calculated velocity with using a higher frequency.}
% \label{figure:geom_atten_improve_freq_1d}
% \end{figure}

% \begin{figure} [t]
% \includegraphics[width=\linewidth]{figures/7_applications/velocity_calculation_high_freq_cornell_v1.3.pdf}
% \caption{
% In presence of global illumination (multi bounce), the velocity calculation algorithms fail. However, we observed that the global illumination effects are significantly diminished if we use higher modulation frequency illumination.}
% \label{figure:velocity_calculation_high_freq}
% \end{figure}

% However, these methods are still not enough.
% We did not consider other factors that affects velocity calculation such as strong indirect illumination or largely varying normal and BSDF.
% In this paper, we do not propose a scalable velocity calculation method for multi-bounce case.
% However, this could be further studied in a future work using our physically based simulator framework.

% \subsection{Velocity Reconstruction on More Scenes}
% Finally, we show the experimental results for more  scenes.
% \cref{figure:bunny_translate_rotate_scale} shows the result of translate, rotate and scale.
% Using our renderer with higher illumination frequency shows accurate velocity reconstruction.
% Fig~\ref{figure:more_scenes} shows the results on more interesting scenes.
% Please check the videos in supplementary material for results on a wide-range of scenes. 

% Let's start from a simple case that an object is moving  collocated light.

% In this section, we will review the limitation of \cite{Heide:2015:Doppler} and propose some better techniques for velocity estimation.

%% file: sections/conclusions.tex
\section{Conclusion}

%One paragraph summary
We developed a D-ToF Monte Carlo rendering framework and tailored sampling techniques for efficiently simulating D-ToF cameras. 
Using our open source implementation, we showed that our techniques provide orders of magnitude improved performance \jkim{compared to naive sampling techniques}, under various illumination and sensor modulation functions. 
We additionally reproduced in simulation previously-reported results from real D-ToF hardware systems~\citep{Heide:2015:Doppler,Hu:2022:Differential}, and investigated the accuracy of their velocity estimation on a variety of scenes. 

\rev{Our work suggests several directions for future research. Recent advances in path reuse and shift mapping techniques, for both static and animated scenes~\citep{Bitterli:2020:Spatiotemporal,Sawhney:2022:Decorrelating}, could be adapted into temporal mapping techniques for D-ToF rendering, to improve rendering performance for challenging scenes or more general modulation waveforms. Additionally, our D-ToF path integral framework invites the application to D-ToF rendering of other Monte Carlo algorithms, such as Markov chain Monte Carlo algorithms that have proven effective also for the related motion blur rendering problem \citep{Li:2010:Anisotropic,Luan:2020:Langevin}. }

% Differentiable rendering
Differentiable D-ToF rendering is another intriguing future research direction, which could facilitate the design of D-ToF imaging systems and related inverse rendering applications. 
The D-ToF integral in \cref{eq:dynamic_tof_path_integral} is differentiable when the illumination and exposure codes are differentiable. 
However, as the integrand can become negative, building efficient path sampling techniques is non-trivial~\citep{Zhang:2021:Antithetic,Chang:2023:Parameter}. Additional challenges arise for discontinuous illumination and exposure waveforms (e.g., square): differentiating \cref{eq:dynamic_tof_path_integral} then results in singularities in the time-domain, which require sampling on delta manifolds similar to those studied by \citet{Pediredla:2019:Ellipsoidal} and \citet{Wu:2021:Differentiable}. 
% \jkim{Although we did not model specific sensor noise characteristics, further studies could be done on realistic sensor model}

\rev{
As light-based velocity sensing becomes commonplace in critical applications (autonomous vehicles, robotic navigation, remote sensing), we expect that our work will inspire rendering research for other common or emerging technologies for this sensing modality. An example is rendering frequency-modulated continuous-wave (FMCW) ToF sensors~\citep{Qian:2022:Videorate}: even though their operation is also based on the Doppler effect and correlation measurements, these sensors are interferometric~\citep{Fercher:2003:Optical,Kotwal:2020:Interferometric}, thus simulating them requires rendering challenging wave effects~\cite{Steinberg:2021:Generic,Bar:2019:Monte}.
}

% \jkim{
% Our work is on the intersection of hardware and software, it could be further researched in each direction.
% For imaging system researchers, considering more realistic Poisson noise characteristics and incorporating it into our renderer would be an interesting following work.
% On the other hand, rendering researchers would be interested in figuring out better sampling techniques for more generalized modulation signals, or better path correlation strategies as ~\citet{Lin:2022:Generalized} proposed.
% }

% What applications can a D-ToF renderer be useful for? 
%generic 
Last but not least, our open-source simulator can facilitate research and engineering efforts toward designing and optimizing all aspects of future D-ToF computational imaging systems. Examples include the design of new sensor architectures, D-ToF modulation functions, and velocity estimation algorithms that are robust to noise and global illumination. In this context, our simulator can act as a digital twin that enables quick prototyping, supervised data generation, quantitative evaluation, and even end-to-end optimization of both hardware and software components of a real D-ToF computational imaging pipeline. Towards realizing these applications, it will be important to research and incorporate into our renderer realistic sensor noise models for D-ToF imaging.

% it can help generate datasets for 
% Sensor design is another field where our physically based framework can help.
% One of the challenges in developing deep neural networks for futuristic computational cameras is the lack of real data. 
% Our simulator can act as a digital twin in these cases enabling end-to-end training. 

\begin{acks}
We thank the anonymous reviewers for their feedback and especially Anonymous Reviewer 4 for the idea to use scene scaling in \cref{section:high_frequency_term_precision} to mitigate the ringing artifacts due to limited floating-point precision. We also thank the authors of \citet{Hu:2022:Differential} for helpful discussions on \cref{section:7_applications}, and in particular correcting an error in \cref{figure:velocity_calculation_demo_plot}. Wojciech Jarosz was supported by NSF award 1844538, Ioannis Gkioulekas by NSF awards 1730147, 1900849, and a Sloan Research Fellowship, and Adithya Pediredla by a Burke research initiation award.
\end{acks}